\documentclass[final]{appolb}

    \usepackage{latexsym,bm,amsmath,amssymb,amsfonts}
    \usepackage{epsfig,graphics,graphicx}
    \usepackage{makeidx}
    \usepackage{citesort}
    \usepackage{slashed}

    \newcommand{\PLB}{{\it Phys. Lett. }{\bf B}}
    \newcommand{\NPA}{{\it Nucl. Phys. }{\bf A}}
    \newcommand{\NPB}{{\it Nucl. Phys. }{\bf B}}
    \newcommand{\PRep}{{\it Phys. Rept.}}
    \newcommand{\PPNP}{{\it Prog. Part. Nucl. Phys.}}
    \newcommand{\CPC}{{\it Comput. Phys. Commun.}}

    \newcommand{\PRL}{{\it Phys. Rev. Lett.}}
    \newcommand{\PRD}{{\it Phys. Rev. }{\bf D}}
    
    \newcommand{\PR}{{\it Phys. Rev.}}

    \newcommand{\JHEP}{{\it JHEP}}
    \newcommand{\EPJC}{{\it Eur. Phys. J. }{\bf C}}
    \newcommand{\NCA}{{\it Nuovo Cim. }{\bf A}}
    \newcommand{\ZP}{{\it Z. Phys.}}
    \newcommand{\ZPC}{{\it Z. Phys. }{\bf C}}

    \newcommand{\SJNP}{{\it Sov. J. Nucl. Phys.}}
    \newcommand{\SPJETP}{{\it Sov. Phys. JETP}}
    \newcommand{\JETPL}{{\it JETP Lett.}}
    \newcommand{\APB}{{\it Acta Phys. Polon. }{\bf B}}

    \newcommand{\AP}{{\it Ann. Phys.}}

    \newcommand{\Jour}[4]{#1 {\bf #2}, #4 (#3)}
    \newcommand{\arxiv}[1]{{\tt #1}}

    \newcommand{\dif}{{\rm d}}
    \newcommand{\abar}{\bar{\alpha}_s}
    \newcommand{\atpi}{\frac{\bar{\alpha}_s}{2\pi}}
    \newcommand{\del}{\partial}
    \newcommand{\dY}{\dif Y}
    \newcommand{\lan}{\langle}
    \newcommand{\ran}{\rangle}
    \newcommand{\rme}{{\rm e}}
    \newcommand{\rmi}{{\rm i}}
    \newcommand{\tr}{{\rm tr}}
    \def\Tr{{\rm Tr}}

    \newcommand{\Lam}{\Lambda_{{\rm QCD}}}
    \newcommand{\lap}[1]{\nabla_{\bm{#1}}^2}
    \newcommand{\nn}{\nonumber\\}
    \newcommand{\aln}{&&\!\!\!\!\!\!}
    
    \def\i{\imath}
    \def\j{\jmath}
    \def\cal{\mathcal}
    \def\dot{\! \cdot \!}
    \def\-{\! - \!}

    \newcommand{\beq}{\begin{eqnarray}}
    \newcommand{\eeq}{\end{eqnarray}}

    \def\thenames{D.N.~Triantafyllopoulos}
    \def\thetitle{Pomeron Loops in High Energy QCD}

\begin{document}

\title{POMERON LOOPS IN HIGH ENERGY QCD\thanks{Based on lectures given
at the XLV Course of the Cracow School of Theoretical Physics,
Zakopane, Poland, 3-12 June 2005 [To appear in Acta Physica Polonica
B].}}
\author{D.N.~Triantafyllopoulos
\thanks{{\it Present address:} ECT*, Villa Tambosi, Strada delle
Tabarelle 286, I-38050, Villazzano (TN), Italy}
\address{Service de Physique Th\'{e}orique, Saclay, CEA\\
F-91191, Gif-sur-Yvette, France\\
{\tt dionysis@dsm-mail.cea.fr, dionysis@ect.it}}} \maketitle

\begin{abstract}
We discuss the QCD evolution equations governing the high energy
behavior of scattering amplitudes at the leading logarithmic level.
This hierarchy of equations accommodates normal BFKL dynamics,
Pomeron mergings and Pomeron splittings. Pomeron loops are built in
the course of evolution and the scattering amplitudes satisfy the
unitarity bound.
\end{abstract}
\PACS{11.15.Kc, 12.38.Cy, 13.60.Hb}

\begin{flushright}
    \vspace*{-10.3cm}
    {\footnotesize SACLAY--T05/182\\ECT*--05--18\\}
    \vspace*{9.4cm}
\end{flushright}

\section{Introduction}\label{SecIntro}

Over the last three decades, one of the main active fields of
research within QCD has been the study of its behavior in the high
energy limit. In general, a scattering process will be considered as
a high energy one, when the (square of the) total energy $s$ of the
colliding (hadronic) objects is much larger than the momentum
transfer $Q^2$ between them. At the same time one hopes to approach
the problem via analytical methods, since in this limit there is the
possibility of a large kinematical window $s \gg Q^2 \gg \Lam^2$
where one can apply weak coupling methods.

The first approach to the problem was done in the mid seventies when
the BFKL (Balitsky, Fadin, Kuraev, Lipatov) equation \cite{BFKL},
one of the central equations governing the approach to high energy,
was derived. It was understood that certain Feynman diagrams in
perturbation theory are enhanced by logarithms of the energy and
therefore they have to be resummed. This equation was solved in a
special case (for the forward amplitude), and a total cross section
growing as a power of the energy emerged. This growth may not be so
surprising (at least) a posteriori, since at high energies the
wavefunction of a hadron can contain a large number of partons,
mainly gluons, due to the available phase space for virtual
fluctuations and due to the triple-gluon coupling in QCD. In
general, QCD at high energy will be characterized by high densities
and increasing (but constraint) cross sections.

Afterwards, in the early to mid eighties, it was realized that one
needs to find a mechanism to tame the too steep increase of the
partonic densities as predicted by the BFKL equation and the concept
of saturation was introduced as a dual description of unitarity in
high energy scattering. The gluon density at a given momentum should
never exceed a value of order $\cal{O}(1/\alpha_s)$ and, as the
mechanism to fulfill this saturation bound, a non-linear term was
proposed \cite{GLR} to be added to the (linear) BFKL equation. Later
on a proof of that equation in a special limit (the double
logarithmic limit, which nowadays is known not to be a good
approximation for a high density system) was given \cite{MQ86}. At
the same time another step of progress was made, as the solution of
the BFKL equation for non-forward scattering, or equivalently at
fixed impact parameter, was obtained \cite{Lip86}.

In the mid nineties, one can say that there was a major
breakthrough. It was also the time when various subfields were
created, as different approaches to address and approach the high
energy problem started to develop. The color dipole picture
\cite{Mue94,MP94,Mue95,CM95} was formulated as a description for the
wavefunction of an energetic hadron, a picture which goes beyond the
BFKL equation in the sense that it also contains transitions between
different number of Pomerons in the multi-color limit (with a
Pomeron defined, more or less, as the object which evolves with
energy according to the BFKL equation). This allowed one to
calculate higher moments of the gluon densities while at the same
time the approach to unitarity limits could be studied. A program to
calculate the vertices for these Pomeron transitions (beyond the
large-$N_c$ limit) was started in \cite{BW95}. Roughly at the same
time a somewhat different problem was studied, namely the saturation
of densities in the wavefunction of a fast moving large nucleus, due
to the strong coherent classical fields generated by the large
number of its valence partons \cite{MV}. Even though there was no
QCD evolution in that model, the ideas introduced proved of great
significance for what followed in the forthcoming years. Moreover,
that period faced the first attempts to attack the high energy
problem by the method of effective actions
\cite{VV93,KLS9495,Lip95}. Finally a Hamiltonian, equivalent to an
integrable system, was given for a particular configuration of $n$
``reggeized'' gluons \cite{Lip9394,FK95}.

Even more progress was seen during the last years of the previous
decade and the beginning of the current one. The idea introduced
earlier in \cite{MV} that the wavefunction of an energetic hadron
can be described in terms of strong classical color fields, was used
properly in order to derive an equation describing the evolution to
higher and higher energies. This functional equation, called the
JIMWLK (Jalilian-Marian, Iancu, McLerran, Weigert, Leonidov, Kovner)
equation \cite{JKLW97,ILM01,FILM02,Wei02}, gives the evolution of
the probability to find a given configuration of a color field
associated with the hadronic wavefunction. In general, the physical
system which is supposed to be described by such an equation, like a
fast moving proton, nucleus or a quarkonium, was called a Color
Glass Condensate (CGC) \cite{ILM01}, for reasons to be explained
later. In turn, the JIMWLK equation can be used to derive equations
for the scattering amplitudes of given projectiles off the evolved
hadron, and the outcome is what is known by today as the Balitsky
hierarchy \cite{Bal9601}. This is a set of non-linear equations,
which under a mean field and/or large-$N_c$ approximation collapses
to a single one \cite{Kov9900} giving the evolution of the amplitude
for a color dipole to scatter off the CGC. As a byproduct, the
odderon problem (scattering with an odd number of gluon exchanges)
introduced in the early eighties \cite{Bar80,KP80,Jar80} was
reformulated and extended to its non-linear version
\cite{KSW04,HIIM05}. During that period another important
accomplishment was the calculation of the next to leading order
correction to the BFKL equation, which was finally completed in
\cite{FL98,CC98}, while there was an effort to calculate the
vertices for Pomeron transitions, their properties and consequences
\cite{BLW96,BE99,BV99,ES04,BLV05}, a task which is still ongoing.

Until the beginning of the last year it was thought, at least in
some part of the ``high energy community'', that the JIMWLK equation
was a more or less ``complete'' and self-consistent description of
the high energy limit of QCD (at the leading logarithmic level).
However, a calculation done within the dipole picture resulted in
some large corrections \cite{MS04} for the saturation momentum (the
momentum scale at a given energy at which gluonic modes saturate).
This result was clearly different from what was known up to that
time, and the discrepancy was initially attributed to the difference
between the JIMWLK equation and its mean field version. Shortly it
was understood that these corrections are an effect of the low
density fluctuations in the dilute, high-momentum, tail of the
wavefunction \cite{IMM05} and in fact the significance of these
fluctuations in the evolution of the system had been also observed
and realized much earlier from numerical simulations within the
dipole picture \cite{Sal9596,MS96}. The understanding that the
JIMWLK equation does not describe properly the low density region of
the hadronic wavefunction came soon \cite{IT05a,IT05b}, and a
generalization of the Balitsky equations at large-$N_c$ was given.
This new hierarchy does not allow any kind of mean field
approximations and contains loops of Pomerons, in contrast to the
Balitsky-JIMWLK one which contains only ``one-way'' Pomeron
transitions. And in fact, the possibility to allow for the formation
of Pomeron loops in the course of evolution is of crucial importance
for a self-consistent approach to unitarity. Not surprisingly,
within this QCD description and under certain logical
approximations, the results in \cite{MS04,IMM05} were reproduced.
Triggered by these observations, facts and derivations, there has
been an ongoing effort to the direction of constructing a
Hamiltonian formulation and a generalization of the equations at
finite-$N_c$ \cite{MSW05,KL05a,KL05c,BIIT05,HIMST05,Bal05}.

In these lectures we will start by introducing in the next section
the BFKL Pomeron and the BFKL equation in coordinate space as was
derived in the dipole picture. In Sec.~\ref{SecBFKLPath} we will
discuss its pathologies and in Sec.~\ref{SecSat} the concepts of
saturation and Pomeron mergings will be proposed as the resolution
to the problem, while at the same time a non-linear equation will
naturally emerge. In Sec.~\ref{SecCGC} the Color Glass Condensate,
the JIMWLK equation and the Balitsky equations will be presented,
and in Sec.~\ref{SecSatMom} the calculation for the energy
dependence of the saturation momentum within this formulation will
be given. In Secs.~\ref{SecDefBal} and \ref{SecMissing} we will try
to explain which are the main problems encountered within the JIMWLK
evolution and which is the most crucial missing element. In
Sec.~\ref{SecDip} the dipole picture will be reviewed and the
significance of fluctuations will be stressed, while in
Secs.~\ref{SecPomSplit} and \ref{SecPloop} the new hierarchy will be
derived, and then the way Pomerons loops are generated will be quite
obvious. In the next three sections we will discuss the approach to
formulate the problem at the Hamiltonian level, a remarkable
(possible) property of the Hamiltonian and the approach to the
generalization of the theory beyond the multi-color limit. In the
last section the saturation momentum will be revisited, since its
energy dependence will be influenced by the low density behavior of
the effective theory. As it is clear the presentation will be mostly
based on the dipole picture and the Color Glass Condensate
formulation of the high energy problem, and over the years there
have been nice lectures and reviews for both approaches
\cite{Mue97,McL99,Mue99b,Ven99,Mue01b,ILM02,IV03,Sta04,Wei05,Ko05,KJ06}.
We shall not discuss at all any phenomenological aspects of the
theory and the possible importance of high density phenomena in the
relevant experiments. Nevertheless, we need to say that there have
been successful qualitative and sometimes quantitative descriptions
of the the low-$x$ data in deep inelastic lepton hadron scattering
(DIS) \cite{GBW99,GBKS01,GLLM03,IIM04} and of the observed particle
spectra in heavy-ion collisions
\cite{KLN03,GJ03,KKT03,AAKSW04,IIT04,KKT04,BGV04,KJ06}.

\section{The BFKL Pomeron and the BFKL Equation}\label{SecBFKL}

We start these lectures by introducing the concept and the
significance of the BFKL Pomeron and by giving a heuristic, but also
intuitive, derivation of the BFKL equation. Imagine that we want to
measure the gluon distribution of a generic hadron. One way to do
this, is by probing the hadron with a small, in size, color dipole.
If the dipole has a size $r =|\bm{x}-\bm{y}|\ll \Lam^{-1}$, where
$\bm{x}$ and $\bm{y}$ are the coordinates of its quark and antiquark
legs respectively, it will probe the gluonic components of the
hadron with momenta such that $Q \sim 1/r \gg \Lam$. Clearly, the
condition that the dipole be small, is dictated by the requirement
that the QCD coupling constant be small and thus the problem can be
approached by analytical methods. Such a probe is not only a good
theoretical object due to its overall color neutrality, but it can
also be ``created'' as a fluctuation of the virtual photon in deep
inelastic lepton-hadron scattering (DIS). Therefore, if one is able
to calculate the dipole-hadron scattering amplitude, one can obtain
the cross section for DIS, since the probability for the creation of
the dipole is determined by a lowest order calculation in QED. The
latter can be easily performed \cite{NZ91} and therefore we shall
restrict ourselves to the analysis of the dipole-hadron scattering.

Let us assume that the hadron, to which we shall frequently refer as
the target, is right-moving, while the projectile dipole is left
moving. We shall also assume that the target is energetic enough, so
that a partonic description of its wavefunction is meaningful, while
the dipole is slow enough so that it is ``bare''; that is, there are
no additional components, through higher order radiative
corrections, in the dipolar wavefunction.

\begin{figure}[t]
    \centerline{\epsfig{file=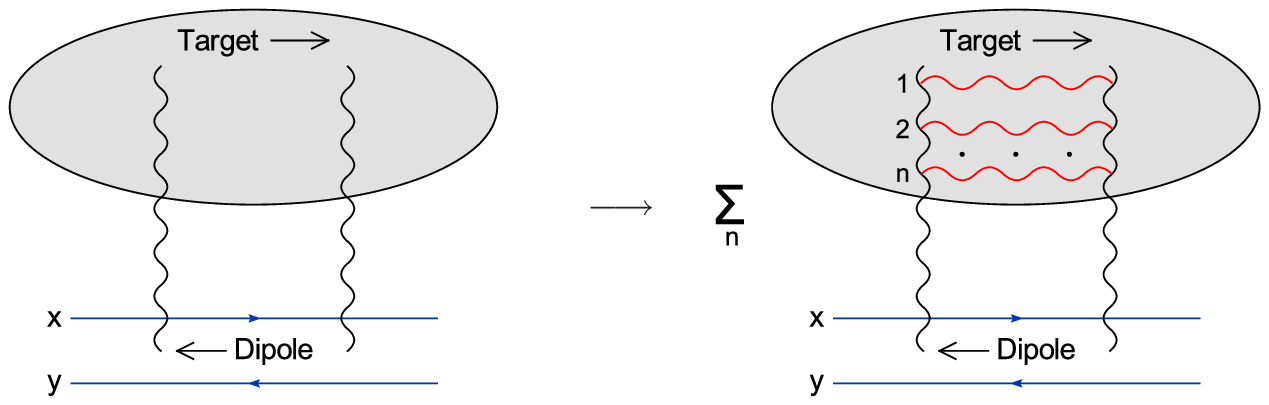,width=12.5cm}}
    \caption{\sl From lowest order in pQCD to the BFKL Pomeron.}
    \label{FigPomeron}
\end{figure}

At lowest order in QCD perturbation theory the dipole and the hadron
will interact via a two-gluon exchange as shown in the left part of
Fig.~\ref{FigPomeron}. Now let the hadron be boosted at very high
energy and let $p^+$ be the (light-cone) longitudinal momentum of a
``valence'' parton of the hadronic wavefunction\footnote{Since we
are interested in the high energy limit, any possible masses will be
always taken to be zero.}. Then, and as we show in Appendix
\ref{AppDipKer} in more detail, the probability to emit a soft gluon
with longitudinal momentum in the interval from $k^+$ to $k^+\!+\dif
k^+$, with $k^+ = x p^+ \ll 1$, is proportional to $\alpha_s \,\dif
k^+\!/k^+ = \alpha_s \,\dif x\!/x$. When $x$ is small enough and due
to the QCD triple-gluon coupling, the gluon with momentum fraction
$x$ can be generated through the intermediate emission of one or
more gluons which have their longitudinal momenta strongly ordered,
i.e.~they satisfy $x \ll x_n \ll \dots \ll x_1 \ll 1$, where
$x_{\i}$ is the momentum fraction of the $\i$-th gluon in this
sequence of emissions of $n$ intermediate gluons. This process, when
compared to the direct emission of the soft gluon at $k^+$ from the
valence parton at $p^{+}$, is of order\footnote{We shall always let
the coupling be fixed (only in Sec.~\ref{SecSatMom} we shall briefly
discuss some extensions to running coupling) and for the reasons
explained earlier we shall consider it to be small.}
    \beq\label{seq}
    \alpha_s^n
    \int\limits_{x}^{1} \frac{\dif x_1}{x_1}\,\,
    \dots\!
    \int\limits_{x}^{x_{n-1}} \frac{\dif x_n}{x_n}
    = \frac{1}{n!}\,\left(\alpha_s \ln\frac{1}{x}\right)^n,
    \eeq
where the logarithms have been clearly generated from the
integration over the available longitudinal phase space for the
intermediate emissions. At high energies, the rapidity $Y$, defined
as $Y=\ln(1/x)$, can compensate the smallness of the coupling
$\alpha_s$, so that $\alpha_s Y \gtrsim 1$. Thus one needs to resum
all these $(\alpha_s Y)^n$ enhanced terms and such a resummation
gives rise to the so-called BFKL Pomeron. It is obvious in
Eq.~(\ref{seq}), that this procedure will lead to an exponential, in
$Y$, increase of the gluon density and this will be verified shortly
and in more detail when we also take into account the degrees of
freedom in the transverse phase space. The interaction of an
energetic hadronic target with a color dipole is shown on the right
part of Fig.~\ref{FigPomeron}, where the summation over the gluon
ladder represents the BFKL Pomeron or equivalently the small-$x$
components of the hadronic wavefunction. Notice that this figure
corresponds already to the square of a diagram in perturbation
theory, since we are interested in determining the probability to
find a given mode inside the hadronic wavefunction.

\begin{figure}[t]
    \centerline{\epsfig{file=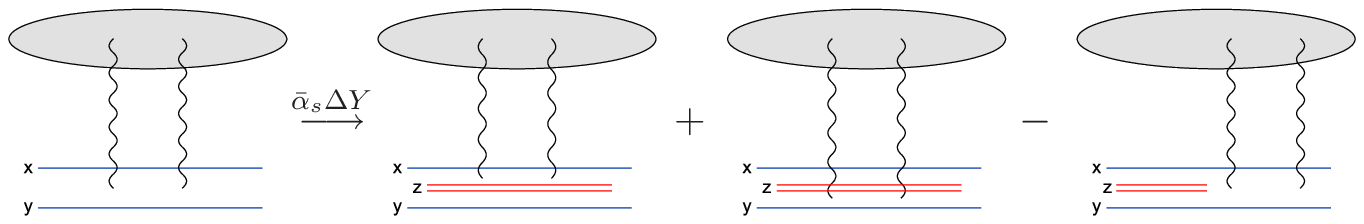,width=12.5cm}}
    \caption{\sl The BFKL equation for the dipole-hadron scattering amplitude.}
    \label{FigTBFKL}
\end{figure}

Instead of performing the resummation of diagrams, one can
equivalently study the evolution of the target wavefunction or of
the scattering amplitude under a step $\Delta Y$ in rapidity.
Furthermore, one can also view the last soft gluon as being emitted
from the projectile dipole, which is a simple object and therefore
its evolution can be easily studied. In the large-$N_c$ limit the
soft gluon can be represented by a quark-antiquark pair, and thus
the final system is composed of two dipoles $(\bm{x},\bm{z})$ and
$(\bm{z},\bm{y})$, with $\bm{z}$ the transverse coordinate of the
emitted soft gluon. As we show in Appendix \ref{AppDipKer} the
differential probability for this splitting process is \cite{Mue94}
    \beq\label{SplitProb}
    \dif P =
        \atpi\,
        \frac{(\bm{x}-\bm{y})^2}
        {(\bm{x}-\bm{z})^2(\bm{z}-\bm{y})^2}\,
        \dif^2 \bm{z}\, \dY
        \equiv
        \atpi\,
        \mathcal{M}_{\bm{x}\bm{y}\bm{z}}\,
        \dif^2 \bm{z}\, \dY,
    \eeq
where $\abar = \alpha_s N_c/\pi$, with $N_c$ the number of colors.
In the above equation, $\dif Y = \dif \ln(1/x) = -\dif k^+/k^+$
represents the differential enhancement in the longitudinal phase
space as shown in detail in Appendix \ref{AppDipKer}, while the
kernel $\cal{M}_{\bm{x}\bm{y}\bm{z}}$ contains all the QCD dynamics
in the transverse phase space. Now, each of the two final dipoles
can scatter off the target and therefore the evolution equation for
the imaginary part of the scattering amplitude $T$ reads
    \beq\label{BFKL}
        \frac{\del \lan T_{\bm{x}\bm{y}} \ran}{\del Y}\Big|_{\rm BFKL}\!=
        \frac{\abar}{2\pi}\!\int\limits_{\bm{z}}\!
        \mathcal{M}_{\bm{x}\bm{y}\bm{z}}
        \left[ \lan T_{\bm{x}\bm{z}} \ran
        +\lan T_{\bm{z}\bm{y}} \ran
        -\lan T_{\bm{x}\bm{y}} \ran \right]
        %\nn
        \equiv \abar \cal{K}_{\rm BFKL}\!
        \otimes \lan T_{\bm{x}\bm{y}} \ran,
    \eeq
where the last term is the virtual contribution arising from the
normalization of the dipole wavefunction, and with the average to be
taken over the target wavefunction, even though this is irrelevant
for the moment. This is the BFKL equation \cite{BFKL} in coordinate
space \cite{Mue94,NZZ94} and its diagrammatic representation is
shown in Fig.~\ref{FigTBFKL}. It is free of any divergencies, since
the potential singularities arising from the poles of the dipole
kernel cancel. For example, when $\bm{z} = \bm{x}$ the last two
terms in the square bracket cancel each other, while the first term
vanishes due to color transparency. In Appendix \ref{AppBFKL} we
give a more rigorous derivation of Eq.~(\ref{BFKL}), by using the
corresponding BFKL equation for the density of dipoles which is
derived in Sec.~\ref{SecDip}.

\begin{figure}[t]
    \centerline{\epsfig{file=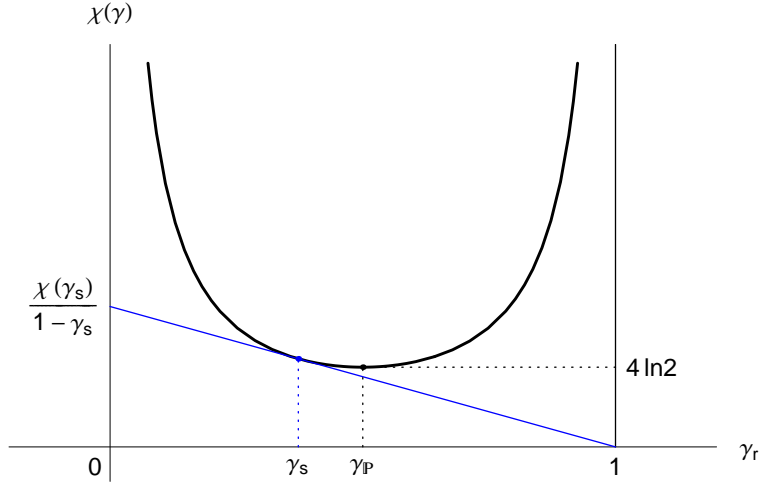,width=10.5cm}}
    \caption{\sl The BFKL eigenvalue
    $\chi(\gamma)$ as a function
    of $\gamma_{\rm r}$ when $\gamma_{\rm i} =0$,
    with $\gamma=\gamma_{\rm r} + \rmi \gamma_{\rm i}$. The single poles
    at $\gamma = 0,1$ correspond to the collinear (double logarithmic) limits.
    The saddle point for the Pomeron intercept occurs at
    $\gamma_{\mathbb{P}}=1/2$, while $\gamma_s = 0.372$
    corresponds to the relevant anomalous dimension for saturation
    (see Sec.~\ref{SecSatMom}).}
    \label{FigChi}
\end{figure}

The BFKL equation is a linear one, and therefore the solution can be
obtained by studying the corresponding eigenvalue problem. Let us
first look at the simplified case where the scattering amplitude is
integrated over the impact parameter $\bm{b} \equiv
(\bm{x}+\bm{y})/2$ to give the total cross section, and is invariant
under rotations in $\bm{r}$. Then it depends only on the magnitude
$r\equiv |\bm{x}-\bm{y}|$ of the projectile dipole. As we show in
Appendix \ref{Appeigen} the solution to this eigenvalue problem is
    \beq\label{Keigen}
    \cal{K}_{\rm BFKL}\otimes r^{2(1-\gamma)}
    = \chi(\gamma)\,r^{2(1-\gamma)},
    \eeq
where
    \beq\label{chifun}
    \chi(\gamma)=
    2\, \psi(1) - \psi(\gamma) - \psi(1-\gamma),
    \eeq
with $\psi(\gamma)$ the logarithmic derivative of the
$\Gamma$-function. The eigenvalue function $\chi(\gamma)$ is plotted
in Fig.~\ref{FigChi}. Thus, one can write the solution to the BFKL
equation as the superposition of the evolved with rapidity
eigenfunctions, namely
    \beq\label{BFKLsol}
    \lan T(r,Y) \ran=
    \frac{2 \pi \alpha_s^2}{\mu^2}
    \int\limits_{\cal{C}}
    \frac{\dif \gamma}{2\pi \rmi}\,
    T_0(\gamma)
    \exp \left[
    \abar \chi(\gamma) Y + (1-\gamma) \ln \left(r^2\mu^2\right)
    \right].
    \eeq
In the above equation, $\mu$ is a momentum scale related to the
target, the integration contour $\cal{C}$ is parallel to the
imaginary axis with $0< {\rm Re}(\gamma) <1$, and $T_0(\gamma)$ is
proportional to the Mellin transformation of the initial cross
section at $Y=0$. For instance, when we consider dipole-dipole
scattering this initial condition is given by Eq.~(\ref{sigmadd}) in
Appendix \ref{Appdipdip} and in this case $T_0(\gamma)$ is equal to
$1/[2 \gamma^2 (1\-\gamma)^2]$.

At high energies, i.e.~when $Y\to \infty$, and with the dipole size
$r$ being fixed, one can perform a saddle point integration in
Eq.~(\ref{BFKLsol}). The saddle point of the $\chi$-function occurs
at $\gamma_{\mathbb{P}}=1/2$, with corresponding value $\chi(1/2) =
4 \ln2$ as shown in Fig.~\ref{FigChi}, and then the dominant
asymptotic energy dependence of the amplitude is given by
    \beq\label{Growth}
    \lan T \ran
    \sim \alpha_s \lan \varphi \ran
    \sim \alpha_s^2 \lan n \ran
    \sim \alpha_s^2 \exp [\omega_{\mathbb{P}} Y],
    \eeq
where $\omega_{\mathbb{P}} = 4 \abar \ln 2 $ is the so-called hard
Pomeron intercept\footnote{The expression for $\lan T \ran$ in
Eq.~(\ref{Growth}) vanishes in the strict large-$N_c$ limit ($\abar$
= fixed, $\alpha_s^2 \to 0$), because of its prefactor which is
proportional to the initial value of the amplitude. But there is no
real problem with that. First of all, and as we will see later on,
the BFKL equation remains valid at finite-$N_c$. Furthermore, one
can always assume a ``modified'' large-$N_c$ limit where all
quantities, like the r.h.s.~of Eq.~(\ref{Growth}), which are
suppressed by $1/N_c^2$ factors, are taken as leading order effects
provided they are enhanced by appropriate powers of the energy
\cite{MP94,Mue95}. For instance $\alpha_s^4 \exp [2
\omega_{\mathbb{P}} Y]$ is a leading effect, while $\alpha_s^4 \exp
[\omega_{\mathbb{P}} Y]$ is subdominant.}. In Eq.~(\ref{Growth}),
$\varphi$ represents the gluon density of the hadron, while $n$ will
be its dipole density if we assume that it is composed of dipoles
and this is the case in the large-$N_c$ limit. The exponential
increase is not surprising, since the BFKL equation is a linear one;
in each step of the evolution, the change in the amplitude is
proportional to its previous value. After all, this is what we had
already expected from the summation of the series whose $n$-th term
is given in Eq.~(\ref{seq}). Notice also that, even though this is
not explicitly written in Eq.~(\ref{Growth}) but easily inferred
from Eq.~(\ref{BFKLsol}), the dominant $r$-dependence of the
amplitude will be proportional to $r$. This is to be contrasted with
the result of fixed order perturbation theory which is proportional
to $r^2$. The fact that the anomalous dimension differs from this
perturbative result by a finite pure number is a unique
characteristic feature of BFKL dynamics.

The BFKL equation has also been solved in its most general case,
i.e.~without assuming rotational symmetry and for a given, but
arbitrary, impact parameter \cite{Lip86}. In that case the system
evolves ``quasi-locally'' in impact parameter space, and therefore
the dominant energy dependence of the solution, as $Y \to \infty$
and with $\bm{r}$ and $\bm{b}$ fixed, is still given by
Eq.~(\ref{Growth}), since the same eigenvalue dominates the
corresponding integration.

\section{Pathologies of the BFKL Equation}\label{SecBFKLPath}

There are two major problems associated with the BFKL equation. The
first is the violation of unitarity bounds. As we saw in
Sec.~\ref{SecBFKL} the amplitude at a fixed impact parameter
increases exponentially with the rapidity $Y$, or equivalently as a
positive power of the total energy $s$ in the process, since
$Y=\ln(s/s_0)$ (the precise value of the scale $s_0$ is not
important for our arguments). However, it should satisfy the
unitarity bound\footnote{This will be totally clear later on. See
Eq.~(\ref{SofV}), where the precise definition of $T(\bm{r},\bm{b})$
is given in terms of Wilson lines, thus including an arbitrary
number of gluon exchanges with the target.}
    \beq\label{Tbound}
    T(\bm{r},\bm{b}) \leq 1.
    \eeq
One should stress that this bound is different from the well-known
Froissart bound \cite{Fro61,Mar65,LM67,Hei52}, which states that any
hadronic total cross section $\sigma_{\rm tot}$  should not grow
faster than the square of the logarithm of the energy, more
precisely
    \beq\label{Froissart}
    \sigma_{\rm tot} \leq \frac{\pi}{m_{\pi}^2}\,
    \ln^2 s,
    \eeq
with $m_{\pi}$ the pion mass. Of course we should not expect to
satisfy this bound by weak coupling methods. The BFKL equation, and
all the equations that we will present later on in this article,
will never treat properly the ``edges'' of the hadron, where long
range forces become important and therefore the problem is genuinely
non-perturbative. Nevertheless, there is no reason a priori which
would imply that Eq.~(\ref{Tbound}) cannot be fulfilled in
perturbation theory. Notice that Eq.~(\ref{Tbound}) is equivalent to
the condition that the maximal allowed gluon density in QCD should
be of order $1/\alpha_s$. Indeed, one has,
    \beq\label{phimax}
    \varphi \sim a^{\dag} a
    \sim A^2
    \lesssim \frac{1}{g^2}
    \sim \frac{1}{\alpha_s},
    \eeq
where $a^{\dag}$ and $a$ are gluonic creation and annihilation
operators, while $A$ is the gauge field associated with the
wavefunction of the hadron. This maximal value of order $1/\alpha_s$
can be obtained in a heuristic way by setting the cubic and the
quartic in $A$ terms in the QCD Lagrangian to be of the same order.

The second problem is the sensitivity to non-perturbative physics.
As already said, the longitudinal momenta are strongly ordered in
the course of the evolution. However the transverse coordinates of
the dipoles (or equivalently the transverse momenta of the gluons)
are not strongly ordered, and therefore the dipole kernel is
non-local as clearly seen in Eq.~(\ref{SplitProb})\footnote{In DGLAP
\cite{DGLAP} evolution, where one resums enhanced powers of
$\alpha_s \ln(Q^2/\mu^2)$ one encounters the ``opposite'' situation;
the transverse momenta are strongly ordered, while the longitudinal
ones are not. Therefore the corresponding DGLAP kernel is local in
$Q$, but non-local in $Y$.}. This non-locality results in a
diffusion factor which accompanies the dominant asymptotic behavior
given in Eq.~(\ref{Growth}), and which can be easily obtained by
performing the Gaussian integration over $\gamma$ in
Eq.~(\ref{BFKLsol}) around the saddle point
$\gamma_{\mathbb{P}}=1/2$. This factor reads
    \beq\label{psipom}
    \psi_{\mathbb{P}} = \frac{1}{\sqrt{\pi D_{\mathbb{P}} Y}}\,
    \exp\left[
    -\frac{\ln^2(r^2 \mu^2)}{D_{\mathbb{P}} Y}
    \right],
    \eeq
with $D_{\mathbb{P}} = 2 \abar \chi''(1/2) = 56 \abar \zeta(3)$. The
form of the function $\psi_{\mathbb{P}}$ suggests that it can be
written as the solution to the one-dimensional diffusion equation.
That is, after the dominant exponential behavior has been isolated,
the evolution can be viewed as a random walk in $\ln(r^2\mu^2)$.
Thus, independently of how small the initial dipole is, and after
some critical value of rapidity, there will be diffusion to the
infrared; dipoles with sizes bigger than $\Lam^{-1}$ will be created
and the weak coupling assumption will not be valid any more. One
needs to emphasize that this is true even if we want to calculate
the amplitude in the perturbative region. Since the diffusion
equation is of second order, one can formulate the problem as a path
integral from the initial dipole size to the final one. Then,
clearly there are paths which go through the non-perturbative
region. In this sense the BFKL equation is not self-consistent.

Before moving to the next section, where we will discuss the
solution to these problems, let us mention that the next to leading
BFKL equation \cite{FL98,CC98} shares the same pathological
features. In that case one resums powers of the form $\alpha_s
(\alpha_s Y)^n$, and this procedure gives rise to a contribution of
order $\cal{O}(\alpha_s)$ to the Pomeron intercept
$\omega_{\mathbb{P}}$, as compared to the leading one.

\section{Unitarity, Saturation and Mergings of Pomerons}\label{SecSat}

\begin{figure}[t]
    \centerline{\epsfig{file=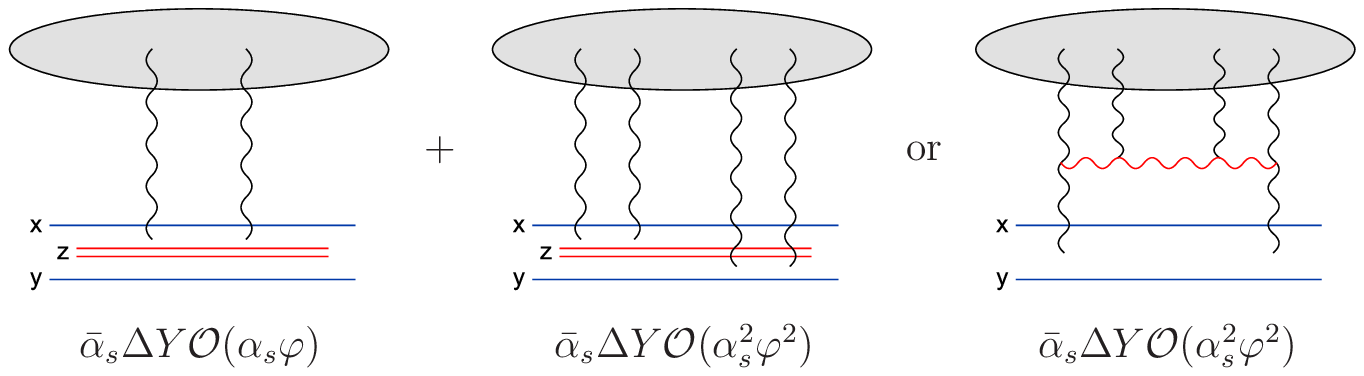,width=12.5cm}}
    \caption{\sl Estimation of diagrams contributing
    to the first Balitsky equation.}
    \label{FigMerging}
\end{figure}

Let us continue our heuristic approach and try to find out which is
the basic element that is missing from the procedure we have
followed so far. The first diagram in Fig.~\ref{FigMerging} shows
one of the contributing diagrams to the BFKL equation. As indicated,
this is of order $\abar \Delta Y \cal{O}(\alpha_s \varphi)$, where
the factor $\abar \Delta Y$ comes from the evolution step, the
factor $\alpha_s$ comes from the two couplings at the lowest part of
the diagram and the factor $\varphi$, the target gluon density, is
simply the upper part of the diagram. It is clear that the other two
BFKL diagrams, which are not shown here, are of the same order.
Following the same reasoning, the second diagram is of order $\abar
\Delta Y \cal{O}(\alpha_s^2 \varphi^2)$, since now there are four
vertices at the lowest part, while one also probes the gluon pair
density of the hadron. This diagram is suppressed with respect to
the BFKL ones when $\varphi \ll 1/\alpha_s$, i.e.~at low densities,
or equivalently when $T\ll 1$ and in that case it can be neglected.
However it is equally important in the high density
limit\footnote{In that case, this is a just typical diagram, since,
in general, diagrams with more than two-gluon exchanges per dipole
will be equally important.}, precisely in the region where the
unitarity problem of the BFKL evolution appears. Therefore,
following this ``active'' point of view in the projectile evolution
as before, i.e.~the r.h.s.~of the equation contains what we have
after the evolution step, we see that we should add to the BFKL
equation the term which corresponds to the simultaneous scattering
of the two dipoles $(\bm{x},\bm{z})$ and $(\bm{z},\bm{y})$, that is
    \beq\label{Tmerge}
    \frac{\del \lan T_{\bm{x}\bm{y}} \ran}
    {\del Y}\Big|_{\rm merge} =
    -\frac{\abar}{2\pi}\int\limits_{\bm{z}}
    \mathcal{M}_{\bm{x}\bm{y}\bm{z}}\,
    \lan T_{\bm{x}\bm{z}}T_{\bm{z}\bm{y}} \ran.
    \eeq
The resulting evolution equation is the first Balitsky equation
\cite{Bal9601}. There are a couple of points which need to be
clarified in the above equation. The first is the notation ``merge''
for this particular term. This comes from the fact that the second
diagram in Fig.~\ref{FigMerging} is equivalent to the third one. The
latter corresponds to the evolution of the hadron in a ``passive''
point of view; now the red gluon is supposed to be emitted in the
wavefunction of the target, and the two Pomerons, which existed
before the evolution step, merge to give rise to a contribution to
the single dipole scattering amplitude. However, one should be very
careful about the terminology and the interpretation of this merging
process. When the hadron is boosted to higher and higher energies,
there is no process which would lead to the ``mechanical''
recombination of the partons that it is composed of. The term
merging corresponds to the $4 \to 2$ vertex connecting the upper
part of the diagram, the hadron, and the lower part, the dipole.
The second point that needs to be explained is the minus sign in
Eq.~(\ref{Tmerge}). Intuitively, one would indeed expect this term
to be negative, since, in almost any physical system, a merging
process is supposed to tame the growth of the particle density. A
better way to justify the sign, is to notice that we can express the
first Balitsky equation in a more compact form, by introducing the
$S$-matrix  $S_{\bm{x}\bm{y}} = 1 - T_{\bm{x}\bm{y}}$ (recall that
$T$ is the imaginary part of the scattering amplitude). Then we have
    \beq\label{Bal1}
    \frac{\del \lan S_{\bm{x}\bm{y}} \ran}{\del Y}=
    \atpi\,
    \int\limits_{\bm{z}}
    \cal{M}_{\bm{x}\bm{y}\bm{z}}
    \left[\lan S_{\bm{x}\bm{z}}S_{\bm{z}\bm{y}} \ran
    -\lan S_{\bm{x}\bm{y}} \ran \right],
    \eeq
which has a clear interpretation; the $S$-matrix for the dipole-pair
to scatter off the target multiplied by the splitting probability
gives the change in the $S$-matrix for the dipole-hadron scattering
(after we subtract the term proportional to $\lan S_{\bm{x}\bm{y}}
\ran$ which gives the survival probability for the parent dipole).

One can proceed and perform certain approximations to this equation.
When the target hadron is a large nucleus, and for not too high
energies, one can perform a sort of mean field approximation, by
assuming that the two projectile dipoles scatter independently off
the target, namely
    \beq\label{meanfield}
    \lan T_{\bm{x}\bm{z}}T_{\bm{z}\bm{y}} \ran
    \simeq
    \lan T_{\bm{x}\bm{z}} \ran
    \lan T_{\bm{z}\bm{y}} \ran,
    \eeq
and similarly for the $S$-matrices. Then one obtains a closed
equation, the so-called BK (Balitsky, Kovchegov) equation
$\cite{Bal9601,Kov9900}$ (see also \cite{Bra00a}). We immediately
notice that this equation has two fixed points; ({\bf i}) $\lan T
\ran =0 \Leftrightarrow \lan S \ran =1$, which is an unstable fixed
point, since no matter how small the initial amplitude is, it will
start to grow and ({\bf ii}) $\lan T \ran =1 \Leftrightarrow \lan S
\ran =0$, which is the black-disk limit, and which is a stable fixed
point for any perturbation in $\lan T \ran$. Therefore the BK
equation is much better behaved than the BFKL equation. It seems
that the amplitude will never exceed the unitarity bound $T_{\rm
max} =1$, and the gluon density will saturate at a value of order
$\cal{O}(1/\alpha_s)$. Furthermore the non-linear term cuts all the
diffusion to the infrared \cite{GBMS02,MT02} and there is no
sensitivity to non-perturbative physics any more. In fact, all
diffusive paths that go beyond the saturation line (see below), will
be eliminated by the non-linear evolution.

Here it is appropriate to introduce the concept of the saturation
momentum $Q_s$. This is a line in the $\ln (r^2 \mu^2)\!-\!Y$ plane
(in general, the saturation momentum depends also on the impact
parameter), along which the amplitude satisfies $\lan T(r = 1/Q_s)
\ran= {\rm const} <1$. It is simply the border between the region
where BFKL dynamics can be safely applied and the region where
saturation has been reached and unitarity corrections have to be
taken into account. At this moment, it is not hard to understand
that more and more gluonic modes in the wavefunction of the hadron
will be saturated, as we keep increasing its energy. As we shall see
later on in a detailed analysis, the saturation momentum increases
exponentially with rapidity, i.e.~$Q_s^2 \approx \Lam^2
\exp(\lambda_s Y)$. Thus, when the rapidity is large enough, one
will have $\alpha(Q_s) \ll 1$, which means that the use of weak
coupling techniques is justified. It is conceptually important to
realize that this is a ``QCD phase'' where the physical system is
dense but the coupling constant is weak.

By no means we have given a rigorous derivation of the first
Balitsky equation in this section. For example, when high density
effects become significant, one should not restrict oneself to the
two-gluon exchange approximation (even though we didn't really make
any use of it) when considering the scattering of a single dipole
with the hadron. Indeed, each coupling of either the quark or the
antiquark of the dipole with the classical field $A$ associated with
the target is of order $g A \sim \sqrt{\alpha_s \varphi}$, and thus
in the high density limit $\varphi \sim 1/\alpha_s$ all multi-gluon
exchanges are equally important. Nevertheless, as we shall see in
the next section, the first Balitsky equation remains as given here,
even when we include multiple gluon exchanges.

\section{The Color Glass Condensate, the JIMWLK Equation and
the Balitsky Hierarchy}\label{SecCGC}

Perhaps the most elegant, modern and compete approach to describe
the merging of Pomerons is the Color Glass Condensate (CGC) and its
evolution according to a Renormalization Group Equation (RGE). This
is an effective theory within QCD, and its name is not accidental,
since it corresponds to some of the basic features that it
accommodates. {\it Color} stands for the color charge carried by the
gluons, {\it Glass} stands for a clear separation of time scales
between the ``fast'' and ``slow'' degrees of freedom in the
wavefunction, and {\it Condensate} stands for the high density of
gluons which can reach values of order $\mathcal{O}(1/\alpha_s)$.

The essential motivation for the formulation of the CGC is the
separation of scales in the longitudinal momenta between the fast
partons and the emitted soft gluons. Denoting by $p^{\mu}$ and
$k^{\mu}$ the corresponding light-cone 4-momenta, one has $k^{+} \ll
p^{+}$. This translates to an analogous separation in light cone
energies $k^{-}=\bm{k}^2/2k^{+} \gg p^{-}=\bm{p}^2 /2p^{+}$, which
in turn leads to a separation of time-scales; the lifetime $\Delta
x^{+} \sim 1/k^{-}$ of the soft gluons will be very short in
comparison with the typical time scale $\sim 1/p^{-}$  for the
dynamics of the fast partons. Thus, even though the fast partons are
virtual fluctuations in reality, they appear to the soft gluons as
being a ``source'' which is $x^{+}$-independent, i.e.~as a static
source. Furthermore, the source is random, since it corresponds to
the color charge seen by the soft gluons at the short period of
their virtual fluctuation; this happens at an arbitrary time and it
is instantaneous compared to the lifetime of the source. The color
charge density $\rho_{a}(x^{-},\bm{x})$ associated with this source
at the scale $p^{+}$, propagates along the light cone $x^{-} \simeq
0$ and the corresponding current has just a $+$ component. This
source is localized near the light cone within a small distance
$\Delta x^{-} \sim 1/p^{+}$, which is non-zero, but much smaller
than the longitudinal extent $\sim 1/k^{+}$ of the slow partons. One
of the consequences is that the size of the hadron will extend in
the longitudinal direction with increasing energy.

Based on the above kinematical considerations, one can represent the
fast color sources by a color current $J_a^{\mu}(x^-,\bm{x}) =
\delta^{\mu +} \rho_a(x^-,\bm{x})$, and the small-$x$ gluons
correspond to the color fields as determined by the Yang-Mills
equation in the presence of this current, namely
    \beq\label{YMEq}
    \left(D_{\nu} F^{\nu\mu}\right)_a(x)=\delta^{\mu+} \rho_a(x^-,\bm{x}).
    \eeq
In principle, and in reality in certain gauges, one can solve this
classical equation and obtain the gauge field $A(\rho)$ as a
function of a given source. Then, any observable $\cal{O}$ which is
related to the field, for example the gluon occupation number or the
amplitude for an external dipole to scatter off the hadron, can be
expressed in terms of $\rho$, and its expectation value will be
given by the functional integral
    \beq\label{Oave}
    \langle \cal{O} \rangle =
    \int \cal{D}\rho\, Z_Y[\rho] \cal{O}(\rho).
    \eeq
It is obvious that $Z_Y[\rho]$ serves as a weight measure and it
gives the probability density to have a distribution $\rho$ at a
given rapidity $Y$ (we assume that $Z_Y[\rho]$ is normalized to
unity). This probabilistic interpretation for the source relies on
its randomness mentioned above; there is no quantum interference
between different $\rho$ configurations. In other words, one
performs a classical calculation for a fixed configuration of the
sources, and then one averages over all the possible configurations
with a classical probability distribution.

So far, in Eqs.~(\ref{YMEq}) and (\ref{Oave}) we have not used at
all the small-$x$ QCD dynamics, and it is obvious that this must be
encoded in the probability distribution $Z_Y[\rho]$, which is the
only quantity in Eq.~(\ref{Oave}) depending on rapidity. But before
analyzing this aspect in more detail, we have to say that this idea
to describe the soft modes of the hadronic wavefunction in terms of
classical fields and probability densities was introduced in the MV
(McLerran, Venugopalan) model \cite{MV}, where a nucleus with large
atomic number ($A \gg 1$) was considered. In this ``static'' model
the only color sources are assumed to be the $A \times N_c$ valence
quarks, which are taken to be uncorrelated for transverse
separations such that $|\Delta\bm{x}| \lesssim \Lam^{-1}$, so that
the probability density is given by the Gaussian \cite{MV,Kov96}
    \beq\label{MV}
    Z_{\rm MV}[\rho] \approx \exp
    \left[
    -\frac{1}{2} \int^{\Lambda^{-1}}\!\!\!\!\dif^2{\bm{x}}\,
    \frac{\rho^a(\bm{x})\, \rho^a(\bm{x})}
    {\mu^2(\bm{x})}
    \right],
    \eeq
where $\mu^2 \sim \Lam^2 A^{1/3}$ is the average color charge
density squared. Even though there is no $Y$-dependence in the
model, a strong coherent color field of order $\cal{O}(1/g)$ can be
created due to the large number of nucleons and the solution of, the
non-linear, Eq.~(\ref{YMEq}) will lead to the saturation of the
gluon density at a value of order $\cal{O}(1/\alpha_s)$, and with a
saturation scale $Q_s^2(A) \approx \Lam^2 A^{1/3} \ln A$.

\begin{figure}[t]
    \centerline{\epsfig{file=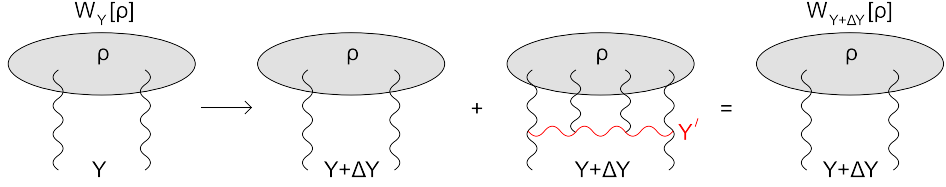,width=12.5cm}}
    \caption{\sl Evolution of the probability distribution of color charges.}
    \label{FigWevol}
\end{figure}

Now let us see how and why the evolution of the probability
distribution arises. Indeed, at rapidity $Y$, one should include in
the source all those modes with longitudinal momenta (much) larger
than $k^+$, where $Y=\ln(1/x)=\ln(P^{+}/k^{+})$ with $P^{+}$ the
total longitudinal momentum of the hadron. Now let us imagine that
we increase the rapidity from $Y$ to $Y + \Delta Y$. Then, some
modes that previously were ``slow'', now they become ``fast'' and
they need to be integrated over in order to be included in the
source. This is a procedure which will crucially depend on the
actual QCD dynamics and will lead to a change in the probability
distribution. Still, we should emphasize that the theory at the new
``scale'' $Y+\Delta Y$ is again defined through Eqs.~(\ref{YMEq})
and (\ref{Oave}), if we simply let $Y \to Y+\Delta Y$. A simple
pictorial interpretation of the evolution with $Y$ of the
probability distribution is shown in Fig.~\ref{FigWevol}, where the
red gluon at the rapidity $Y'$ is a representative of the
``semi-fast'' modes which are integrated, i.e.~$Y \ll Y' \ll
Y+\Delta Y$.

We shall not present here the derivation of this RGE, which is
called the JIMWLK equation (even though a ``sketch'' of such a
derivation is given in Appendix \ref{AppJIMWLK}). It can be found in
the original papers \cite{JKLW97,ILM01,FILM02,Wei02} (see also
\cite{JKW98,KMW00}), while rather simple derivations can be found in
\cite{Mue01a,HIMST05,Ko05}.  The basic element is the resummation of
$\abar \ln(1/x)$ enhanced contributions in the presence of a
background color field. The latter is allowed to be strong in
general, since one aims to describe properly the physics in the high
density region. This equation can be given an elegant Hamiltonian
formulation, and its most compact representation arises when it is
expressed in terms of the color field $\alpha(x^-,\bm{x}) \equiv
A^+(x^-,\bm{x})$ in the Coulomb gauge. It reads
    \beq\label{Wevol}
    \frac{\del}{\del Y}\, Z_{Y}[\alpha] =
    H\left[\alpha, \frac{\delta}{\delta \alpha}\right]
    Z_{Y}[\alpha],
    \eeq
where the explicit form of the Hamiltonian is
\cite{ILM01,FILM02,HIIM05}
    \beq\label{HJIMWLK}
    H = -\, \frac{1}{16 \pi^3}\!
    \int\limits_{\bm{u}\bm{v}\bm{z}}\!
    \cal{M}_{\bm{u}\bm{v}\bm{z}}
    \left[ 1
    +\tilde{V}_{\bm{u}}^{\dag} \tilde{V}_{\bm{v}}
    -\tilde{V}_{\bm{u}}^{\dag} \tilde{V}_{\bm{z}}
    -\tilde{V}_{\bm{z}}^{\dag} \tilde{V}_{\bm{v}}
    \right]^{ab}
    \frac{\delta}{\delta \alpha^a_{\bm{u}}}\,
    \frac{\delta}{\delta \alpha^b_{\bm{v}}}.
    \eeq
The dipole kernel\footnote{The dipole kernel appears in the above
equation when one assumes that the Hamiltonian will act on the
observables defined below in Eq.~(\ref{Ogen}). In general one should
replace the dipole kernel $\cal{M}_{\bm{u}\bm{v}\bm{z}}$ by
$2(\bm{u}-\bm{z})\dot(\bm{z}-\bm{v})/(\bm{u}-\bm{z})^2(\bm{z}-\bm{v})^2$.}
is readily recognized, while the Wilson lines $\tilde{V}$ in the
adjoint representation arise from the gluon propagator of the
integrated modes and they are given by
    \beq\label{WilsonVa}
    \tilde{V}^\dagger_{\bm{x}}[\alpha]=
    {\rm P}\,\exp\left[
    \rmi\, g \int\limits_{-\infty}^{\infty}
    \dif x^- \alpha^a(x^-,{\bm{x}})\, T^a
    \right].
    \eeq
The longitudinal coordinate in the functional derivatives is to be
taken at $\infty$, since $x^- \sim 1/k^+ \sim \rme^Y/P^+ \to
\infty$; as we pointed out earlier, the hadron extends in the
longitudinal direction, and all the ``action'' is expected to take
place at the last layer of rapidity. Finally, the two functional
derivatives appearing in Eq.~(\ref{HJIMWLK}) correspond precisely to
the two outgoing legs in the bottom of the diagrams in
Fig.~\ref{FigWevol}. The appearance of only two such legs is related
to the certain class of diagrams which are effectively resummed by
the JIMWLK equation; during a single evolution step, only the two
point correlation function $\lan \rho \rho \ran$ changes, and this
comes from the fact that the color field, or equivalently the
sources, are assumed to have values (much) larger than $g$. This is
certainly a well-defined approximation, but with a decisive
influence on the outcomes of the theory. We shall not discuss more
details here, but we shall return to this issue and to its proper
treatment, in the next sections.

Given a Hamiltonian, the natural question that one may ask is which
the observables are. They can be nothing else than the quantities
already appearing in the Hamiltonian, and to be more precise, they
will be gauge invariant operators built from Wilson lines. For our
purposes, we shall only consider Wilson lines in the fundamental
representation and then the most generic form of such an operator
will read
    \beq\label{Ogen}
    \mathcal{O[\alpha]} =
    \tr\big(V_{\bm{x}_1}^{\dagger} V_{\bm{x}_2}
    V_{\bm{x}_3}^{\dagger} V_{\bm{x}_4} \dots \big)
    \tr\big(V_{\bm{y}_1}^{\dagger} V_{\bm{y}_2} \dots \big)
    \dots
    \eeq
Notice that a Wilson line has a direct physical interpretation
\cite{Nac91}. Let us consider the scattering of a left-moving quark
off the classical field created by the (right-moving) hadron. The
general expression for the $S$-matrix in the interaction picture is
    \beq\label{Smatrix}
    S = {\rm T}
    \exp
    \left[-\rmi
    \int\limits_{-\infty}^{\infty} \dif^4x'\,
    H_{\rm I}(x') \right],
    \eeq
where T stands for a time ordered product, and the relevant part of
the QCD Hamiltonian for this process is $H_{\rm I} = -g \bar{\psi}
\gamma^{-} A^{+} \psi$. Here we have used the fact that the quark is
a fast left mover and therefore the $\gamma^{-}$ matrix dominates
the inner product. Now, since the trajectory is ``eikonal'', the
transverse and $+$ coordinates of the particle are fixed, and they
are chosen to be $\bm{x}'=\bm{x}$, and (by convention) $x'^+=x^+=0$.
Therefore, so long as the color independent part is concerned, one
can make the substitution
    \beq\label{eikonal}
    \bar{\psi}(x')\, \gamma^{-} A^{+}(x')\, \psi(x')
    \to
    \delta(\bm{x}'-\bm{x})\, \delta(x'^+\,) A^{+}(x).
    \eeq
Since $x^-$ is increasing along the trajectory, one can replace the
T-product by a Path ordered product. Therefore the $S$-matrix is
equal to the Wilson line $V_{\bm{x}}^{\dag}$ given in
Eq.~(\ref{WilsonVa}), but in the fundamental representation,
i.e.~one needs to do the replacement $T^a \to t^a$. Notice that all
kind of multiple exchanges are included in this procedure, and they
can be recovered by expanding the Wilson lines to a given order in
the coupling constant. It is not hard to understand that the
$S$-matrix for a dipole $(\bm{x},\bm{y})$ to scatter off the target
will be given by the gauge invariant expression
    \beq\label{SofV}
    S_{\bm{x}\bm{y}} =
    \frac{1}{N_c}\,
    \tr\big(V_{\bm{x}}^{\dagger} V_{\bm{y}}\big)
    = 1- T_{\bm{x}\bm{y}}
    = 1 - \frac{g^2}{4N_c}
    ( \alpha^a_{\bm{x}} - \alpha^a_{\bm{y}})^2 +
    \mathcal{O}(g^3),
    \eeq
where we have also expanded to second order to obtain the scattering
amplitude in the two-gluon exchange approximation for later
convenience, and with the field $\alpha^a$ in this expansion being
integrated over $x^-$.

As discussed earlier, the evolution of the expectation value of a
generic operator $\cal{O}$ will come from the evolution of the
probability distribution $Z_{Y}[\rho]$. Using Eqs.~(\ref{Oave}),
(\ref{Wevol}) and (\ref{HJIMWLK}), and after a functional
integration by parts we find
    \beq\label{dOdY}
    \frac{\partial
    \left\langle
    \mathcal{O}
    \right\rangle}
    {\partial Y}=
    \int \mathcal{D}\alpha \, Z_Y[\alpha]\, H\, \mathcal{O} =
    \left\langle H \,\mathcal{O}\right\rangle.
    \eeq
Now we are ready to write the equations obeyed by the scattering
amplitudes. Naturally, we start from the scattering of a single
projectile dipole. Using Eqs.~(\ref{HJIMWLK}), (\ref{SofV}) and
(\ref{dOdY}), and as we show explicitly in Appendix
\ref{AppBalitsky}, we arrive at the first Balitsky equation which we
rewrite here for convenience
    \beq\label{Bal1b}
    \frac{\del \lan S_{\bm{x}\bm{y}} \ran}{\del Y}=
    \atpi\,
    \int\limits_{\bm{z}}
    \cal{M}_{\bm{x}\bm{y}\bm{z}}
    \lan S_{\bm{x}\bm{z}}S_{\bm{z}\bm{y}}
    -S_{\bm{x}\bm{y}} \ran.
    \eeq
Notice that the above equation is valid for a finite number of
colors and the same will be true for the BFKL equation which arises
in the limit $\lan T \ran = 1- \lan S \ran \ll 1$. As we have
already discussed, this is not a closed equation and one needs to
find how $\lan S_{\bm{x}\bm{z}}S_{\bm{z}\bm{y}}\ran$ evolves with
rapidity. Following the same procedure, and as we show again in
Appendix \ref{AppBalitsky}, one obtains the second Balitsky
equation, which reads
    \beq\label{Bal2}
    \frac{\del \lan  S_{\bm{x}\bm{z}}S_{\bm{z}\bm{y}} \ran}
    {\del Y}=
    \atpi\,
    \int\limits_{\bm{w}}
    \aln \cal{M}_{\bm{x}\bm{z}\bm{w}}\,
    \lan (S_{\bm{x}\bm{w}} S_{\bm{w}\bm{z}}- S_{\bm{x}\bm{z}})
    S_{\bm{z}\bm{y}} \ran
    \nn
    +\atpi\,
    \int\limits_{\bm{w}}
    \aln
    \cal{M}_{\bm{z}\bm{y}\bm{w}}\,
    \lan S_{\bm{x}\bm{z}}
    (S_{\bm{z}\bm{w}} S_{\bm{w}\bm{y}} - S_{\bm{z}\bm{y}} )\ran
    \nn
    \aln\hspace{-2.2cm}
    +\frac{1}{2 N_c^2}\,
    \atpi\,
    \int\limits_{\bm{w}}
    (\cal{M}_{\bm{x}\bm{y}\bm{w}}
    \!-\!\cal{M}_{\bm{x}\bm{z}\bm{w}}
    \!-\!\cal{M}_{\bm{z}\bm{y}\bm{w}})
    \lan
    Q_{\bm{x}\bm{z}\bm{w}\bm{y}}+
    Q_{\bm{x}\bm{w}\bm{z}\bm{y}}
    \ran,
    \eeq
where we have defined the ``quadrupole'' operator
    \beq\label{Qop}
    Q_{\bm{x}\bm{z}\bm{w}\bm{y}}
    \equiv
    \frac{1}{N_c}
    \left[
    \tr \big( V^{\dag}_{\bm{x}} V_{\bm{w}} V^{\dag}_{\bm{z}}
    V_{\bm{y}} V^{\dag}_{\bm{w}} V_{\bm{z}} \big)
    -\tr \big( V^{\dag}_{\bm{x}} V_{\bm{y}} \big)
    \right].
    \eeq
At this point a few comments need to follow. We immediately notice
that the first two terms in the second Balitsky equation can be
obtained just by applying the Leibnitz differentiation rule to the
first Balitsky equation. This has a natural interpretation in terms
of projectile evolution as the two dipoles $(\bm{x},\bm{z})$ and
$(\bm{z},\bm{y})$ can evolve independently; either of these two
dipoles can split into two new dipoles which subsequently scatter
off the unevolved target hadron. However, there is the third term
which goes beyond this simple dipolar evolution. This is not
surprising since the small-$x$ gluon emitted by the one of the two
original dipoles can be absorbed by the other dipole (more
precisely, emitted by the other dipole in the complex conjugate
amplitude when calculating the wavefunction of the projectile) and
therefore no dipole survives after the evolution step. Thus, the
projectile system is lead to a more complicated multipolar state,
which can be naturally called a color quadrupole because of its
dependence on four transverse coordinates. For the next step one
will have to write not only the evolution equation for the
three-dipoles operator, but also the one for this quadrupole
operator. It becomes clear that the complexity in the structure of
this hierarchy of equations, which is called the Balitsky hierarchy,
will rapidly increase as we proceed to describe the evolution of
``higher-point'' functions. Following this ``active'' point of view
in the projectile evolution, we see that is wavefunction will be
successively composed of
    \beq
    {\rm one \,\,dipole} \aln \, \to\, {\rm two\,\, dipoles}
    \,\to\, {\rm three\,\, dipoles}\, +\, {\rm quadrupole} \nn
    \aln \,\to \,\dots \, \to\, {\rm dipoles} \,
    +\, {\rm higher\,\, multipolar\,\, states}. \nonumber
    \eeq
Furthermore, we should emphasize that no factorization like the one
in Eq.~(\ref{meanfield}) will solve the infinity hierarchy.
Nevertheless, we see that the two fixed points of the BK equation,
when generalized properly, satisfy this system of equations. ({\bf
i}) When the classical color field $\alpha^a$ vanishes there is no
scattering; the expectation values of all the Wilson lines become
equal to unity, i.e.~$\lan S \ran = \lan SS \ran =\lan Q \ran =\dots
=1$ and ({\bf ii}) when the color field becomes large, we reach the
black-disk limit; the Wilson lines oscillate rapidly \cite{IM01} and
their expectation values vanish; i.e.~$\lan S \ran = \lan SS \ran
=\lan Q \ran =\dots =0$.

One expects drastic simplifications in the large-$N_c$ limit where
the degrees of freedom can be chosen to be the color dipoles.
Indeed, the last (quadrupole) term in Eq.~(\ref{Bal2}) is of order
$\cal{O}(1/N_c^2)$, negligible at large-$N_c$ when compared to the
first two (dipolar) terms which are of order $\cal{O}(1)$ and
therefore the evolution will proceed only through dipolar states.
Now one can see that the hierarchy becomes consistent with a
factorization of the form \cite{LL04,Jan04}
    \beq\label{fact}
    \lan S_{\bm{x}_1\bm{y}_1} \dots
    S_{\bm{x}_{\kappa}\bm{y}_{\kappa}}\ran
    = c^{\kappa-1} \lan S_{\bm{x}_1\bm{y}_1} \ran
    \dots \lan S_{\bm{x}_{\kappa}\bm{y}_{\kappa}}\ran,
    \eeq
with $c$ an arbitrary constant. Then the whole large-$N_c$ hierarchy
collapses to a single equation, the BK equation but with a
coefficient $c$ in front of the $\lan S_{\bm{x}\bm{z}}\ran \lan
S_{\bm{z}\bm{y}}\ran$ term. The asymptotic fixed point of this
equation is clearly $1/c$, but it is not clear whether a value $c
\neq 1$ has a simple physical interpretation or not. In any case, if
the initial conditions at a rapidity $Y_0$ are of factorized form,
this factorization will be preserved by the large-$N_c$ evolution.
But even if the initial conditions are not of this type, no new
``correlations'' will be generated by this large-$N_c$ evolution;
only the initial ones will be propagated to higher rapidities.

\section{The Saturation Momentum}\label{SecSatMom}

With the theory of the Color Glass Condensate being theoretically
established, one of the central problems has been the determination
of the saturation momentum $Q_s$ which, as we discussed at the end
of Sec.~\ref{SecSat}, is the momentum scale at which we start to
approach unitarity limits. In principle one should be able to
calculate the energy (rapidity) dependence of $Q_s$ (at least
asymptotically), since all the dynamics is contained in the JIMWLK
equation, however its precise value cannot be determined since it
will depend on initial conditions and thus on details of
non-perturbative physics. Recall that the saturation momentum is an
intrinsic property of the target hadron, since it is also the scale
where the gluon density saturates. And the only scale associated
with the hadron is $\Lam$. Given the above considerations, and the
fact that the pure BFKL evolution leads to an exponential growth,
one may guess that the saturation momentum will be of the form (for
fixed coupling, neglecting the impact parameter dependence and at
asymptotic energies)
    \beq\label{Qs}
    Q_s^2 \simeq c\, \Lam^2 \frac{\exp(\lambda Y)}{Y^{\beta}},
    \eeq
where the constants $\lambda$ and $\beta$ should be calculable,
while the constant $c$ would require the knowledge of the
non-perturbative structure of the hadron.

Even though the problem seems difficult at a first sight due to the
complicated structure of the Balistky-JIMWLK equations, one can do
certain logical simplifications that render the calculation
tractable. As we saw, the hierarchy reduces to the much simpler BK
equation in the large-$N_c$ limit and under a mean-field
approximation, and then, presumably, the most that one may lose is
$1/N_c^2$ corrections. Furthermore, one can in fact use BFKL
dynamics, if one is careful enough to treat properly the effects of
the linear terms. In turn, this means that we may not even lose the
possible $1/N_c^2$ corrections, since the BFKL equation is valid at
finite-$N_c$.
\begin{figure}[t]
    \centerline{\epsfig{file=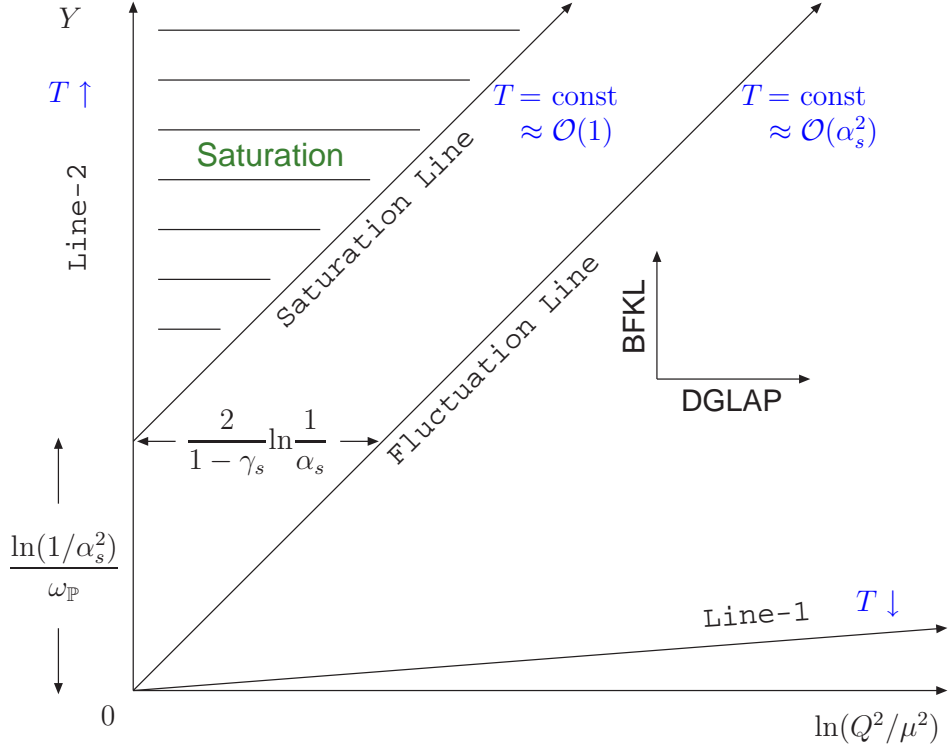,width=12.5cm}}
    \caption{\sl Scattering amplitude and saturation momentum
    in the logarithmic plane.}
    \label{FigPlane}
\end{figure}

Now let us have a look at Fig.~\ref{FigPlane} where the expected
qualitative behavior of the scattering amplitude in the logarithmic
plane $\ln(Q^2/\mu^2)-Y$ is shown. Here $Q$ should be thought as the
inverse of the dipole size or as the transverse momentum of a gluon
in the hadron. Along Line-1, which corresponds to an anomalous
dimension $\gamma \to 0$, the amplitude decreases since the momentum
increases too fast. This corresponds to DGLAP evolution (in the
double logarithmic limit) where one resums powers of $\alpha_s
\ln(Q^2/\mu^2)$ ($\times  \ln(1/x)$). In this case there is again a
cascade of partons inside the hadron, but these partons are very
small in size and they will never overlap to form a high density
system. Along Line-2, which corresponds to an anomalous dimension
$\gamma_{\mathbb{P}}=1/2$, the amplitude increases, since the energy
increases while the momentum remains fixed. This is the hard Pomeron
intercept line and corresponds to the saddle point of the BFKL
eigenvalue function. Starting from a value of order
$\cal{O}(\alpha_s^2)$ and after covering a rapidity interval $\sim
\ln(1/\alpha_s^2)/\omega_{\mathbb{P}}$, the amplitude will
eventually ``hit'' the unitarity/saturation region.

Therefore, there must be some ``critical'' line between those two
lines which belongs to the region of linear evolution and along
which the amplitude remains constant, for example of order
$\cal{O}(\alpha_s) \ll 1$. Clearly this line must correspond to an
anomalous dimension, say $\gamma_s$, such that $0 < \gamma_s <1/2$.
Then the saturation line which corresponds to constant amplitude of
order $\cal{O}(1)$ but smaller than 1, for example 1/2, will be
parallel to the critical line, and therefore characterized by the
same anomalous dimension and energy dependence. The reason why the
two lines are parallel, may not be so clear at the moment, but we
will try to justify it in a while.

The question is whether or not we can use the BFKL dynamics to
determine this critical line. Even though the line belongs to the
linear region, the answer is negative if one wants to get the
prefactors correct, i.e.~the value of $\beta$ in Eq.~(\ref{Qs}). The
reason is that as the system evolves along this line, and after we
have isolated the leading exponential behavior, there we will be
some paths, really in the functional integral sense, that go through
the saturation region and then return to the linear one. These
diffusive paths are absent when the full BK equation is equation is
considered, and therefore should be cut. This can be done by using
an absorptive boundary just beyond the saturation line, which will
mimic the effects of the non-linear terms for the problem under
consideration. Put it another way, this procedure is equivalent to a
self-consistent solution of the BFKL equation with the boundary
condition $\lan T (Q=Q_s(Y)) \ran = c < 1$.

Now we are ready to determine the ``slope'' of the critical, and
therefore of the saturation, line. We impose two conditions for the
exponent in the solution to the BFKL equation as given in
Eq.~(\ref{BFKLsol})\footnote{For our convenience, here we consider a
case where the amplitude is averaged over impact parameter and
therefore it is dimensionless. Then it is still given by
Eq.~(\ref{BFKLsol}) but the factor $\sim 2\pi/\mu^2$ is not
present.}. First we require the exponent to be zero so that the
amplitude be constant. Second we impose a saddle point condition,
which is a valid approximation in the asymptotic limit $\abar Y \gg
1$. Then we find that the anomalous dimension is determined by the
saddle-point of $\chi(\gamma)/(1\-\gamma)$ and is given by
\cite{GLR,Mue99a,IIM02,MT02,MP04a}
    \beq\label{gammas}
    \chi(\gamma_s) + (1\-\gamma_s)\, \chi'(\gamma_s)=0
    \Rightarrow \gamma_s = 0.372.
    \eeq
The energy dependence of the saturation momentum is better expressed
in terms of its logarithmic derivative which reads \cite{MT02,MP04a}
(with the leading term already known from \cite{Mue99a,IIM02})
    \beq\label{lambda}
    \lambda_s \equiv \frac{\dif \ln Q_s^2}{\dif Y} =
    \abar\,\frac{\chi(\gamma_s)}{1\-\gamma_s} -
    \frac{3}{2 (1\-\gamma_s)} \,\frac{1}{Y} =
    4.88 \,\abar - \frac{2.39}{Y},
    \eeq
while for the scattering amplitude one obtains \cite{MT02,MP04a}
(and with the absence of the logarithmic modification, already known
from \cite{IIM02})
    \beq\label{Tscale}
    \lan T \ran= \left( \frac{Q_s^2}{Q^2}\right)^{1-\gamma_s}\!\!
    \left(\ln \frac{Q^2}{Q_s^2} + \rm{c} \right)
    \exp \left[ - \frac{\ln^2(Q^2/Q_s^2)}{D_s Y}\right],
    \eeq
a form which is valid in the region $Q_s^2 \ll Q^2 \ll Q_s^2
\exp(D_s Y)$, and where the diffusion coefficient is $D_s = 2 \abar
\chi''(\gamma_s) = 97 \abar$.

Now let us discuss the results. As we see in Eq.~(\ref{gammas}),
whose graphical solution is shown in Fig.~\ref{FigChi}, the relevant
value of the anomalous dimension for saturation lies indeed in the
interval $(0,1/2)$. We should say that the eigenvalue
$\chi(\gamma_s)$ will be selected, so long as the initial condition
contains the corresponding eigenfunction and this will be true for
all interesting cases. Notice that $\gamma_s$ is a pure number, and
this would have never been obtained by applying pure DGLAP
evolution. The latter always gives anomalous dimensions which start
at order $\cal{O}(\alpha_s)$.

In Eq.~(\ref{lambda}) we see that the leading contribution to the
``intercept'' of the saturation line is totally  fixed by BFKL
dynamics. It is not too difficult to see how the subleading
correction arises, and to this end let us go a few steps back in the
derivation of Eqs.~(\ref{lambda}) and (\ref{Tscale}). After we have
performed the Gaussian integration around the saddle point
$\gamma_s$, the solution reads $T \propto(Q_0^2/Q^2)^{1-\gamma_s}
\psi_s$, with $Q_0$ containing only the leading behavior of $Q_s$,
and where $\psi_s$ satisfies the diffusion equation. The
``standard'' solution to the diffusion equation behaves as $Y^{1/2}$
(times the exponential diffusion factor), but in the presence of an
absorptive boundary the survival probability of the ``particle''
becomes smaller and $\psi_s$ is proportional to $1/Y^{3/2}$. Thus,
by combining this prefactor and the leading behavior $Q_0$, one
obtains the correction written in Eq.~(\ref{lambda}). Notice that
the saturation effects lead to a slower increase of the saturation
momentum, as they should. Even though the second term vanishes when
$Y \to \infty$, it cannot be neglected, since upon integration of
Eq.~(\ref{lambda}) it will generate a $Y$-dependent prefactor in
$Q_s$. This $1/Y$ term in the logarithmic derivative of the
saturation momentum should be interpreted as $1/R_{D}^2$, where $R_D
\sim \sqrt{Y}$ is the diffusion radius. This radius is practically
the available phase space in logarithmic units of transverse
momentum. In reality this phase space extends to infinity, since
there is no boundary to the ultraviolet, but in practice it is only
the space inside the diffusion radius which will contribute to the
subsequent steps of evolution.

Finally, as we see in Eq.~(\ref{Tscale}), the scattering amplitude
has a scaling form \cite{IIM02,MT02}, so long as $\ln(Q^2/Q_s^2) \ll
\sqrt{D_s Y}$ so that the exponential diffusion factor can be set
equal to unity. That is, the amplitude does not depend on $Q^2$ and
$Y$ separately, but only through the variable $Q^2/Q_s^2$. The pure
power is not an unexpected result; the BFKL evolution generates the
anomalous dimension $\gamma_s$ which modifies the behavior $\sim
1/Q^2$ of fixed order perturbation theory. The logarithm is
generated by the absorptive boundary when solving the diffusion
equation. Notice that both terms in Eq.~(\ref{Tscale}), the power
and the power modified by the logarithm, are exact degenerate
solutions to the BFKL equation \cite{Tri03}, and thus the final
solution is just their linear combination. Such a scaling behavior,
which will be preserved even in the running coupling case (but in a
more narrow window in $Q^2$), but which will be violated by the
evolution equations that we will discuss later on in
Sec.~\ref{SecPomSplit}, is consistent with fits
\cite{GBW99,GBKS01,IIM04} of the low-$x$ data in DIS.

Here it is appropriate to mention that the solution of the BK
equation close to the unitarity limit $\Lam^2 \ll Q^2 \ll Q_s^2$ is
also of scaling form and as we show in Appendix \ref{AppBKSol} it
reads \cite{MS96,LT00,IM01}
    \beq\label{Tsat}
    \lan S \ran= 1-\lan T \ran  \approx
    \exp \left[
    -\frac{1-\gamma_s}{2 \chi(\gamma_s)}\,
    \ln^2 \frac{Q_s^2}{Q^2}
    \right].
    \eeq
Thus, one naturally expects the scattering amplitude to satisfy
scaling everywhere from deep inside the saturation region up to
momenta which belong in the region of linear evolution. This
explains why the critical and saturation lines are parallel to each
other.

Eqs.~(\ref{gammas}), (\ref{lambda}) and (\ref{Tscale}) have been
confirmed by studying the analogy to the (nonlinear) FKPP (Fisher,
Kolmogorov, Petrovsky, Piscounov) equation \cite{MP03,MP04a}.
Furthermore, the last asymptotic term of $\lambda_s(Y)$ which is
independent of the initial conditions, and whose behavior is $ \sim
1/Y^{3/2}$ has been obtained in the same fashion in \cite{MP04b}.
The full JIMWLK hierarchy has been solved numerically on the lattice
\cite{RW03} and the results agree with the ones we presented in this
section. Moreover it was found that violations of factorization in
the scattering amplitude correlations are extremely small.

Before closing this section, let us see how these results are
modified when we consider BFKL dynamics at the next to leading
level. The calculation of the next to leading order (NLO) correction
to the BFKL kernel was completed in \cite{FL98,CC98}. However, this
negative correction turned out to be larger in magnitude than the
leading contribution for reasonable values of the coupling, say
$\abar \approx 0.25$. Even worse, when $\abar \gtrsim 0.05$ the full
kernel has two complex saddle points which lead to oscillatory cross
sections \cite{Ros98}. But it was immediately recognized that these
large corrections emerge from the collinearly enhanced physical
contributions \cite{Sal98,CC99,CCS99,Sal99}. A method was developed
to resum collinear effects to all orders in a systematic way and the
resulting renormalization group (RG) improved BFKL equation was
consistent with the leading order DGLAP \cite{DGLAP} equation by
construction.

But before going to the NLO case, let us consider the leading BFKL
kernel, but with a running coupling $\alpha_s(Q^2)$. One expects two
major changes with respect to the fixed coupling case. Since the
saturation momentum increases with rapidity, the coupling will
decrease as we evolve close to the saturation line, and $Q_s$ will
increase much slower. Moreover, the integration of the quadratic
fluctuations around $Q_s$ will be affected by the fact that the
coupling is not a constant quantity, and therefore the mechanism of
diffusion will be modified. An analytic expression can be given and
it reads \cite{MT02,Tri03} (with the leading terms already known
from \cite{GLR,IIM02})
    \beq\label{lambdarun}
    \lambda_s =
    \frac{\chi(\gamma_s)}{b(1-\gamma_s)}
    \left(
    \frac{1}{\tau}
    -\frac{|\xi_1|D_r}{4}\,
    \frac{1}{\tau^{5/3}}
    \right) =
    \frac{1.80}{\sqrt{Y+Y_0}} -
    \frac{0.893}{(Y+Y_0)^{5/6}}
    \eeq
for the logarithmic derivative of the saturation momentum, and
    \beq\label{Tscalerun}
    \lan T \ran =
    \left(\frac{Q_s^2}{Q^2}\right)^{1-\gamma_s}\!
    \tau^{1/3}
    {\rm Ai}\left(\xi_1 + \frac{\ln(Q^2/Q_s^2)+c}{D_r \tau^{1/3}}
    \right)
    \exp\left[
    -\frac{2 \ln^2(Q^2/Q_s^2)}{3 \chi''(\gamma_s) \tau}
    \right]
    \eeq
for the scattering amplitude. Here we have defined $b=(11N_c -
2N_f)/(12N_c)$, $\tau = \sqrt{2 \chi(\gamma_s)(Y+Y_0)/[b
(1\-\gamma_s)]}$, $D_r=\{\chi''(\gamma_s)/[2
\chi(\gamma_s)]\}^{1/3}=1.99$, Ai is the Airy function and $\xi_1 =
-2.33$ is the location of its leftmost zero. The second expression
for $\lambda_s$ corresponds to $N_f=3$ flavors. Notice that the
first term in $\lambda_s$ is equal to the fixed coupling result, in
the sense that it may be written as
$\chi(\gamma_s)\abar(Q_s^2)/(1\-\gamma_s)$. The second term is
negative since it accounts for the contribution of the boundary (and
the prefactors) and it has a parametric form $\alpha_s^{5/3}$
\cite{Tri03}, a well-known type of correction in NLO BFKL dynamics
\cite{KM98,CCS99}. We should notice that in this case, $Q_s$ will be
always proportional to $\Lam$ at high rapidities $Y \gg Y_0$
\cite{MT02,Tri03,Mue03}. This is in contrast to the fixed coupling
case where $Q_s$ is proportional to the initial saturation scale, if
such one exists. For example if the target hadron is a large
nucleus, the square of the initial saturation momentum and thus the
constant $c$ in Eq.~(\ref{Qs}) is proportional to $A^{1/3} \ln A$.
Finally we note that the expression for the scattering amplitude
takes a scaling form in the window $\ln(Q^2/Q_s^2) \ll D_r
\tau^{1/3}$, and this form is exactly the one found in the fixed
coupling case.
\begin{figure}[t]
    \centerline{\epsfig{file=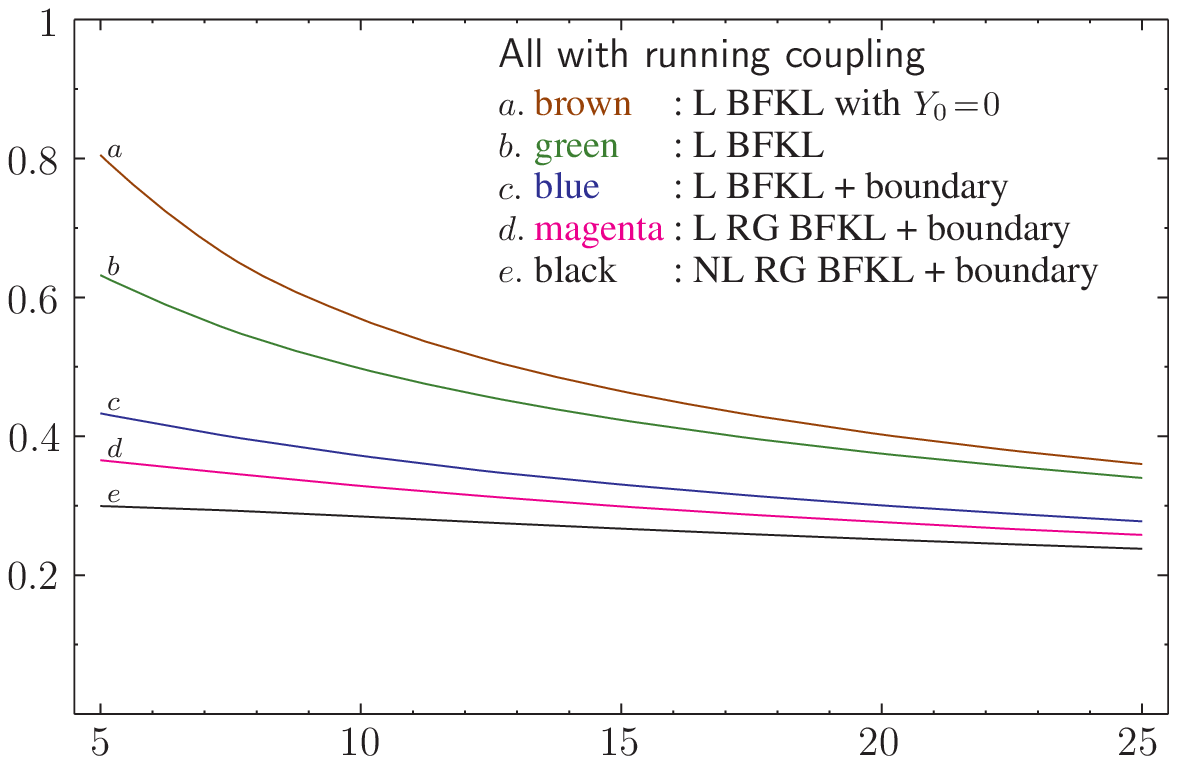,width=12.5cm}}
    \caption{\sl The logarithmic derivative of the
    saturation momentum $\lambda_s$ as a function of the rapidity $Y$
    for running coupling.}
    \label{FigLambda}
\end{figure}

Even though one cannot have such nice analytic expressions when
considering the improved kernels, the results are not so hard to
obtain since they involve the numerical solution of algebraic
transcendental equations \cite{Tri03}. We summarize all cases in the
plots in Fig.~\ref{FigLambda}. In order to understand properly all
the results, we have also plotted the first three lines which
correspond to the analytic expression given in
Eq.~(\ref{lambdarun}). Line-a is the first term with $Y_0 = 0$ while
Line-b corresponds to the same term but with a typical value for
$Y_0$ which is of order $\cal{O}(1)$. Line-c stands for the full
expression in (\ref{lambdarun}). Line-d and Line-e represent the
improved kernels at leading and next to leading order respectively.
Notice that all the lines will ``merge'' together at very high
values of rapidity, since the leading term will become more and more
dominant over the corrections as the coupling decreases along $Q_s$.
The NLO result is stable in the sense that it adds a small
correction to the leading one. We should mention that the NLO result
is practically constant, $\lambda_s \simeq 0.3$ in the region
$Y=5-10$. This is in good agreement with the phenomenology
\cite{GBW99,GBKS01,GLLM03,IIM04}. However, it is not clear whether
this theoretical value should be trusted or not. As we shall see in
the forthcoming sections, the JIMWLK equation misses important
contributions, which result in corrections that are much more
significant than any NLO BFKL correction.

\section{Deficiencies of the Balitsky-JIMWLK Hierarchy}\label{SecDefBal}

Even though the Balitsky-JIMWLK hierarchy encompasses nicely the
BFKL evolution and the merging of Pomerons at finite-$N_c$, it faces
certain crucial problems. These problems, which we will immediately
discuss, are related to each other, and therefore it seems that some
unique element is missing from the CGC effective theory in the form
it has been developed so far.

({\bf i}) The first problem is the extreme sensitivity of the JIMWLK
equation to the ultraviolet. Since in the high momentum region the
non-linearities are unimportant, we understand that this issue can
be analyzed within the BFKL evolution. Let us try to reconstruct the
solution in two (or more) global steps by completeness. To be more
precise, let us first assume that we find the solution $T(Q,Y)$ by
evolving the system from zero rapidity to rapidity $Y$. Now let us
imagine that we evolve from zero to, say, $Y/2$. Then we can
consider the solution $T(Q,Y/2)$ as the initial distribution and
subsequently evolve to $Y$ to obtain $T(Q,Y)$. As we show in
Appendix \ref{Apptwosteps}, the solution obtained from this
procedure will coincide with the one obtained from the single global
evolution step, so long as we include (at least) the contributions
from all momenta such that $\ln(Q^2/Q_s^2) \lesssim \sqrt{D_s Y/2}$,
in the initial condition at $Y/2$. Of course there is no reason to
cut the momenta that lie outside the diffusion radius, but this
algorithm reveals the width of the momentum phase space which is
important for a self-consistent solution. This feature brings us in
a quite embarrassing situation; as $Y$ increases, this phase space
will open up more and more to momenta above the saturation line, and
moreover the big numerical value of the coefficient $D_s$ will make
the problem even worse. For instance, when one finds the saturation
momentum to be a few GeV, at the same time one is sensitive to
momenta a few orders of magnitudes above. Notice, that this explains
why in the numerical solutions of both the BK
\cite{GBMS02,AB01a,Lub01,AAKSW04,AAMSW05,MS05} and the JIMWLK
equation \cite{RW03}, one had to go very far to the ultraviolet in
order to obtain a reasonably accurate solution. In the running
coupling case the situation is somewhat better, since the coupling
decreases at higher momenta, and thus the effects of these seemingly
non-physical contributions are reduced. Indeed, as we saw in
Sec.~\ref{SecSatMom}, the ``diffusion'' radius increases much
slower, namely $R_D \sim Y^{1/6}$. Nevertheless, the theoretical
problem still exists.

({\bf ii}) Quite surprisingly, the second problem is the violation
of unitarity. Here we shall not analyze the argument in detail
\cite{MS04}, but only indicate its essence. Assume that we want to
calculate the amplitude close to, but above, the saturation line in
the two ways we described in the previous paragraph. Then one will
have
    \beq\label{Tviolation}
    1 > c = T \sim \frac{1}{\alpha_s^2}\, T_1\, T_2,
    \eeq
where $T_1$ and $T_2$ denote the contributions of the two successive
steps. It is clear that for $T_1 < \alpha_s^2$ the above equation
imposes that the second step satisfy $T_2
>1$. Thus, all the paths which go through the region to the right of
the fluctuation line (the terminology will be understood in a while)
in Fig.~\ref{FigPlane} will violate unitarity in the intermediate
steps of the evolution. Returning to the problem we discussed in
({\bf i}), and noticing that the diffusion radius extends to the
region where the amplitude can be much smaller than $\alpha_s^2$, we
see that these contributions from the ultraviolet region must be
indeed non-physical.

These unitarity violating paths were ``discovered'' when
dipole-dipole scattering was studied in the presence of saturation
with the additional, but clearly natural, condition that the
amplitude be Lorentz invariant \cite{MS04}, even though at that time
it was not realized that they are part of JIMWLK evolution. A
calculation of the saturation momentum was performed by cutting
these paths with an absorptive UV boundary (in addition to the IR
one), and the corrections found were huge. In Sec.~\ref{SecSatRev}
we shall review/obtain this result in a very simple way. We should
say here, that in \cite{IMM05} a spectacular analogy with the
physics and the results of the stochastic FKPP (sFKPP) equation was
observed, and the significance of fluctuation effects in the low
density region $T \sim \alpha_s^2 \Leftrightarrow \varphi \sim
\alpha_s$ was understood. Still, the fact that one needs to go
beyond the JIMWLK equation was not yet realized.

({\bf iii}) The third problem in the evolution equations we have
discussed so far, is the absence of Pomeron splittings \cite{IT05a}.
As we have seen, the mechanism of Pomeron mergings, which was
essential to describe properly the physics near the unitarity limit,
is described by the JIMWLK equation. If we expand the Hamiltonian in
powers of $g$, then each term will involve at least two factors of
the color field $\alpha$ and exactly two functional derivatives with
respect to $\alpha$. Then a typical term in the evolution equation
of the $n$-th point correlator of the color fields will have the
structure (suppressing the coordinates' dependence, which is not
important for the forthcoming argument)
    \beq\label{dandy}
    \frac{\del \lan \alpha^n \ran}{\del Y} =
    \lan H \,\alpha^n \ran
    \sim
    \bigg\langle \underbrace{\alpha\, \alpha ... \alpha}_{ \geq 2}
    \frac{\delta}{\delta \alpha}\,
    \frac{\delta}{\delta \alpha}\,
    \alpha^n \bigg\rangle
    \sim \lan \alpha^m \ran
    \quad {\rm with} \quad  m \geq n.
    \eeq
So, as we already knew, the JIMWLK Hamiltonian can describe BFKL
dynamics and Pomeron mergings. But the natural question at this
point is, ``how could we have two or more ladders in the first
place?'' Clearly JIMWLK cannot do that, since one would need $m <n$
in Eq.~(\ref{dandy}). One option would be to consider a large
nucleus target, which contains many valence quarks and antiquarks.
These sources can evolve with rapidity and produce many BFKL
Pomerons, which will merge when saturation becomes important.
However, this is just a special case because of its particular
initial condition, and the dynamical problem is not solved.
Furthermore, even in the nucleus, there will always be some dilute
``tail'' corresponding to the high momentum modes. After some
evolution, and since the saturation momentum increases, these modes
will need to saturate. But still there is no dynamics to produce the
corresponding Pomerons which will eventually merge. Thus, the only
solution to this problem is to find how QCD will give rise to the
Pomeron splittings. Then indeed, one could start, for example, even
from a single bare dipole, and end up with a fully saturated
wavefunction. As we shall see in the following sections, when
``completing'' the theory by including the diagrams which were
``forgotten'' \cite{IT05a,IT05b}, and which lead to the splittings
of Pomerons, we will also automatically solve the two problems
presented in ({\bf i}) and ({\bf ii}).

\section{The Missing Diagrams}\label{SecMissing}

\begin{figure}[t]
    \centerline{\epsfig{file=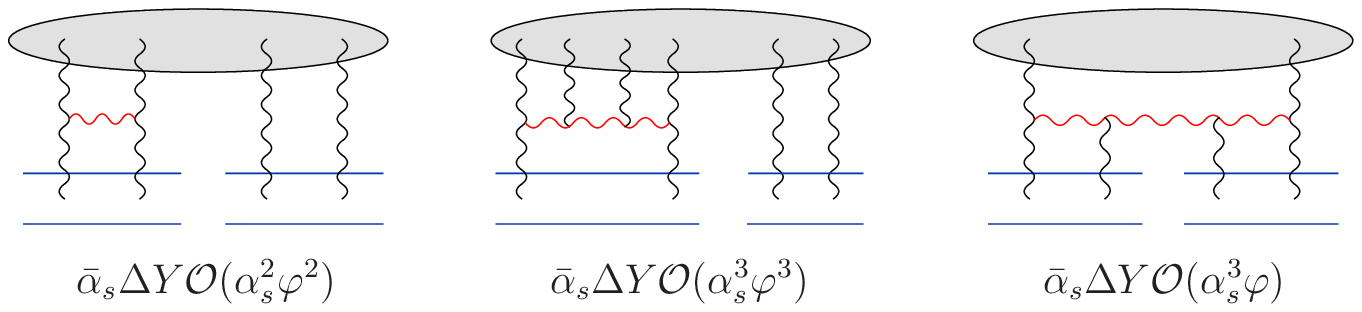,width=12.5cm}}
    \caption{\sl Diagrams contributing to the evolution of two BFKL Pomerons.
    The third diagram was ``missed'' in JIMWLK.}
    \label{FigSplitting}
\end{figure}

Since one of the basic problems of the JIMWLK equation is the
absence of Pomeron Splittings, one should consider diagrams like the
third one in Fig.~\ref{FigSplitting} in order to resolve the issue.
Indeed, this diagram corresponds to a transition from one to two
Pomerons. If one wants to ``measure'' two Pomerons, and in the
two-gluon exchange approximation, one needs to probe the hadron wave
function with two projectile dipoles, and this is what is shown in
the typical diagrams in Fig.~\ref{FigSplitting}. The first diagram
corresponds to normal BFKL evolution, with the one of the Pomerons
being a spectator. As indicated, it is of order $\abar \Delta
Y\cal{O}(\alpha_s^2 \varphi^2)$; $\abar \Delta Y$ for the evolution
step, $\alpha_s^2$ for the four fermion-gluon vertices and
$\varphi^2$ since there were two Pomerons before the step. Similarly
the second diagram, which corresponds to a $2 \to 1$ Pomeron
transition, with the third Pomeron being a spectator, is of order
$\abar \Delta Y\cal{O}(\alpha_s^3 \varphi^3)$. Both diagrams are
described by the JIMWLK equation\footnote{Also diagrams like the
first one, but with the soft gluon connecting the two ladders are
included; these diagrams are of the same order in $\alpha_s$ and
$\varphi$, but they are suppressed at large $N_c$.}. The third
diagram is of order $\abar \Delta Y\cal{O}(\alpha_s^3 \varphi)$,
since there are six vertices and one Pomeron before the step.
Clearly this diagram is suppressed with respect to at least one of
the first two diagrams when $\varphi \gg \alpha_s$, and this is the
reason why it was neglected in the derivation of the JIMWLK
equation, which aimed to describe a high density system. However,
due to the non-locality of the evolution kernel which leads to the
ultraviolet diffusion, this diagram will give significant
contributions through the intermediate evolution steps. It is rather
crucial to notice that this diagram becomes important when $\varphi
\sim \alpha_s$ or equivalently when $T \sim \alpha_s^2$, which
corresponds precisely to the line at which JIMWLK faces its
unitarity problem. This is the fluctuation line we have already
shown in Fig.~\ref{FigPlane}, and which divides the region of normal
BFKL evolution from the low density region. The latter is
characterized by fluctuations and this explains in a way why we need
(at least) two projectiles dipoles.
By probing with one dipole we will measure only the ``one-point''
function, say the gluon occupation number. But this will not release
any information about the two-point function, since in a dilute
system the pair density is affected by the fluctuations and is not
simply the square of the single density. Thus one needs to probe
with two dipoles in order to measure the non-trivial low density
correlations. This analysis also implies that, presumably, the first
Balitsky equation will not change. But of course, any mean field
approximation to this equation will not be valid any more, since the
presence of fluctuations will have a significant impact on its
non-linear term. In the next section we shall shortly describe some
elementary but essential, for our purposes, features of the color
dipole picture. We shall return to give a more quantitative analysis
of what we discussed here, in Sec.~\ref{SecPomSplit}.

\section{Evolution of Dipoles}\label{SecDip}

Since we have realized that what we need to correct in the JIMWLK
equation is its low-density limit, let us look more carefully at the
non-saturated part of the wavefunction of a hadron. We will first
consider the large-$N_c$ limit case, that is, the hadron is supposed
to be composed of color dipoles. When the dipole density is not too
high, i.e~when $n \ll 1/\alpha_s^2$, the emission of a soft gluon
from a dipole will not be affected by the remaining surrounding
dipoles. We have already written the (positive and well-defined)
probability for this emission in Eq.~(\ref{SplitProb}) with the
derivation given in Appendix \ref{AppDipKer}. From now on we shall
not present the original formulation of this dipole picture
\cite{Mue94,MP94,Mue95,CM95}, rather we will follow the procedure
developed in \cite{IM04a} which is better suited to our purposes.

\begin{figure}[t]
    \centerline{\epsfig{file=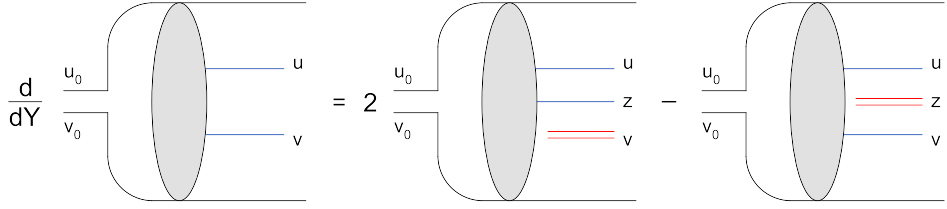,width=12.5cm}}
    \caption{\sl Evolution of the dipole density for a dilute hadron
    at large $N_c$.}
    \label{FigDipEvol}
\end{figure}

One can write a master equation describing the evolution of the
probabilities to find a given configuration. To be more specific, a
given configuration is characterized by the number of dipoles $N$
and by $N-1$ transverse coordinates $\{\bm{z}_{\i}\}=\{\bm{z}_1,
\bm{z}_2,\dots,\bm{z}_{N-1}\}$, such that the coordinates of the $N$
dipoles are $(\bm{z}_0,\bm{z}_1)$,
$(\bm{z}_1,\bm{z}_2)$,\dots,$(\bm{z}_{N-1},\bm{z}_N)$, with
$\bm{z}_0 \equiv \bm{u}_0$ and $\bm{z}_N \equiv \bm{v}_0$, assuming
that the initial state was a dipole $(\bm{u}_0,\bm{v}_0)$. The
probability $P_N(\{\bm{z}_{\i}\};Y)$ to find a given configuration
at rapidity $Y$ obeys
    \beq\label{Pevol}
    \frac{\del P_N(\bm{z}_1,...,\bm{z}_{N-1};\!Y)}
    {\del Y}
    =\aln \frac{\abar}{2\pi}
    \!\sum_{\i=1}^{N-1}
    \!\mathcal{M}(\bm{z}_{\i-1},\bm{z}_{\i+1},\bm{z}_{\i})
    P_{N-1}
    (\bm{z}_1,...,\slashed{\bm{z}}_{\i},...,\bm{z}_{N-1};\!Y)
    \nn
    \aln - \frac{\abar}{2\pi}
    \sum_{\i=1}^N \int\limits_{\bm{z}}
    \mathcal{M}(\bm{z}_{\i-1},\bm{z}_{\i},\bm{z})\,
    P_N(\bm{z}_1,...,\bm{z}_{N-1};\!Y),
    \eeq
where a slashed variable is omitted. The interpretation is quite
obvious; while the gain (first) term describes the formation of an
$N$-dipoles state through the splitting of a dipole in a
pre-existing state with $N-1$ dipoles, the second term describes the
emission of a soft gluon from the $N$-dipoles state, which leads to
a state with $N+1$ dipoles and therefore to a loss of probability.
It is straightforward to show that the total probability is
conserved, which is one of the requirements that Eq.~(\ref{Pevol})
be well-defined. The expectation value of an operator $\mathcal{O}$
which depends only on dipole coordinates is given by
    \beq\label{opeave}
    \lan \mathcal{O}(Y) \ran \,=\, \sum_{N=1}^\infty\int \dif
    \Gamma_N\,P_N(\{\bm{z}_{\i}\};Y)\, \mathcal{O}_N(\{\bm{z}_{\i}\}),
    \eeq
where the phase space integration is simply $\dif\Gamma_N\,=\,{\rm
d}^2\bm{z}_1{\rm d}^2\bm{z}_2\dots{\rm d}^2\bm{z}_{N-1}$. Then by
using the master equation (\ref{Pevol}) one can show that
    \beq\label{DODYb}
    \frac{\del \lan \mathcal{O}(Y) \ran}{\del Y}=
    \frac{\abar}{2\pi}
    \sum_{N=1}^\infty \int \aln \dif \Gamma_N \,
    P_N(\{\bm{z}_{\i}\};Y)\,
    \sum_{i=1}^N \int\limits_{\bm{z}}\,
    {\mathcal{M}}({\bm{z}}_{\i-1},{\bm{z}}_{\i},{\bm z})
    \nn
    \aln \times \left[- {\cal O}_N(\{\bm{z}_{\i}\})
    + {\cal O}_{N+1}(\{\bm{z}_{\i},\bm{z}\})\, \right],
    \eeq
where the $\bm{z}$ argument in ${\cal O}_{N+1}$ is to be placed
between $\bm{z}_{\i-1}$ and $\bm{z}_{\i}$.  In what follows, we
shall use Eq.~(\ref{DODYb}) to derive evolution equations for the
dipole number densities. Consider first the average number density
of dipoles at $(\bm{u},\bm{v})$. The corresponding part for an
$N$-dipole configuration is
\begin{equation}\label{densityN}
    n_{N}(\bm{u},\bm{v})=
    \sum_{\j=1}^N
    \delta(\bm{z}_{\j-1}-\bm{u})
    \delta(\bm{z}_\j-\bm{v}),
\end{equation}
so that $\lan n_{\bm{u}\bm{v}} \ran$ will be given by the r.h.s.~of
Eq.~(\ref{opeave}) if we let $O_N \to n_N$. Then by making use of
Eq.~(\ref{DODYb}) and after relatively simple manipulations one
arrives at the evolution equation for the average of the dipole
number density $\lan n_{\bm{u}\bm{v}} \ran$ which reads
    \beq\label{nBFKL}
    \frac{\del \lan n_{\bm{u}\bm{v}} \ran}{\del Y}
    \aln =
    \frac{\abar}{2\pi} \int\limits_{\bm{z}}
    \left[
    {\cal M}_{{\bm{u}}{\bm{z}}{\bm{v}}}
    \lan n_{\bm{u}\bm{z}} \ran
    + {\cal M}_{{\bm{z}}{\bm{v}}{\bm{u}}}
    \lan n_{\bm{z}\bm{v}} \ran
    -{\cal M}_{{\bm{u}}{\bm{v}}{\bm{z}}}
    \lan n_{\bm{u}\bm{v}} \ran\right]
    \nn
    \aln \equiv \frac{\abar}{2\pi}
    \int \limits_{\bm{z}}
    \cal{K}_{\bm{u}\bm{v}\bm{z}} \otimes \lan n_{\bm{u}\bm{v}} \ran.
    \eeq
This is simply the BFKL equation for the dipole density, and its
pictorial representation is given in Fig.~\ref{FigDipEvol}. This
equation should be read in the ``passive'' point of view, in
contrast to Eq.~(\ref{BFKL}). That is, as it was obvious in the
derivation, the r.h.s.~contains what was existing before the
evolution step. The first term is proportional to the probability
for a dipole $(\bm{u},\bm{z})$ to split into two new dipoles
$(\bm{u},\bm{v})$ and $(\bm{v},\bm{z})$ times the initial density at
$(\bm{u},\bm{z})$. Since we want to find the change in the density
at $(\bm{u},\bm{v})$ only the first of these two child dipoles is
measured. Similarly for the second term. The last term is
proportional to the probability for a dipole $(\bm{u},\bm{v})$ to
split into two new dipoles times the initial density at
$(\bm{u},\bm{v})$ and naturally gives a negative contribution.
Notice that the loss term in Eq.~(\ref{Pevol}) was crucial in order
to obtain this virtual contribution.
As we show in Appendix \ref{AppBFKL}, this equation leads to the
BFKL equation, Eq.~(\ref{BFKL}), for the scattering amplitude of a
projectile dipole off the dilute hadron, when we assume the two
gluon exchange approximation which is appropriate in this dilute
limit.

\begin{figure}[t]
    \centerline{\epsfig{file=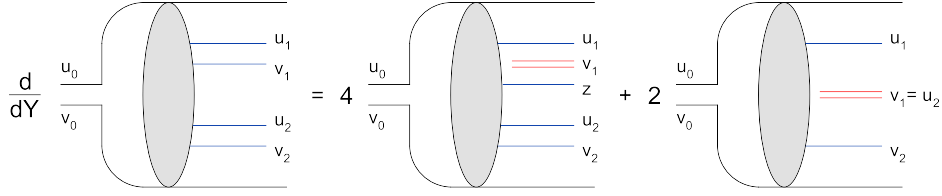,width=12.5cm}}
    \caption{\sl Evolution of the dipole pair density for a dilute hadron
    at large $N_c$. The loss term of the normal BFKL evolution is not shown.}
    \label{FigPairEvol}
\end{figure}

Now we can follow the same procedure to derive the analogous
equation for the dipole pair density \cite{IT05a}. The corresponding
part for a given $N$-dipole configuration is
    \beq\label{pairdensityN}
    n_N^{(2)}(\bm{u}_1\bm{v}_1;\bm{u}_2,\bm{v}_2)=
    \!\sum_{\substack{\j,k=1\\ \j \neq k}}^N\!
    \delta(\bm{z}_{\j-1}\-\bm{u}_1)
    \delta(\bm{z}_\j\-\bm{v}_1)
    \delta(\bm{z}_{k-1}\-\bm{u}_2)
    \delta(\bm{z}_k\-\bm{v}_2),
    \eeq
and following the same procedure as in the case of the dipole
density, we arrive at the evolution equation
    \beq\label{n2BFKL}
    \frac{\del \langle n^{(2)}_{\bm{u}_1\bm{v}_1;\bm{u}_2\bm{v}_2} \rangle}
    {\del Y} =
    \frac{\abar}{2\pi}\,
    \bigg[
    \aln\int_{\bm{z}}
    \cal{K}_{\bm{u}_1\bm{v}_1\bm{z}}
    \otimes
    \langle n^{(2)}_{\bm{u}_1\bm{v}_1;\bm{u}_2\bm{v}_2} \rangle
    \nn
    \aln +\,\delta_{\bm{u}_2\bm{v}_1}
    \mathcal{M}_{\bm{u}_1\bm{v}_2\bm{u}_2}
    \langle n_{\bm{u}_1\bm{v}_2} \rangle
    \bigg]
    \,+\, 1 \leftrightarrow 2.
    \eeq
with the notation introduced in Eq.~(\ref{nBFKL}). The two terms on
the r.h.s.~correspond to the two typical contributing diagrams in
Fig.~\ref{FigPairEvol}. The first term (which in turn is a sum of
three terms) corresponds to the BFKL evolution of the first dipole
while the second dipole remains a spectator. The second term
corresponds to a single dipole initial density and the mother dipole
splits into two new dipoles both of which are ``measured''. We shall
refer to this term as the ``splitting'' term, and this in the sense
that a lower moment of the density gives rise to a higher moment of
the density.

Of course one can continue and write the evolution equation for the
$\kappa$-th density \cite{LL05a}. We shall not do it here, since it
is just a matter of proper combinatorics. It is clear that there
will be $\kappa$ terms corresponding to normal BFKL evolution where
only one dipole is evolving and $\kappa\-1$ are spectators.
Furthermore, there will be $\kappa (\kappa-1)$ splitting terms,
according to the terminology we just defined, which will be
proportional to the $\kappa\-1$-th density where the $\kappa\-2$
dipoles will be spectators.

These splitting contributions, or to be more precise their analogue,
were not included in the JIMWLK equation. We already start to see
the consequences, since it is trivial to show that the hierarchy of
equations obeyed by the density moments is not consistent with any
sort of factorization, for example
    \beq\label{nofactn}
    \langle n^{(2)}_{\bm{u}_1\bm{v}_1;
    \bm{u}_2\bm{v}_2}\rangle
    \neq c
    \langle n_{\bm{u}_1\bm{v}_1}\rangle
    \langle n_{\bm{u}_2\bm{v}_2}\rangle.
    \eeq
This factorization is ``broken'' because of the splittings terms
which become important in the region $n\sim 1$ (or equivalently when
$T \sim \alpha_s^2$), that is, when fluctuations in the number
density of particles become a significant effect. We need to say
here that in the original formulation of the dipole picture the
equation for the dipole-pair density (and the higher moments) was
not written as given in Eq.~(\ref{n2BFKL}), but in an equivalent
form in which the fluctuations were more difficult to recognize.
Nevertheless, based on that picture, these low-density fluctuations
and some of their consequences were in fact ``seen'' in the
numerical simulations of the wavefunction of an evolved dipole and
of the approach to unitarity\footnote{In Sec.~\ref{SecDual} we will
briefly explain how unitarity comes in that picture.}
\cite{Sal9596,MS96}.

\section{Splittings of Pomerons, Large-$N_c$ Equations
and the Langevin Equation}\label{SecPomSplit}

\begin{figure}[t]
    \centerline{\epsfig{file=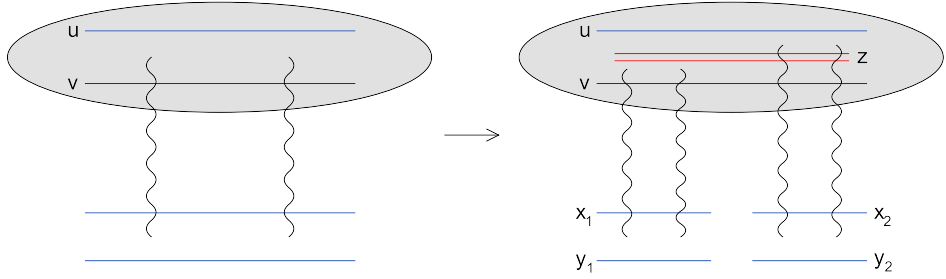,width=12.5cm}}
    \caption{\sl Splitting contribution to the dipole pair scattering.}
    \label{Fig2Dipoles}
\end{figure}

Now that we have derived the equation for the dipole-pair density,
it is not hard to transform to the corresponding equation for the
amplitude of two given external dipoles to scatter off the target.
To the order of accuracy it is enough to consider that dipoles
scatter in the two-gluon exchange approximation. This elementary
scattering amplitude for two dipoles $(\bm{x},\bm{y})$ and
$(\bm{u},\bm{v})$ is calculated in Appendix \ref{Appdipdip} and it
reads (see e.g.~\cite{NW98,BB03})
    \beq\label{T0Nc}
    T_0(\bm{x}\bm{y}|\bm{u}\bm{v}) =\aln
    \frac{\alpha_s^2}{8}\,
    \frac{N_c^2-1}{N_c^2}\,
    \ln^2 \left[\frac{(\bm{x}-\bm{v})^2 (\bm{y}-\bm{u})^2}
    {(\bm{x}-\bm{u})^2 (\bm{y}-\bm{v})^2}
    \right]
    \nn \equiv \aln\,
    \alpha_s^2\,
    \frac{N_c^2-1}{N_c^2}\,
    \cal{A}_0(\bm{x}\bm{y}|\bm{u}\bm{v}).
    \eeq
For what follows, we shall set the fraction involving the color
factor equal to one, since we have already assumed the large-$N_c$
limit in our analysis. To this end, the amplitude for the projectile
dipole $(\bm{x},\bm{y})$ to scatter off the target will be
    \beq\label{Tconvn}
    \lan T_{\bm{x}\bm{y}} \ran = \alpha_s^2
    \int\limits_{\bm{u}\bm{v}}
    \cal{A}_0(\bm{x}\bm{y}|\bm{u}\bm{v})\,
    \lan n_{\bm{u}\bm{v}} \ran
    \Rightarrow
    \lan n_{\bm{u}\bm{v}} \ran+  \lan n_{\bm{v}\bm{u}} \ran=
    \frac{4}{g^4}\,
    \lap{u} \lap{v} \lan T_{\bm{u}\bm{v}} \ran,
    \eeq
where in the second part, valid for $\bm{u} \neq \bm{v}$, we have
inverted the equation to obtain the symmetrized dipole density in
terms of the amplitude for our later convenience. Now let us
consider the scattering of a pair of dipoles $(\bm{x}_1,\bm{y}_1)$
and $(\bm{x}_2,\bm{y}_2)$ off the target. Then the extension of
Eq.~(\ref{Tconvn}) reads
    \beq\label{T2convn2}
    \lan T_{\bm{x}_1\bm{y}_1}T_{\bm{x}_2\bm{y}_2} \ran=
    \alpha_s^4
    \int\limits_{\bm{u}_\i\bm{v}_\i}
    \cal{A}_0(\bm{x}_1\bm{y}_1|\bm{u}_1\bm{v}_1)\,
    \cal{A}_0(\bm{x}_2\bm{y}_2|\bm{u}_2\bm{v}_2)\,
    \lan n^{(2)}_{\bm{u}_1\bm{v}_1;\bm{u}_2\bm{v}_2} \ran.
    \eeq
Clearly, in writing the above equation, we have assumed that two
dipoles do not interact with the same dipole. We need to say that
neither a large-$N_c$ nor a small coupling argument justifies this
assumption. Nevertheless, such processes will be suppressed at
higher energies, as they will grow like a single BFKL Pomeron. Now
by differentiating Eq.~(\ref{T2convn2}) and using the last, linear
in $\lan n \ran$, term in Eq.~(\ref{n2BFKL}) for the evolution of
the dipole pair density (the bilinear terms give rise to the normal
BFKL evolution of $\lan TT \ran$, which we already know how to
write), we obtain the splitting contribution to the evolution of the
dipole-pair amplitude, which reads \cite{IT05a,IT05b,MSW05,BIIT05}
(see also \cite{LL05b,Bra05})
    \beq\label{T2split}
    \frac{\del \langle
    T_{\bm{x}_1\bm{y}_1}T_{\bm{x}_2\bm{y}_2} \rangle}
    {\del Y}\,
    \bigg|_{\rm split} =
    \left(\frac{\alpha_s}{2\pi}\right)^2
    \frac{\abar}{2 \pi}\!
    \int\limits_{\bm{u}\bm{v}\bm{z}}
    \aln\mathcal{M}_{\bm{u}\bm{v}\bm{z}}\,
    \cal{A}_0(\bm{x}_1\bm{y}_1|\bm{u}\bm{z})\,
    \cal{A}_0(\bm{x}_2\bm{y}_2|\bm{z}\bm{v})\,
    \nn \aln\times
    \lap{u} \lap{v}\,
    \langle T_{\bm{u}\bm{v}} \rangle.
    \eeq
Notice that the poles of the kernel cancel with the zeros of the
dipole-dipole scattering amplitude. The pictorial interpretation of
this equation is given in Fig.~\ref{Fig2Dipoles}. A dipole
$(\bm{u},\bm{v})$ of the target splits into two new dipoles
$(\bm{u},\bm{z})$ and $(\bm{z},\bm{v})$, leading to the dipole
kernel in Eq.~(\ref{T2split}). The two daughter dipoles scatter with
the two projectile dipoles, as represented by the two $\cal{A}_0$'s
in the equation. Finally the last term is proportional to the
initial dipole density, which has been expressed in terms of the
single scattering amplitude by making use of Eq.~(\ref{Tconvn}).

This is the term that we would like to add to the r.h.s.~of the
second (large-$N_c$) Balitsky equation. It corresponds to the
splitting of one Pomeron into two, and this is really the mechanism
for the generation of correlations in higher-point functions which
have important consequences to the subsequent evolution of the
system. For example, through Eq.~(\ref{T2split}) the single
scattering amplitude will give rise to correlations in the
dipole-pair scattering amplitude and their significance will be
realized when they feedback to the single amplitude through the
first Balitsky equation.

It is trivial to add the corresponding splitting term in the
$\kappa$-th equation of the large-$N_c$ Balitsky hierarchy. In the
r.h.s.~of the evolution equation for the amplitude for $\kappa$
dipoles there will be $\kappa(\kappa\-1)/2$ terms proportional to
the amplitude for $\kappa\-1$ dipoles, and where in each term
$\kappa\-2$ dipoles will be simply spectators. Therefore, the new
hierarchy can be easily inferred from Eqs.~(\ref{BFKL}),
(\ref{Tmerge}) and (\ref{T2split}), since in the $\kappa$-th
evolution equation the relevant processes are $1 \to 1$, $1\to 2$
and $2\to 1$ Pomeron transitions, with all other Pomerons being just
spectators, and one just needs to take into account all the possible
permutations. Thus, it is not necessary to present the full set of
equations here (we shall give an equivalent compact form in
Eq.~(\ref{langevin}) below), rather we shall indicate only its
structure by suppressing the transverse coordinates. It reads
    \beq\label{Thier}
    \frac{\del \lan T^\kappa \ran}{\del Y}
    = \kappa\,\abar\, \lan T^{\kappa} \ran
    - \kappa\, \abar\,\lan T^{\kappa+1} \ran
    + \frac{\kappa (\kappa-1)}{2}\, \abar \,\alpha_s^2\,
    \lan T^{\kappa\-1} \ran,
    \eeq
where the $\kappa$-dependent coefficients stand for the number of
terms arising from the permutations.

For the reasons analyzed in Sec.~\ref{SecDip}, it is clear that our
hierarchy is not consistent with any sort of factorizing solution;
the low density behavior of the theory has totally changed with
respect to the theory without Pomeron splittings. Therefore, one has
to find alternative ways to deal with this set of equations. One way
to do that, is to reformulate the problem as a single stochastic
equation. Indeed, the Langevin equation
    \beq\label{langevin}
    \!\!\!\!\!\frac{1}{\abar}\frac{\del T_{\bm{x}\bm{y}}}{\del Y}\aln =
    \frac{1}{2\pi} \int\limits_{\bm{z}}
    \cal{M}_{\bm{x}\bm{y}\bm{z}}\,
    [T_{\bm{x}\bm{z}} + T_{\bm{z}\bm{y}} - T_{\bm{x}\bm{y}}
    -T_{\bm{x}\bm{z}}T_{\bm{z}\bm{y}}]
    \nn
    \!\!\!\!\!\aln + \,\frac{\alpha_s}{2\pi}\, \sqrt{\frac{1}{2 \pi}}
    \!\int\limits_{\bm{u}\bm{v}\bm{z}}\!
    \cal{A}_0(\bm{x}\bm{y}|\bm{u}\bm{z})\,
    \frac{|\bm{u}-\bm{v}|}{(\bm{u}-\bm{z})^2}\,
    \sqrt{\lap{u}\lap{v} T_{\bm{u}\bm{v}}}\,\,
    \nu(\bm{u}\bm{v}\bm{z};Y),
    \eeq
where the noise satisfies
    \beq\label{noise}
    \lan \nu(\bm{u}_1\bm{v}_1\bm{z}_1;Y)
    \nu(\bm{u}_2\bm{v}_2\bm{z}_2;Y') \ran=
    \delta_{\bm{u}_1\bm{v}_2}
    \delta_{\bm{v}_1\bm{u}_2}
    \delta_{\bm{z}_1\bm{z}_2}
    \delta(\abar Y\-\abar Y'),
    \eeq
and where all other noise correlators vanish, gives an equivalent
description \cite{IT05b}. We show this equivalence in Appendix
\ref{Applangevin} for a simple zero-dimensional model, while the
generalization to the QCD problem at hand is
straightforward\footnote{We should note here, that also the JIMWLK
equation can be reformulated as a Langevin problem \cite{BIW03}.
However, in that case the physics is totally different since the
noise describes color fluctuations, rather than particle number
fluctuations which is the case here.}. However, because of the
complexity of the noise correlation, this form may not be the best
option in the search for numerical solutions. Nevertheless, one can
rely on certain approximations to gain a first idea on the new
features of the evolution. Assuming that the elementary
dipole-dipole scattering amplitude is local in transverse
coordinates, performing a coarse-graining in impact parameter space
and defining the Bessel transformation $\varphi$ of the scattering
amplitude one arrives at \cite{IT05a}
    \beq\label{langsimple}
    \frac{1}{\abar}
    \frac{\del \varphi(\bm{k})}{\del Y} =
    \aln
    \frac{1}{2\pi}
    \int \limits_{\bm{p}}
    \frac{\bm{k}^2}{\bm{p}^2(\bm{k}-\bm{p})^2}
    \left[2\, \frac{\bm{p}^2}{\bm{k}^2}\, \varphi(\bm{p})
    -\varphi(\bm{k})\right]-
    \varphi^2(\bm{k})
    \nn
    \aln + \sqrt{2\, c \,\alpha_s^2\, \varphi(\bm{k})}\,\,
    \nu(\bm{k,Y}),
    \eeq
with $\nu(\bm{k},Y)$ a Gaussian white noise, i.e.~the only
non-vanishing correlator is $\lan \nu(\bm{k}_1,Y)
\,\nu(\bm{k}_2,Y')\ran = \delta(\abar Y - \abar Y')\,
\delta(\bm{k}_1^2 - \bm{k}_2^2)\, \bm{k}_1^2$, and where $c$ is a
constant of order $\cal{O}(1)$. Notice that, up to an overall
normalization factor of order $\cal{O}(1/\abar)$, $\varphi(\bm{k})$
is the unintegrated gluon distribution. A numerical solution to this
equation has been given in \cite{Soy05}. Furthermore, if one
performs a saddle point approximation to the BFKL kernel in
Eq.~(\ref{langsimple}), the resulting equation is the stochastic
FKPP equation which has been studied numerically in \cite{EGBM05}.
So long as the energy dependence of $Q_s$ is concerned, the results
from these numerics are consistent with the ones from the analytical
approach that we will present in Sec.~\ref{SecSatRev}.

It is appropriate to make here a few important comments on the
hierarchy we have derived, and which is compactly written in the
``exact'' Langevin Eq.~(\ref{langevin}). The first is with respect
to the strength of each term. In the low density region where $ T
\sim \alpha_s^2$, one can see in the Langevin equation that both the
noise (splitting) term and the BFKL terms are of the same order,
while the merging term is suppressed by a factor
$\cal{O}(\alpha_s^2)$. In the intermediate region where $\alpha_s^2
\ll T \ll 1$, the BFKL terms dominate, and in the region near the
unitarity limit $T \sim 1$, the BFKL terms and the merging term are
of the same order, while the noise term is subdominant by a factor
$\cal{O}(\alpha_s^2)$\footnote{Terms which are of order
$\cal{O}(\abar T^3)$ when inserted to the r.h.s. of
Eq.~(\ref{langevin}) have also been calculated \cite{BB02}. One can
see that such term are subdominant in all regions.}. The second
comment we should make is that, as in the case of the Balitsky
equations, $T=1$ is still a fixed point of the evolution; the noise
term in the Langevin equation vanishes for constant $T$ due to the
presence of the Laplacians. We should also say that we do not expect
the term we added to describe properly the Pomeron splittings in the
high density region, since we have heavily relied on the two-gluon
exchange approximation.  In fact, even though we expect on general
grounds the Pomeron splittings to be positive, the term we added can
be negative in some regions \cite{IST05}. Nevertheless, this is not
very worrisome, since Pomeron splittings are supposed to dominate in
the low density region, and to this end the leading contribution has
been taken into account. In that region the r.h.s.~of
Eq.~(\ref{T2split}) is clearly positive, since
$\lap{u}\lap{v}T_{\bm{u}\bm{v}}$ is in fact proportional to the
dipole density which is positive, while away from that region the
term is anyway suppressed with respect to the other contributions as
we have just seen.

Before closing this section, and in order to cross-check that our
approach was correct, let us compare with a well-known result
obtained from the construction of Pomeron vertices in perturbative
QCD. By neglecting the non-linear evolution terms, it is clear that
one arrives at a solvable problem. Thus, by first solving the BFKL
equation, and subsequently solving the second equation in the
hierarchy, which is an inhomogeneous one, and assuming the initial
condition $\lan TT \ran =0$, we find \cite{IT05b}
    \beq\label{3Pom}
    \lan T_{\bm{x}_1\bm{y}_1}T_{\bm{x}_2\bm{y}_2} \ran=
    \left(\frac{\alpha_s}{2\pi}\right)^2
    \frac{\abar}{2 \pi}\,
    \int\limits_0^Y \dif y
    \int\limits_{\bm{u}\bm{v}\bm{z}}
    \aln
    \mathcal{M}_{\bm{u}\bm{v}\bm{z}}\,
    \cal{A}_{Y-y}(\bm{x}_1\bm{y}_1|\bm{u}\bm{z})\,
    \cal{A}_{Y-y}(\bm{x}_2\bm{y}_2|\bm{z}\bm{v})\,
    \nn
    \aln \times \lap{u}\lap{v}
    \lan T_{\bm{u}\bm{v}} \ran.
    \eeq
This coincides with the large-$N_c$ triple Pomeron vertex as given
in \cite{BV99}.

\section{Loops of Pomerons}\label{SecPloop}

\begin{figure}[t]
    \centerline{\epsfig{file=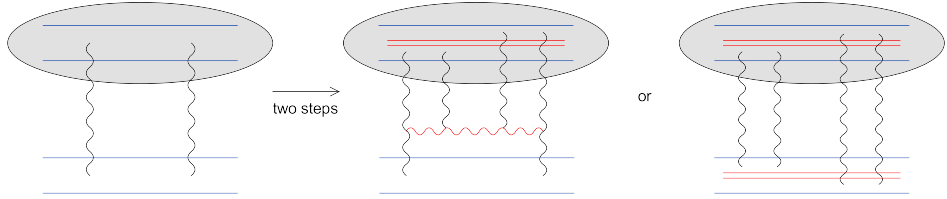,width=12.5cm}}
    \caption{\sl Formation of a loop under a two step evolution.}
    \label{FigPloop}
\end{figure}

Since the hierarchy presented in the previous section contains both
splittings and mergings of Pomerons, it is clear that Pomeron loops
will be generated in the course of evolution. For example, let us
find the minimal loop, which will be formed after a two step
evolution. This is depicted in the middle diagram of
Fig.~\ref{FigPloop}. In the first step the dipole emits a soft gluon
which, as always, can be considered as a quark-antiquark pair and
the two daughter dipoles ``radiate'' two ``elementary'' Pomerons. In
the second step, the two Pomerons merge into one through the
emission of another soft gluon.

Of course one understands that a ``real'' Pomeron loop will be
formed after a large number of steps, if we dress all parts of the
process with normal BFKL evolution. Following this procedure we can
construct the Pomeron loop vertex and let us make a short digression
here, to say that one can do the same for all kinds of such Pomeron
vertices (i.e.~vertices integrated over a large rapidity interval)
that one can imagine, e.g.~for the triple Pomeron vertex given at
the end of the previous section. Of course this is a very tedious
procedure, and at the end one has to embed all possible allowed
combinations of these vertices into a huge diagram, which has the
two initial colliding objects at its two ends. Furthermore, given
the non-linearities of the system, it is not clear whether this
approach to the problem can be successful or not. Therefore it seems
that, instead of trying to calculate building blocks of the theory,
it is more advantageous to deal with the full theory as a whole,
i.e.~the hierarchy of equations we just presented, or its
generalization to finite-$N_c$ (when the latter becomes available).
Of course, in order to construct the evolution equations it is
necessary to know the ``reggeized'' gluon transition vertices
(i.e.~the vertices for a single step in rapidity) \cite{BW95,BE99}.
In the large-$N_c$ limit, and for the specific observables that we
were interested in, we only needed to have the $2 \to 4$ vertex in
coordinate space which is simply the dipole kernel
$\cal{M}_{\bm{x}\bm{y}\bm{z}}$.

Let us now return to the calculation of the minimal Pomeron loop,
since the result will reveal some very interesting features. In
terms of the evolution equations, and starting from the single
scattering amplitude $\lan T \ran$, the diagram in
Fig.~\ref{FigPloop} corresponds to the generation of $\lan TT \ran$
through the fluctuation (splitting) term given in
Eq.~(\ref{T2split}), which subsequently gives feedback to $\lan T
\ran$ through the non-linear term of the first Balitsky
equation\footnote{In terms of the language used in the forthcoming
Sec.~\ref{SecDual}, this procedure corresponds to the successive
operation $H^{\dag}_{1\to 2} H^{\dag}_{2 \to 1}$ on the single
amplitude.}. Assuming that the initial state of the target was
described by a single dipole at $(\bm{u},\bm{v})$ and the projectile
dipole is at $(\bm{x},\bm{y})$, it is easy to show that
\cite{BIIT05}
    \beq\label{PL0}
    \mathbb{P L}^0 = - 2 \,\alpha_s^4
    \left(\atpi \right)^2
    \int\limits_{\bm{z}\bm{w}}
    \cal{M}_{\bm{x}\bm{y}\bm{z}}\,
    \cal{M}_{\bm{u}\bm{v}\bm{w}}\,
    \cal{A}_0(\bm{x}\bm{z}|\bm{u}\bm{w})\,
    \cal{A}_0(\bm{z}\bm{y}|\bm{w}\bm{v})
    \eeq
times a factor of order $(\Delta Y)^2=({\rm step})^2$ which we have
suppressed. We immediately see that this result is free of any
possible divergencies, since the poles of the dipole kernels are
canceled by the zeros of the $\cal{A}_0$'s. Furthermore, and as
expected, the contribution of the Pomeron loop is negative and leads
to a decrease of the scattering amplitude(s) in the course of
evolution.

Even though we were thinking that we put the ``whole'' evolution in
the wavefunction of the target dipole $(\bm{u},\bm{v})$, it is clear
that Eq.~(\ref{PL0}) serves for an alternative interpretation, as
shown in the last diagram in Fig.~\ref{FigPloop}. Both the original,
target and projectile, dipoles split into two child dipoles (each of
them), processes which are represented by the two dipole kernels
times $\abar^2$. Afterwards, the ensuing dipoles scatter with each
other at the two-gluon exchange level, and this gives rise to the
two $\cal{A}_0$'s accompanied by a factor $\alpha_s^4$. Therefore
one arrives at the conclusion, that a merging of Pomerons in one of
the scattering objects (here the target) can be equivalently
considered as a splitting of Pomerons in the other object (here the
projectile). Of course, one should keep in mind that whichever is
the way we want to view the process, the soft gluon (the red
horizontal gluon in the figures) is always emitted in the
wavefunction of one of the two objects. This feature of exchanging
splittings with mergings by jumping from the one wavefunction to the
other, was already valid at the level of the Balitsky-JIMWLK
evolution, where the mergings \`{a} la JIMWLK could be viewed as
splittings \`{a} la Balitsky. But Eq.~(\ref{PL0}) contains more than
that, since it allows for both type of processes inside the same
wavefunction. Putting all these properties together, we shall see in
the next section that we can formulate a precise duality condition,
which is satisfied by the Hamiltonian governing the evolution of the
hadronic system.

\section{COM Frame, Duality and a Pomeron Effective Theory}\label{SecDual}

Now we will discuss a remarkable duality property which is
presumably satisfied by the complete Hamiltonian at high energy and
at the leading logarithmic level. Here we shall describe the
simplified case where the two colliding hadrons are described by
non-saturated wavefunctions \cite{BIIT05}, while an approach to the
general proof has been given in \cite{KL05c}, where the idea of the
duality was explicitly written for the first time.

So let us consider the scattering of two evolved dipoles. Assuming
that the right (R) moving object is evolved by $Y-y$ and the left
(L) moving one by $y$, one can write the average $S$-matrix as
\cite{IM04a}
    \beq\label{SLR}
    \lan S \ran =
    \int \mathcal{D} \alpha_{\rm R}\,
    \mathcal{D} \alpha_{\rm L}\,
    Z_{Y\!-y}[\alpha_{\rm R}]\,
    Z_{y}[\alpha_{\rm L}]\,
    \exp \left[ \rmi \int\limits_{\bm{z}}
    \rho^a_{\rm L}(\bm{z})\, \alpha^a_{\rm R}(\bm{z})\right].
    \eeq
The last exponential factor stands for the $S$-matrix for a given
state of both the R and L evolved dipoles, and it is just the
eikonal coupling between the color charge in the L object and the
field created by the R object. Notice that it can be brought into a
symmetric form under the exchange R$\leftrightarrow$L, by using the
Poisson equation $\lap{z} \alpha^a_{\rm R/L}(\bm{z})= - \rho^a_{\rm
R/L}(\bm{z})$ and integrating by parts. To obtain the average
$S$-matrix we simply integrate over all possible configurations for
both wavefunctions. This formula would be exact in QED, however it
has a restricted range of validity in QCD. As we have seen in QCD,
and if one wants to include all multiple scattering contributions,
one needs to use path order exponentials to account for the color
matrices non-commutativity in the interaction vertices, so
Eq.~(\ref{SLR}) cannot be correct in general. Nevertheless, it
contains the possibility that any number of projectile dipoles
interact with the same number of target dipoles, even though each
individual dipole interacts, at most, only once. This simultaneous
scattering of many dipoles is equivalent to a ``multiple Pomeron
exchange'' and one expects the final result to respect unitarity
limits. Thus, Eq.~(\ref{SLR}) goes much beyond the BFKL equation. We
can show that all the above approximations and conclusions are
meaningful in a frame which is not fully asymmetric, i.e.~when $y =
\kappa Y$, with $\kappa$ a constant of order $\cal{O}(1)$, but of
course smaller than one, for example in the center of mass (COM)
frame $y=Y/2$. Requiring that the two wavefunctions be unsaturated,
one determines the critical rapidity $Y_c$ up to which the approach
will be valid. In the center of mass frame one will have
    \beq\label{Yc}
    n_{\rm R/L}(Y/2)
    \sim \alpha_s^2 \exp[\omega_{\mathbb{P}} Y/2]
    \lesssim 1
    \Rightarrow Y_c \sim
    \frac{2}{\omega_{\mathbb{P}}} \ln \frac{1}{\alpha_s^2},
    \eeq
while in the case of a not fully symmetric frame one has to replace
the factor of 2 in $Y_c$ with $1/{\rm max}(\kappa,1\-\kappa)>1$. But
the onset of unitarity will come at a much lower rapidity
\cite{Mue95}. Indeed, for $Y=Y_c/2$, with $Y_c$ given by
Eq.~(\ref{Yc}), the contributions of the previously mentioned
simultaneous scatterings is
    \beq\label{unitarity}
    \left[\alpha_s^2 \, n_{\rm R}(Y_c/4)\,
    n_{\rm L}( Y_c/4)\right]^n \sim 1,
    \eeq
while in the case of a not fully symmetric frame unitarity will come
at $Y \sim {\rm max}(\kappa,1\-\kappa)\,Y_c \sim
(1/\omega_{\mathbb{P}})\ln(1/\alpha_s^2) < Y_c$, which is of course
a frame independent value as it should. Simply, the choice of the
COM frame is the optimal one, if one wants to study unitarity
without having to worry about saturation. The integer $n$ in
Eq.~(\ref{unitarity}) refers to the number of Pomeron exchanges, for
example $n=1$ simply gives the BFKL contribution. By proper
summation of all these terms, which are of the same strength and of
order $\cal{O}(1)$, the total $S$-matrix will respect the unitarity
constraints \cite{Mue95}.

After this short digression to justify the use of Eq.~(\ref{SLR}),
let us return to the proof of the duality property, which is based
on two natural conditions \cite{KL05c,BIIT05}. The first is that
both objects are described by the same physics, more precisely both
wavefunctions appearing in Eq.~(\ref{SLR}) obey the same evolution
law in rapidity. The second is that the total averaged $S$-matrix is
independent of $y$, i.e.~independent of the precise separation of
the total rapidity interval $Y$, a condition which is clearly
dictated by Lorentz (boost) invariance. Thus, by setting the
derivative of Eq.~(\ref{SLR}) with respect to $y$ equal to zero, and
assuming that the evolution Hamiltonian has the same functional form
for both hadrons, we easily find that
    \beq\label{dualitytemp1}
    \hspace{-0.6cm} 0 \aln \,= \int  \mathcal{D} \alpha_{\rm R}\,
    \mathcal{D} \alpha_{\rm L}\,
    \exp \left[ \rmi \int\limits_{\bm{z}}
    \rho^a_{\rm L}(\bm{z})\, \alpha^a_{\rm R}(\bm{z})\right]
    \nn
    \hspace{-0.6cm} \aln\times\! \left\{Z_{Y\!-y}[\alpha_{\rm R}]\,
    H\!\! \left[ \alpha_{\rm L}, \frac{\delta}{\rmi \delta \alpha_{\rm L}}\right]
    \!Z_{y}[\alpha_{\rm L}]
    \- Z_{y}[\alpha_{\rm L}]\,
    H\!\! \left[ \alpha_{\rm R}, \frac{\delta}{\rmi \delta \alpha_{\rm R}}\right]
    \!Z_{Y\!-y}[\alpha_{\rm R}]
    \right\}.
    \eeq
Now, by performing an integration by parts with respect to
$\alpha_{\rm R}$ in the second term, using the identity
$H[\alpha,\delta/(\rmi\delta \alpha_{\rm R})]S =
H^{\dag}[\delta/(\rmi\delta \rho_{\rm L}),\rho_{\rm L}]S$, where $S$
is the exponential standing for the eikonal $S$-matrix, and finally
performing another integration by parts again in the second term,
but with respect to $\alpha_{\rm L}$ this time (recall that $\lap{z}
\alpha^a_{\rm L}(\bm{z})= - \rho^a_{\rm L}(\bm{z})$), we find
    \beq\label{dualitytemp2}
    0 = \aln \int  \mathcal{D} \alpha_{\rm R}\,
    \mathcal{D} \alpha_{\rm L}\,
    \exp \left[ \rmi \int\limits_{\bm{z}}
    \rho^a_{\rm L}(\bm{z})\, \alpha^a_{\rm R}(\bm{z})\right]
    \nn
    \aln \times Z_{Y\!-y}[\alpha_{\rm R}]
    \left\{
    H\!\left[ \alpha_{\rm L}, \frac{\delta}{\rmi \delta \alpha_{\rm L}}
    \right]
    -H^{\dag}\!\left[\frac{\delta}{\rmi \delta \rho_{\rm L}}, \rho_{\rm
    L}\right] \right\}
    \!Z_{y}[\alpha_{\rm L}].
    \eeq
The term in the square bracket must vanish and we arrive at the
duality property
    \beq\label{duality}
    H\!\left[\alpha, \frac{\delta}{\rmi\, \delta \alpha} \right] =
    H^{\dagger}\!\left[\frac{\delta}{\rmi\, \delta \rho}, \rho
    \right],
    \eeq
where in $H$ all the functional derivatives stand to the left, while
in its Hermitian conjugate $H^{\dagger}$ they stand to the right.

This duality constraint provides an alternative way to see that the
JIMWLK Hamiltonian given in Eq.~(\ref{HJIMWLK}) can not be the end
of the story. Indeed, while it contains an arbitrary number of color
fields $\alpha^a$ (through the expansion of the Wilson lines), it
contains only two functional derivatives with respect to those
fields. Of course Eq.~(\ref{duality}) cannot determine the
``complete'' Hamiltonian, but it could facilitate the search for it,
or at least help us to construct an ``approximate'' one which
contains all the essential physics. For example, we shall first try
to find out how the evolution equations we presented in
Sec.~\ref{SecPomSplit} can be put in a Hamiltonian form
\cite{BIIT05}.

To this end, we shall recall the Hamiltonian proposed in
\cite{MSW05} in order to describe the splittings of Pomerons. For
our purposes it is more convenient to present it in the form
    \beq\label{H1to2}
    H^{\dagger}_{1 \to 2} =
    -\frac{g^2}{16 N_c^3}\, \atpi
    \int\limits_{\bm{u}\bm{v}\bm{z}}
    \mathcal{M}_{\bm{u}\bm{v}\bm{z}}
    \,\rho^a_{\bm{u}}\, \rho^a_{\bm{v}}
    \left[ \frac{\delta}{\delta \rho^b_{\bm{u}}} -
           \frac{\delta}{\delta \rho^b_{\bm{z}}} \right]^2
    \left[ \frac{\delta}{\delta \rho^c_{\bm{z}}} -
           \frac{\delta}{\delta \rho^c_{\bm{v}}} \right]^2.
    \eeq
The notation $1 \to 2$ is clear, since the Hamiltonian contains two
factors of $\rho$ and four factors of $\delta/\delta \rho$ and
therefore can lead to a transition from one to two Pomerons (and
thus from two to four exchange gluons). Indeed, the dipole density
is bilinear in the charge density, more precisely
    \beq\label{rhofuncn}
    \bar{n}_{\bm{u}\bm{v}} \equiv
    \frac{n_{\bm{u}\bm{v}} + n_{\bm{v}\bm{u}}}{2} =
    -\,\frac{1}{g^2 N_c}\,
    \rho_{\bm{u}}^a\,
    \rho_{\bm{v}}^a,
    \eeq
and then by acting the Hamiltonian on $\bar{n}_{\bm{u}_1\bm{v}_1}
\bar{n}_{\bm{u}_2\bm{v}_2}$, which corresponds to the, symmetrized
under the exchange of quark and antiquark legs, dipole-pair density
(with the appropriate assumptions that the two dipoles have not zero
size and cannot coincide), it is straightforward to show that the
outcome is the splitting term in Eq.~(\ref{n2BFKL}). For this to
happen, one needs to assume the large-$N_c$ limit, and this means
that one neglects the terms arising from the action of two
derivatives with the same color index on two sources of different
color. Similarly, and as a cross-check, we can act on
$T_{\bm{x}_1\bm{y}_1}T_{\bm{x}_2\bm{y}_2}$, which corresponds to the
simultaneous scattering of two external dipoles off the target.
Assuming the two-gluon exchange approximation (for each dipole), and
making use of the Poisson equation, we obtain the splitting term as
given in Eq.~(\ref{T2split}).

Of course this Hamiltonian is not the dual of the JIMWLK one, as can
be seen by just comparing the relevant expressions, two of the
reasons being that the latter does not involve any large-$N_c$
approximation and it can describe any $n \to 2$ vertex, with $n \geq
2$. However, in order to build our approximate Hamiltonian let us
construct the dual part of Eq.~(\ref{H1to2}). By letting $\rho \to
-\rmi \delta/\delta \alpha$ and $\delta /\delta \rho \to \rmi
\alpha$ we find
    \beq\label{H2to1}
    H^{\dagger}_{2 \to 1} =
    \frac{g^2}{16 N_c^3}\, \atpi
    \int\limits_{\bm{u}\bm{v}\bm{z}}
    \mathcal{M}_{\bm{u}\bm{v}\bm{z}}
    \,[\alpha^a_{\bm{u}} - \alpha^a_{\bm{z}}]^2
    [\alpha^b_{\bm{z}} - \alpha^b_{\bm{v}}]^2
    \frac{\delta}{\delta \alpha^c_{\bm{u}}}\,
    \frac{\delta}{\delta \alpha^c_{\bm{v}}}.
    \eeq
By acting on the dipole scattering amplitude $T_{\bm{x}\bm{y}}$,
again at the two-gluon exchange level, we see that we obtain the
non-linear term in the first Balitsky equation. To complete the
construction within the present approximations, let us extract the
BFKL part of the JIMWLK Hamiltonian. By expanding
Eq.~(\ref{HJIMWLK}) to quadratic order in the color field
$\alpha^a$, integrating over the longitudinal coordinate $x^-$ and
``taking the large-$N_c$ limit by an appropriate contraction of
color indices'' \cite{BIIT05}, we obtain
    \beq\label{HBFKL}
    H_0^{\dag}=
    \frac{1}{2 N_c^2}\, \atpi
    \int\limits_{\bm{u}\bm{v}\bm{z}}
    \cal{M}_{\bm{u}\bm{v}\bm{z}}
    [\alpha^a_{\bm{u}} -\alpha^a_{\bm{z}}]
    [\alpha^a_{\bm{z}} -\alpha^a_{\bm{v}}]\,
    \frac{\delta}{\delta \alpha^b_{\bm{u}}}\,
    \frac{\delta}{\delta \alpha^b_{\bm{v}}}.
    \eeq
This part of the Hamiltonian will generate the normal BFKL evolution
for the amplitude for a projectile dipole to scatter off the target
(always at the two-gluon exchange level). Furthermore, it is
self-dual; if we let $\alpha \to -\rmi \delta/\delta \rho$ and
$\delta /\delta \alpha \to \rmi \rho$, the obtained Hamiltonian will
generated the normal BFKL evolution for the target dipole density,
which is equivalent to the evolution of $T_{\bm{x}\bm{y}}$ as we
show in Appendix \ref{AppBFKL}.

By adding the three pieces given in Eqs.~(\ref{H1to2}),
(\ref{H2to1}) and (\ref{HBFKL}), we obtain the total Hamiltonian
    \beq\label{Htotal}
    H^{\dag} =
    H^{\dag}_0
    +H^{\dag}_{1 \to 2}
    +H^{\dag}_{2 \to 1},
    \eeq
which gives rise to the hierarchy of equations we have presented in
Sec.~\ref{SecPomSplit}. It satisfies the self-duality condition
(\ref{duality}) and it describes $\kappa \to \kappa$, $\kappa \to
\kappa  + 1$ and $\kappa + 1 \to \kappa$ Pomeron transitions, where,
in all cases, $\kappa-1$ of the Pomerons are simply spectators. Of
course we should emphasize again that this is not the complete
solution. Nevertheless, it contains the essential features of
``Pomeron dynamics'' and a possible numerical solution to this
Hamiltonian problem would presumably describe the approach to
unitarity in a realistic way.

Before closing this section, and in order for any possible confusion
to be avoided, we should stress that the duality transformation
interchanges Pomeron splittings with Pomeron mergings within the
same wavefunction, i.e.~it changes the physical process. Thus, if
the Hamiltonian contains both processes, then it will be self-dual,
i.e.~invariant under the transformation. On the other hand, a given
process can be always viewed either as a Pomeron splitting in the
wavefunction of the one hadron, or as a Pomeron merging in the
wavefunction of the other hadron.

\section{The Dual of JIMWLK}\label{SecHBar}

\begin{figure}[t]
    \centerline{\epsfig{file=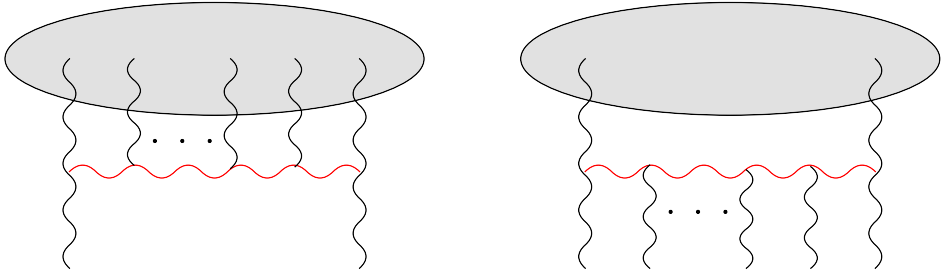,width=12.5cm}}
    \caption{\sl Vertices of JIMWLK and its dual.}
    \label{FigVertices}
\end{figure}

Before proceeding to the quest of a complete Hamiltonian, let us try
to construct a generalization of the splitting Hamiltonian
$H^{\dag}_{1 \to 2}$ \cite{MSW05} written in the previous section.
Recall that the JIMWLK Hamiltonian is written in terms of Wilson
lines and can describe any $n \to 2$, with $n \geq 2$, transition
vertex at finite-$N_c$, as sketched in the left part of
Fig.~\ref{FigVertices}. Therefore it seems reasonable to try and
find a similar way to describe all $2 \to n$ vertices, as shown in
the right part of Fig.~\ref{FigVertices}, without restricting
ourselves to the large-$N_c$ limit. We shall not give the derivation
here, but rather we shall present the answer as the dual of JIMWLK,
which reads \cite{KL05a} (see also \cite{HIMST05})
    \beq\label{Hbar}
    \bar{H} = \frac{1}{16 \pi^3}\,
    \int\limits_{\bm{u}\bm{v}\bm{z}}
    \mathcal{M}_{\bm{u}\bm{v}\bm{z}}\,
    \rho^a_{\bm{u}}\,\rho^b_{\bm{v}}\,
    \left[1 + \tilde{W}_{\bm{u}} \tilde{W}^{\dagger}_{\bm{v}}-
    \tilde{W}_{\bm{u}} \tilde{W}^{\dagger}_{\bm{z}}-
    \tilde{W}_{\bm{z}} \tilde{W}^{\dagger}_{\bm{v}} \right]^{ab},
    \eeq
where the Wilson lines in the adjoint representation are defined
through
    \beq\label{WilsonWa}
    \tilde{W}_{\bm{x}}=
    {\rm P}\,\exp
    \left[ g \int\limits_{-\infty}^{\infty}
    \dif x^+ \frac{\delta}{\delta \rho^a(x^+,\bm{x})}\, T^a
    \right],
    \eeq
and with the sources appearing in $\bar{H}$ evaluated at light-cone
time $x^+ = \infty$. One way to see why the light-cone time $x^+$
appears in the above equations (while at the same time $x^-$ is
absent, or better it has been integrated over), is that the $2 \to
n$ vertex inside the wavefunction of the right mover (target) can be
also viewed as an $n \to 2$ vertex inside the wavefunction of the
left mover (projectile). Then the appropriate longitudinal variable
for this left mover is $x^+$. In general when one applies the
duality transformation, one needs to let $x^+\, \leftrightarrow
x^-$.

Thus, given this Hamiltonian we could write the equations of motion
for our observables. Here we shall do things in a way which is a
little bit different than, but equivalent to, the one we followed in
the JIMWLK case. The variables appearing in Eq.~(\ref{Hbar}) are the
color charges $\rho^a$ and the temporal Wilson lines $\tilde{W}$,
which satisfy the commutation relationships
    \beq\label{rhoWcomm}
    \aln
    \big[ \rho^a_{\bm{u}}, \tilde{W}^{bc}_{\bm{v}}\big] =
    g \big( T^a \tilde{W}_{\bm{u}}\big)^{bc} \delta_{\bm{u}\bm{v}} =
    - \rmi g f^{abd} \tilde{W}_{\bm{u}}^{dc} \delta_{\bm{u}\bm{v}},
    \nn \aln
    \big[\tilde{W}^{ab}_{\bm{u}} , \tilde{W}^{cd}_{\bm{v}}\big] = 0,
    \nn \aln
    \big[\rho^a_{\bm{u}}, \rho^b_{\bm{v}}\big] =
    -\rmi g f^{abc} \rho^c_{\bm{u}}\, \delta_{\bm{u}\bm{v}},
    \eeq
where the first two arise from the definition of the Wilson line in
Eq.~(\ref{WilsonWa}), while the last is imposed dy the Jacobi
identity
    \beq\label{Jacobi}
    \big[\rho^a_{\bm{u}},\big[\rho^b_{\bm{v}},\tilde{W}_{\bm{z}}^{cd}\big]\big]
    +\big[\rho^b_{\bm{v}},\big[\tilde{W}_{\bm{z}}^{cd},\rho^a_{\bm{u}}\big]\big]
    +\big[\tilde{W}_{\bm{z}}^{cd},\big[\rho^a_{\bm{u}},\rho^b_{\bm{v}}\big]\big]
    =0.
    \eeq
Perhaps the non-commutativity of the color charges should not come
as a surprise since we are interested in describing a system whose
density is not high.
In fact, close to the fluctuation line we
defined earlier, the dipole density is of order $\cal{O}(1)$, and
therefore charge densities are of order $\cal{O}(g)$ (of course
$\lan \rho \ran =0$ due to color neutrality, but $\lan \rho \rho
\ran = \cal{O}(g^2)$ and so on). For the same reasons, in the JIMWLK
evolution there was not such an issue, since $\rho$ was always much
bigger than $g$, and the commutator was subleading.  Now the
evolution equation for an appropriate observable $\cal{O}$ may be
written in the form
    \beq\label{Oevolbar}
    \frac{\del \lan \cal{O} \ran}{\del Y} =
    \lan [\bar{H}, \cal{O}] \ran.
    \eeq
Based on the above considerations, it becomes obvious that if we
consider gauge invariant observables built from the temporal Wilson
lines in the fundamental representation, and starting from $\tr
(W_{\bm{x}} W^{\dag}_{\bm{y}})$, we shall generate the dual of the
Balitsky hierarchy. This is not so surprising after all, since the
diagram on the right part of Fig.~\ref{FigVertices} when viewed
upside-down corresponds to the JIMWLK evolution of a left mover,
while the correlators of the temporal Wilson lines correspond to the
amplitudes for right movers to scatter of these evolved left movers.
But of course this trivial aspect is not what we were looking for,
rather we are interested to see how $\bar{H}$ can describe the low
density fluctuations. Before that, and as a first check, one can see
that Eq.~(\ref{Oevolbar}), with $\cal{O}$ the bilinear of the charge
density as defined in Eq.~(\ref{rhofuncn}), will lead to the BFKL
equation for the dipole density of the target hadron. To measure the
fluctuations we need to consider a higher order correlator of
charges, for example the dipole pair density
$\bar{n}_{\bm{u}_1\bm{v}_1} \bar{n}_{\bm{u}_2\bm{v}_2}$ with
$\bar{n}$ defined in Eq.~(\ref{rhofuncn}). Here, one will obtain
Eq.~(\ref{n2BFKL}) but there will be additional terms to the right
hand side of the equation. The same will happen if we consider the
scattering of two projectile dipoles in order to probe these
fluctuations; acting with $\bar{H}$ on
$T_{\bm{x}_1\bm{y}_1}T_{\bm{x}_2\bm{y}_2}$ at the two-gluon exchange
approximation, we obtain the normal BFKL evolution terms, the
splitting term in Eq.~(\ref{T2split}) plus additional terms. Some of
these terms will be suppressed in the large-$N_c$ limit, but some
will not. For example there will be terms which correspond to the
simultaneous scattering off the same target dipole, and this process
is not subdominant in the multi-color limit. More generally the two
dipoles could scatter off more complicated combinations of the
target color sources.

One may have already seen that there are some subtleties associated
with the dual Hamiltonian and the particular observables that we are
interested in. Since the color charges are non-commutative, one has
a difficulty in defining the observables at equal time $x^+$. For
example, even though there is no problem with the single dipole
density, since $[\rho^a_{\bm{u}}, \rho^a_{\bm{v}}] = 0$, there is an
ambiguity in defining the dipole-pair density since
$[\bar{n}_{\bm{u}_1\bm{v}_1}, \bar{n}_{\bm{u}_2\bm{v}_2}] \neq 0$
and similarly for the dipole-pair scattering amplitude, since the
color field $\alpha^a$ is related to the charge $\rho^a$ through the
solution of the Poisson equation. Thus one needs a prescription to
circumvent the problem, for example by considering symmetrized
observables.

Of course these difficulties should disappear when one assumes that
the target be composed of dipoles. The dipoles are color neutral,
and therefore there should be no non-commutativity issues any more.
In that case, one may ask if the dual Hamiltonian $\bar{H}$ will
reduce to $H^{\dag}_{1 \to 2}$ in Eq.~(\ref{H1to2}) since the latter
was derived on the basis of the dipole picture. But one cannot
really answer this question, since the assumption that the relevant
degrees of freedom are dipolar, is a condition to be imposed on the
wavefunction of the system and is not a property of the Hamiltonian.
So let us assume that the wavefunction is of the form
    \beq\label{Zdip}
    Z_Y[\rho] =
    \sum_{N=1}^{\infty} \int \dif \Gamma_N\,
    P_N(\{ z_{\i} \} ; Y)
    \prod_{\i =1}^{N}
    \frac{1}{N_c}\, \tr
    \big(W_{\bm{z}_{\i-1}} W^{\dag}_{\bm{z}_\i}\big)
    \delta(\rho),
    \eeq
with the notation introduced in Sec.~\ref{SecDip}. Now in order to
find the evolution equation of $\lan \cal{O} \ran$, it is much
easier to obtain first the evolution of the wavefunction
$Z_Y[\rho]$, then multiply with $\cal{O}$ and finally integrate over
$\rho$. When we apply this procedure to the dipole-pair density we
obtain the standard evolution equation given in Eq.~(\ref{n2BFKL}),
provided that during the calculation we drop terms which are
suppressed in the multi-color limit \cite{HIMS05,MMSW05,KL05d}.
Still, this description contains more than the one given by
$H^{\dag}_{1\to 2}$. In the latter, only two gluons could be
radiated from each color dipole in the wavefunction of the target,
while in the former there is no such restriction and a dipole can
radiate an arbitrary number of gluons. In particular, this means
that two projectile dipoles are allowed to scatter off the same
target dipole.

\section{Effective Action}\label{SecSeff}

\begin{figure}[t]
    \centerline{\epsfig{file=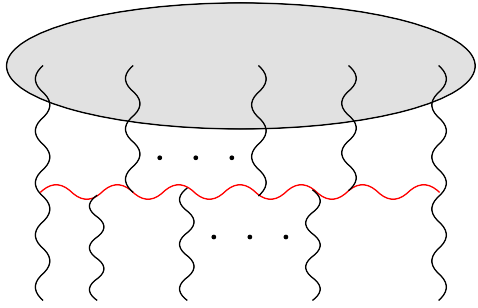,width=6.3cm}}
    \caption{\sl Generic vertex for an $m \to n$ transition.}
    \label{FigVertex}
\end{figure}

The next natural task should be to construct an effective theory
which takes into account all the vertices describing $m \to n$
transitions, with $m$ and $n$ arbitrary (and greater or equal to 2),
like the generic one appearing in Fig.~\ref{FigVertex}. Whether this
is the full set of vertices contributing at the leading logarithmic
level in $\abar \ln(1/x)$ or not (presumably not), is a separate
issue and if not it could turn out that the search for the ``final
solution'' be a formidable and too ambitious task. In any case, it
seems reasonable to avoid the discussion of this problem here.

\begin{figure}[t]
    \centerline{\epsfig{file=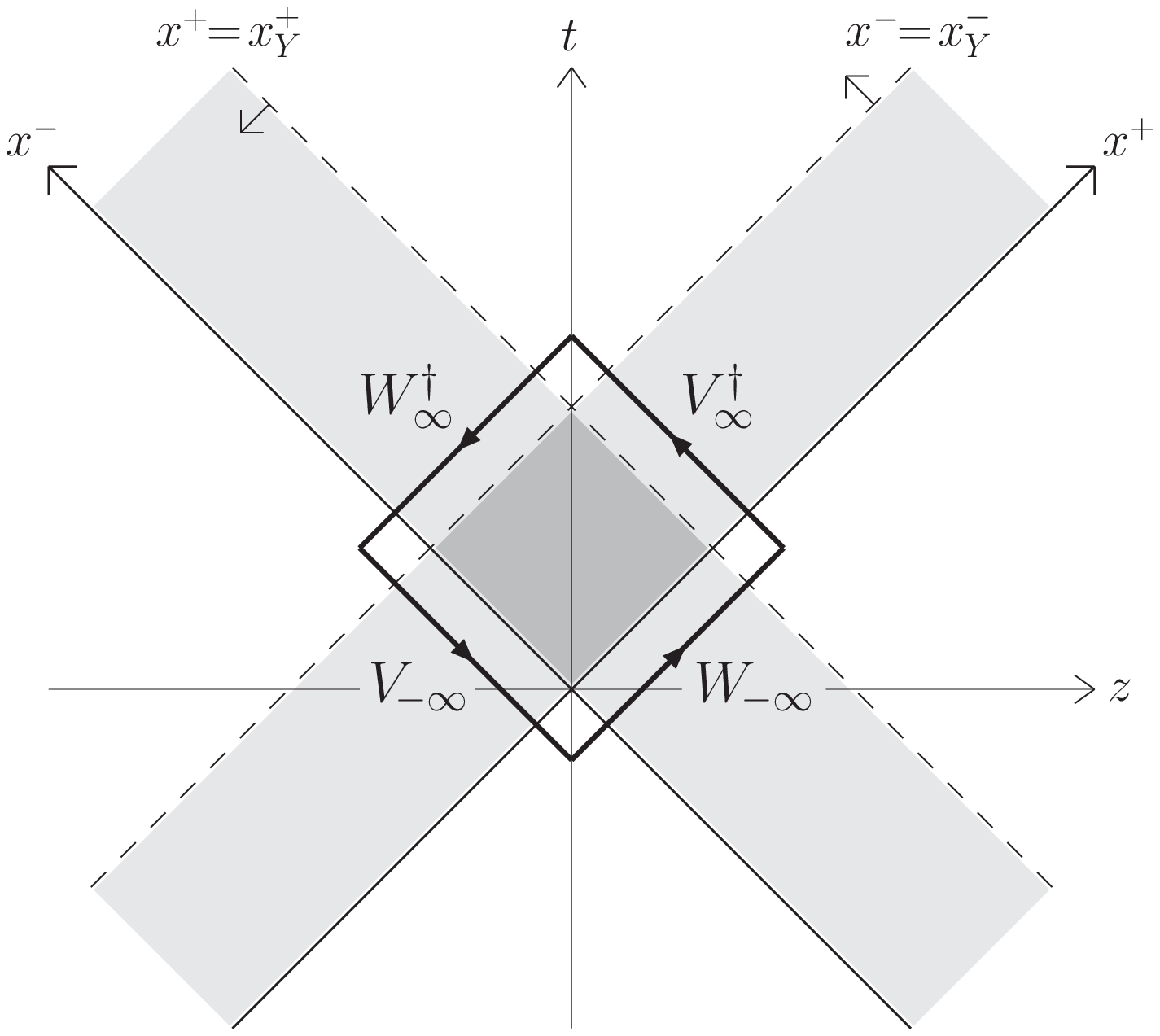,width=8.4cm}}
    \caption{\sl Wilson lines and the distribution of the
    background color fields in the $x^+ - x^-$ plane.}
    \label{FigWilson}
\end{figure}

The diagram in Fig.~\ref{FigVertex}, and in the spirit of the CGC
approach, corresponds to the propagation of a semi-fast gluon (the
horizontal red gluon) in the presence of two types of background
fields, and it represents the modes that we need to integrate as we
go to higher and higher values of rapidity. The two background
fields (in the Coulomb gauge) are the $A^+$ component of the color
field, represented by the vertical gluons above the semi-fast one,
and the radiative part of the $A^{-}$ component, represented by the
vertical gluons below the semi-fast one. The semi-fast gluon will be
slow with respect to the $A^+$ component of the background field, so
it will appear to the latter as moving fast to the positive $x^-$
direction. Thus it will couple to that component through a Wilson
line in the longitudinal direction. Similarly, it will be fast with
respect to the $A^-$ component, it will appear to the latter as
moving fast in the positive $x^+$ direction, and thus it will couple
to that component through a Wilson line in the temporal direction.
Therefore, and not surprisingly, both types of Wilson lines which we
encountered (separately) in the cases of the JIMWLK Hamiltonian and
its dual will appear in the final form of our effective action, and
in fact these will be the only degrees of freedom.
In our current notation these Wilson lines in the adjoint
representation (and by dropping the tilde we have been using so far)
will read
    \beq\label{WilsonVfull}
    \aln V^\dagger_{x^+}(\bm{x})=
    {\rm P}\,\exp\left[
    \rmi\, g \int\limits_{-\infty}^{\infty}
    \dif x^- A^+_a(x^+,x^-,{\bm{x}})\, T^a
    \right],\\
    \aln
    W_{x^-}(\bm{x})=
    {\rm P}\,\exp\left[
    \rmi\, g \int\limits_{-\infty}^{\infty}
    \dif x^+ A^-_a(x^+,x^-,{\bm{x}})\, T^a
    \right].
    \eeq
Again we will just present the final result for the change of the
effective action under a rapidity step, and it can be given in the
form \cite{HIMST05}
    \beq\label{Seff}
    \frac{\Delta S_{\rm eff}}{\Delta Y}= \frac{\rmi}{2 \pi g^2 N_c}
    \int\limits_{\bm{x}} {\rm Tr}
    \left[
    V^{\dagger}_{\infty}
    (\del^i W_{-\infty})
    (\del^i V_{-\infty})
    W^{\dagger}_{\infty}
    \right]
    \,\, + \,\,
    {\rm perm},
    \eeq
where ``perm'' stands for the three more terms arising from the
possible permutations in the position of the spatial derivatives,
and with the indices in the longitudinal and temporal Wilson lines
standing for their $x^+$ and $x^-$ dependence respectively, in
accordance with their definitions given above. This expression for
the effective action has also been obtained in \cite{Bal05} by
studying the scattering of two ``shock waves'' as developed in
\cite{Bal9904}, while an earlier stage of the same expression has
been reached in \cite{KL05c}. The four Wilson lines are not
independent, rather they satisfy the constraint
    \beq\label{TrWilson}
    V^{\dagger}_{\infty}
    W_{-\infty}
    V_{-\infty}
    W^{\dagger}_{\infty} =1 \Rightarrow
    \frac{1}{N_c^2-1}\, \Tr\, W_{\diamondsuit} = 1,
    \eeq
where $W_{\diamondsuit}$ is the Wilson loop shown in
Fig.~\ref{FigWilson}.
This loop encloses the diamond defining the ``interaction'' area;
the color field $A^+$ has support in the strip $0 \leq x^- \leq
x_Y^- \sim x_0^- \rme^Y$ and its width expands with increasing
rapidity, while the color field $A^-$ has support in the strip $0
\leq x^+ \leq x_Y^+ \sim x_0^+ \rme^{-Y}$ and its width contracts
with increasing rapidity. Notice that the value of the effective
action in Eq.~(\ref{Seff}) depends only on the value of the fields
on the boundary of this diamond-shaped area, and therefore it seems
that the high energy problem reduces to a two-dimensional effective
theory. Of course, at the same time, one will not be able to probe
$x^-$ and/or $x^+$ dependent correlations of the system (but this
was the case in both JIMWLK and its dual anyway), since the
construction of $S_{\rm eff}$ results in a coarse-graining along
these longitudinal and temporal directions.

This effective action is invariant under duality transformations as
anticipated \cite{HIMST05}. Indeed, if we let
    \beq\label{dualityW}
    W_{\infty} \rightarrow V^{\dag}_{\infty},
    \qquad
    W_{-\infty} \rightarrow V^{\dag}_{-\infty},
    \eeq
$S_{\rm eff}$ transforms to an expression which is fully equivalent
to the one given in Eq.~(\ref{Seff}). Furthermore, the effective
action will reduce to known expressions in limiting cases. Expanding
the temporal Wilson lines to lowest non-trivial order, integrating
over $x^+$, letting $A^- \!\to -\rmi \delta/\delta \rho$ and using
the Poisson equation to relate $\rho$ to $A^+ \equiv \alpha$, one
obtains the JIMWLK Hamiltonian as given in Eq.~(\ref{HJIMWLK}).
Similarly, an expansion of the longitudinal Wilson lines leads to
the dual of JIMWLK as given in Eq.~(\ref{Hbar}).

However, we should stress that Eq.~(\ref{Seff}) is not the complete
answer to our problem. What we would really like to have is a
Hamiltonian, that is, the effective action should be accompanied by
well-defined commutators satisfied by its degrees of freedom,
i.e.~the Wilson lines. Then we would be able to find how the
wavefunction of the system and the average values of the observables
evolve with rapidity, as we did in the two limiting (JIMWLK and its
dual) cases. For example, we have not been able to do the
``simplest'' but most natural exercise; to write the evolution
equations obeyed by the dipole scattering amplitudes in this general
case where all $m \to n$ transitions are possible, as this would
allow us to generalize the hierarchy presented in
Sec.~\ref{SecPomSplit}. This problem is still ``open'' and therefore
we shall not discuss it here. Let us just mention that the simple
rule $A^- \to -\rmi \delta/\delta \rho$ for promoting an action into
a Hamiltonian, and which is valid in the two limiting cases, will
presumably not work any more. First of all, $A^-$ is not the natural
degree of freedom, rather the latter is given by the corresponding
temporal Wilson line. Furthermore, a replacement like the one
mentioned above would lead to commutators which depend on ``broken''
Wilson lines (lines in, say $x^-$, which run from $-\infty$ to an
upper limit different than $\infty$), a feature that makes the whole
approach ill-defined.

\section{The Saturation Momentum Revisited}\label{SecSatRev}

In this last section we shall try to find the corrections in the
energy dependence of the saturation momentum, which are induced by
the splitting terms in the evolution equations. The result will be
the one found in \cite{MS04}, even though we shall present a much
simpler calculation while at the same time we will attempt to
``improve'' it. As was claimed in \cite{MS04} and in
Sec.~\ref{SecDefBal}, one needs to introduce an ultraviolet
absorptive boundary to cut all the contributions from the region
where the amplitude is such that $T \lesssim \alpha_s^2$, since they
correspond to unitarity violating paths. Put it another way, a
splitting of Pomerons in the target point of view, corresponds to a
merging of Pomerons in the projectile point of view and therefore
this boundary is like a saturation boundary for the projectile. The
result of \cite{MS04} has also been confirmed in \cite{IMM05}, by
identifying and studying the analogy to a similar problem in
statistical physics. Indeed, as we saw in Sec.~\ref{SecPomSplit},
the full large-$N_c$ hierarchy can be reduced, under certain but
justified for our purposes approximations, to a Langevin problem
\cite{IT05a} which is similar to the sFKPP equation. This equation
shares the same features as the, deterministic, FKPP equation, when
the latter is supplied with a cutoff at $\alpha_s^2$.

Therefore we would like to solve the BFKL equation with two
absorptive boundaries\footnote{The BFKL equation, and more precisely
the value of the hard Pomeron intercept, in the presence of momentum
cutoffs has also been studied in the ``remote'' past
\cite{CL92,MFR9597}, but the origin of the cutoffs was much
different.}. The separation, in logarithmic units, between these two
boundaries, the saturation line and the fluctuation line, should be
    \beq\label{boundsep}
    \Delta = \frac{1}{1-\gamma_{\rm r}}\,\ln\frac{1}{\alpha_s^2} +
    \cal{O}(\rm const),
    \eeq
as shown in Fig.~\ref{FigPlane}, since within $\Delta$ the amplitude
will drop from a value of order $\cal{O}(1)$ to a value of order
$\cal{O}(\alpha_s^2)$, as determined from its leading power behavior
in $Q^2$ (in Eq.~(\ref{boundsep}) we have anticipated that the real
part of the anomalous dimension will not be $\gamma_s$ any more and
its value $\gamma_{\rm r}$ is to be determined). The constant in
Eq.~(\ref{boundsep}) is not really under control, since it can be
affected, for example, by the precise value of $T$ on the
boundaries. Even though this constant should not affect the leading
correction, since it is suppressed with respect to the leading term
when $\alpha_s \to 0$, it will turn out to have an important
influence on the results for small, but reasonable, values of the
coupling constant.

By making a change of variables, from $(\ln(Q^2/\mu^2),Y)$ to
$(z,Y)$, with $z\equiv\ln(Q^2/Q_s^2(Y))$, one can write the BFKL
equation as
    \beq\label{BFKLeq0}
    \left[ \chi\left(1 + \frac{\del}{\del z} \right) -
    \lambda_s\, \frac{\del}{\del z} \right] T =0.
    \eeq
The above expression has been set equal to zero, instead of
$(1/\abar)\,\del T /\del Y$, since we are looking for lines of
constant scattering amplitude $T$, that is, $Y$-independent
solutions\footnote{Of course, by making this assumption we will not
be able to find the approach to the asymptotics, i.e.~$Y$-dependent
corrections for $\lambda_s$. Nevertheless, due to the ``strong
absorption'', these corrections drop very fast, more precisely
exponentially, in this two boundary problem.} for $T$. In
Eq.~(\ref{BFKLeq0}), $\lambda_s$ is the logarithmic derivative of
the saturation momentum, which is to be determined. Now let us
recall that the BFKL eigenfunctions are exponentials in $z$, where
the anomalous dimension is in general complex, i.e.~$\gamma =
\gamma_{\rm r} + \rmi\, \gamma_{\rmi}$, with obvious notation. Then
one can see that the unique real linear combination of
eigenfunctions, which satisfies the boundary conditions
$T(z=0)=T(z=\Delta)=0$ is
    \beq\label{Ttwobound}
    T \propto \exp[-(1-\gamma_{\rm r})\,z]
    \sin \frac{\pi z}{\Delta}.
    \eeq
In the above equation the sine has been obviously generated by the
linear combination of the parts of the two degenerate eigenfunctions
that involve the complex piece of the anomalous dimension, which is
uniquely determined in terms of $\Delta$ and reads $\gamma_{\rmi} =
\ \pi/\Delta$. Furthermore, in order for Eq.~(\ref{Ttwobound}) to
satisfy Eq.~(\ref{BFKLeq0}) and since the saturation momentum is a
real quantity, the constant $\lambda_s$ must be given by
    \beq\label{lambdab}
    \lambda_s = \frac{\chi(\gamma)}{1-\gamma}
    \qquad {\rm with} \qquad
    {\rm Im}(\lambda_s) = 0.
    \eeq
This last equation is the solution to our problem, since for a given
value of $\gamma_{\rmi}$ (that is, for a given $\Delta$ or for a
given $\alpha_s$), there will be a unique value of the real part
$\gamma_{\rm r}$ of the anomalous dimension which makes the
logarithmic derivative of the saturation momentum $\lambda_s$ real.
Notice that the solution will not correspond to a saddle point of
$\chi(\gamma)/(1-\gamma)$, in contrast to the single boundary case,
cf.~Eq.~(\ref{gammas}).

\begin{figure}[t]
    \centerline{\epsfig{file=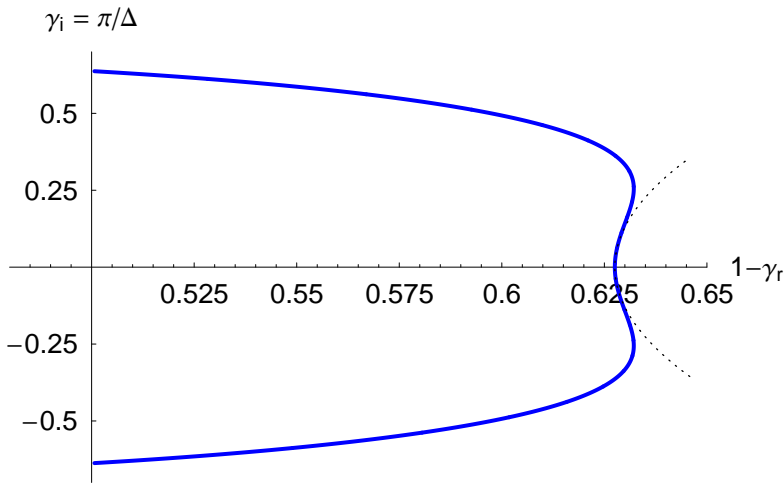,width=10.5cm}}
    \caption{\sl The anomalous dimension
    in the two boundary approximation
    (with the thin dotted line corresponding to Eq.~(\ref{gammacorr})).}
    \label{FigGamma}
\end{figure}

When the separation of the boundaries is large, or equivalently the
coupling is extremely small, i.e.~$\Delta \gg 1 \Leftrightarrow
\alpha_s \ll 1$, we can expand the r.h.s.~of Eq.~(\ref{lambdab})
around the asymptotic anomalous dimension which is $\gamma_s$ as
expected and as shown in Fig.~\ref{FigGamma}. By setting the
imaginary part of this expansion equal to zero, we uniquely
determine $\gamma_{\rm r}$ and then the real part gives $\lambda_s$.
It is not hard to find that in this expansion the logarithmic
derivative of the saturation momentum is given by \cite{MS04}
    \beq\label{lambdacorr}
    \frac{\lambda_s}{\abar} =
    \frac{\chi(\gamma_s)}{1-\gamma_s}
    -\frac{\pi^2 (1-\gamma_s) \chi''(\gamma_s)}
    {2 \ln^2(\alpha_s^2)}=
    4.88 -\frac{150}{\ln^2(\alpha_s^2)},
    \eeq
while the real part of the anomalous dimension reads
    \beq\label{gammacorr}
    \gamma_{\rm r} = \gamma_s
    + \left[
    1
    +\frac{(1-\gamma_s)\chi'''(\gamma_s)}{3 \chi''(\gamma_s)}
    \right]
    \frac{\pi ^2(1-\gamma_s)}{2\ln^2(\alpha_s^2)}
    = 0.372 - \frac{0.562}{\ln^2 (\alpha_s^2)}.
    \eeq
It is obvious that as we let the second boundary to infinity,
i.e.~when $\alpha_s \to 0$, we approach the results of the single
boundary problem. The same is true for the scattering amplitude too;
as $\Delta \to \infty$, Eq.~(\ref{Ttwobound}) reduces to the scaling
part (the first two factors) of Eq.~(\ref{Tscale}). Of course it is
not surprising that the correction in $\lambda_s$ is negative, since
there is absorption from the two boundaries. Moreover, we should
notice that the denominator in the correction of $\lambda_s$ has
again a simple interpretation; it is proportional to the square of
the ``effective'' transverse phase space, in logarithmic units, for
the evolution of the system, which here is $\Delta^2 \sim
\ln^2(\alpha_s^2)$, as it was in the single boundary case
(cf.~Eq.~(\ref{lambda}) and the discussion after that). A rather
important feature of these corrections, which are generated in
reality by the negative contribution of the Pomeron loops formed in
the course of the evolution, is that they appear to be much more
significant that any next to leading order correction; indeed the
latter would simply add a term of order $\cal{O}(\alpha_s)$ with
respect to the leading term.

These corrections have the nice feature that they are universal,
that is, they do not depend on the precise width of the boundary. If
we let $\alpha_s \to \kappa \alpha_s$, with $\kappa = \cal{O}(1)$,
which amounts to the modification of the boundary width by an
additive constant, the induced correction in Eqs.~(\ref{lambdacorr})
and (\ref{gammacorr}) will be of order
$\cal{O}(1/\ln^{3}(\alpha_s^2))$. But within the same reasoning,
these equations are as far as one can go by using analytical
methods, since any higher order correction will depend on the
precise value of the boundary width, and thus one cannot rely on the
BFKL equation any more.

This brings us in a somewhat difficult situation, since the
coefficient in the correction of $\lambda_s$ is huge, mostly due to
the large value of $\chi''(\gamma_s)$. Indeed, for reasonable values
of $\alpha_s$ the correction will dominate the leading contribution;
$\lambda_s$ as given in Eq.~(\ref{lambdacorr}) will be negative so
long as $\alpha_s \gtrsim 0.06$. Of course this does not mean that
there is something wrong neither with Eq.~(\ref{lambdacorr}) nor
with the problem at hand. Eq.~(\ref{lambdacorr}) is really supposed
to be valid at very small $\alpha_s$, while for a realistic value of
$\alpha_s$ we should rely on a (numerical) solution of the full set
of equations, i.e.~including non-linearities and fluctuations.

\begin{figure}[t]
    \centerline{\epsfig{file=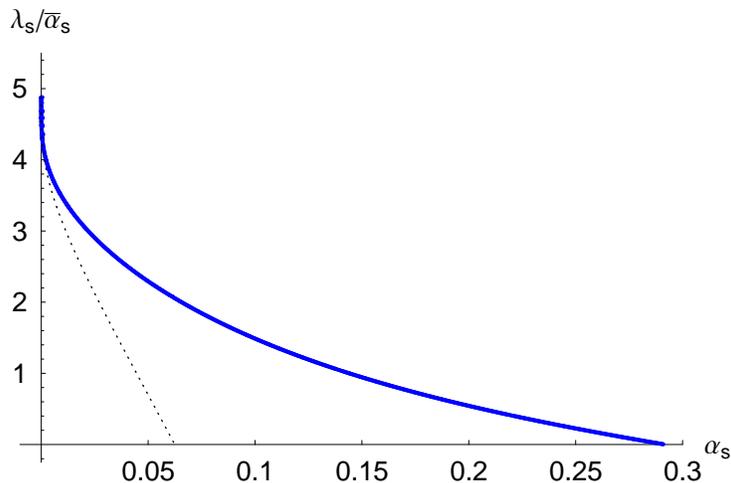,width=10.5cm}}
    \caption{\sl The logarithmic derivative
    of the saturation momentum in the two boundary approximation
    (with the thin dotted line corresponding to Eq.~(\ref{lambdacorr})).}
    \label{FigLambda2}
\end{figure}

One can already see that the situation will be better if we decide
not to expand the BFKL kernel in the diffusion approximation, as we
did while going from Eq.~(\ref{lambdab}) to Eq.~(\ref{lambdacorr}).
Eq.~(\ref{lambdab}) is just an algebraic equation which can be
easily solved numerically for a given, but arbitrary, value of
$\alpha_s$. In Fig.~\ref{FigGamma} we show the solution for the
anomalous dimension in the complex plane, while in
Fig.~\ref{FigLambda2} we show the dependence of $\lambda_s/\abar$ on
$\alpha_s$.

As we see in Fig.~\ref{FigGamma} and also in Eq.~(\ref{gammacorr})
there is an important correction to the real part of the anomalous
dimension, but not as huge as in the case of $\lambda_s$. It is
amusing, but also natural, that there is no solution for values of
$\gamma$ such that $\gamma_{\rm r} \geq \gamma_{\mathbb{P}} = 1/2$.
This is the point where the saturation line becomes identical to the
Pomeron intercept line. That means that the latter becomes a line of
constant amplitude (like the saturation line); the negative
corrections induced by the boundaries will exactly cancel the
leading value $\omega_{\mathbb{P}}= 4 \abar \ln 2$. Since the
Pomeron intercept line corresponds to the line of fastest increase,
we conclude that there will be no high energy growth any more,
therefore no saturation, and thus no solution to our problem!

In Fig.~\ref{FigLambda2} we see that $\lambda_s$ starts from the,
well-known by now, asymptotic value $4.88 \abar$ for vanishing
$\alpha_s$ and it decreases to become zero at a point which is of
course the same as the point in Fig.~\ref{FigGamma} where
$\gamma_{\rm r}=1/2$ and there is no solution any more. The change
in the ``speed'' of the saturation momentum is still significant,
but the total result is positive in a much wider region of possible
values of $\alpha_s$, more precisely one has $\lambda_s(\alpha_s
\lesssim 0.3) >0$. Of course this is not really under control, since
one is free to rescale $\alpha_s$ by a factor of order $\cal{O}(1)$.
Nevertheless, this result behaves much better than
Eq.~(\ref{lambdacorr}), while both are solutions to the same problem
and under the same definition for the boundary width $\Delta$ as a
function of $\alpha_s$. Thus, one expects that in a (numerical)
solution of the full problem, the outcome will presumably be
well-defined for all reasonable values of the coupling constant.

Our presentation would be incomplete, if at this point we would not
mention that the scaling behavior of the amplitude will not persist
at very high values of rapidity \cite{MS04} due to the presence of
the fluctuations terms in the evolution equations
\cite{IMM05,IT05a}. Since the system is of stochastic nature,
different events will lead to different profiles of the scattering
amplitude as a function of $Q^2$ and at a given fixed rapidity $Y$
\cite{IMM05}. These profiles will be of the same form but shifted
with respect to each other according to a probability density, which
at a first approximation can be taken as a Gaussian with a diffusive
radius which scales as $\scriptstyle{\sqrt{\textstyle{\abar
Y/\ln^3(1/\alpha_s^2)}}}$. Averaging over all the events, one finds
that \cite{IMM05}
    \beq\label{Tave}
    \lan T(Q^2,Y) \ran =
    F\left(\frac{\ln(Q^2/\lan Q_s^2 \ran)}
    {\scriptstyle{\sqrt{\textstyle{\abar
    Y/\ln^3(1/\alpha_s^2)}}}}\right),
    \eeq
which clearly violates the geometrical scaling of the amplitude.
Such violations, and their consequences, for instance the breakdown
of the BFKL approximation in the high momentum region \cite{IT05a},
have been seen in the existing numerical solutions
\cite{Soy05,EGBM05}, but the precise value of the critical rapidity
that they set in (which could depend on the initial conditions) is
not under real control yet.

\section{Epilogue}

In these lectures I have tried to give a simple introduction (and
not only) to the basic theoretical developments, during the last
decade or so, in the field of high energy (small-$x$) QCD. To this
end, I have introduced, and relied mostly to, the dipole picture of
high energy evolution and the Color Glass Condensate formulation of
an energetic hadron, since, to my opinion, these are the simplest
and perhaps most promising self-consistent approaches to the
problem. Even though I have not followed a step by step procedure in
the construction of the effective theory governing the dynamics at
this high-energy limit, I have tried to describe as much as possible
the underlying physical picture, and in the Appendices that follow I
give the derivations of some relevant equations in order to have a
more complete presentation.

I feel that we have seen important progress in the field during the
last year, since we were able to derive evolution equations
containing Pomeron loops as one of their building blocks, which is
really the mechanism that eventually leads to the unitarization of
scattering amplitudes and the saturation of hadronic wavefunctions
(even though for an accurate description of the latter one may need
to go beyond the large-$N_c$ limit).

Of course there are still many open questions and things to be done.
For instance, it would be natural to look for the extension of the
evolutions equations to finite-$N_c$ which allow transitions among
arbitrary numbers of Pomerons, even though the solution of such
generalized equations may turn out to be not much different for most
interesting quantities (in the same way that the BK and JIMWLK
equations share, at some level, the same features). So, at the
moment, we would like to have a better analytical insight to the
current set of equations, and presumably this can be achieved by
analyzing their stochastic nature, while at the same time more
accurate solutions are needed. Clearly this is rather important if
one aims to see the implications at the phenomenological level. To
this purpose, and for a more complete analysis, one would really
like to go beyond this hierarchy, which is good for the description
of total cross sections, and extend it, if possible, to the case of
diffractive scattering (such attempts, but restricted to BK and/or
JIMWLK evolution have been already made \cite{KL00,KW01,HSW05}). On
the theoretical side, I believe that it is important to investigate
the relationship with the other approaches to the problem and the
possible equivalence between them, so that at the end of the day one
will have a unified and concrete description of the dynamics in
high-energy QCD.

\section*{Acknowledgments}

I would like to thank the organizers, Andrzej Bialas, Krzysztof
Golec-Biernat, and especially Michal Prasza{\l}owicz, for the
invitation to lecture at the summer school. I am grateful to my
collaborators, and in particular to Edmond Iancu, with whom some of
the ideas developed and results discussed here were obtained. I am
indebted to Jean-Paul Blaizot for a careful reading of the
manuscript and for his critical comments.

\appendix

\section{The Dipole Kernel}\label{AppDipKer}

\begin{figure}[t]
    \centerline{\epsfig{file=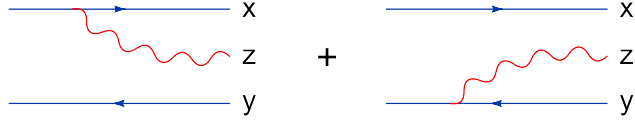,width=8.4cm}}
    \caption{\sl The two diagrams contributing to the splitting amplitude.}
    \label{FigDipole}
\end{figure}

In this Appendix we will derive the expression given in
Eq.~(\ref{SplitProb}) for the dipole emission kernel \cite{Mue94}.
Let us look at the left diagram in Fig.~\ref{FigDipole}, where a
soft gluon is emitted by the quark line. With the 4-momenta of the
quark and the gluon being $p^{\mu}=(p^- \!= \bm{p}^2/2p^+\!\to \!0
,p^+ \! \!\to \! \infty,\bm{p} \!\to \!\bm{0})$ and $k^{\mu}=(k^-=
\bm{k}^2/2 k^+, k^+ = x p^+, \bm{k})$ respectively, the probability
amplitude for this emission in light-cone old fashioned perturbation
theory is given by
    \beq\label{SplitAmp}
    \cal{M}_{\lambda}^a(k^+,\bm{k}) =
    \frac{g\, t^a \epsilon_{\lambda}^-}{\sqrt{(2 \pi)^3 2 k^+}}\,
    \frac{1}{k^-} =
    \frac{2 g\,t^a}{\sqrt{(2 \pi)^3 2 k^+}}\,
    \frac{\bm{\epsilon}_{\lambda} \dot \bm{k}}{\bm{k}^2}.
    \eeq
Here, $\lambda$ stands for the polarization of the gluon, $g\, t^a
\epsilon_{\lambda}^-$ is the quark gluon vertex in the high energy
approximation, $1/k^{-}$ corresponds to the energy denominator which
is dominated by the gluon, and the square root factor arises from
the usual normalization of the color field. We have also used the
fact that, in the light cone gauge $A^+=0$, the condition $k_{\mu}
\epsilon_{\lambda}^{\mu}=0$ implies $\epsilon_{\lambda}^- =
\bm{\epsilon}_{\lambda} \dot \bm{k}/k^+$. It is straightforward to
perform a Fourier transformation in order to obtain the probability
amplitude in transverse coordinate space, which reads
    \beq\label{SplitAmp2}
    \cal{M}_{\lambda}^a(k^+,\bm{x}\!-\!\bm{z}) =
    \aln \int\limits_{\bm{k}}
    \frac{\exp\left[ \rmi \bm{k} \dot (\bm{x}\!-\! \bm{z})\right]}{(2 \pi)^2}\,
    \cal{M}_{\lambda}^a(k^+,\bm{k})
    \nn
    = \aln \frac{1}{2\pi}\,\frac{2 \rmi g\,t^a}{\sqrt{(2 \pi)^3 2 k^+}}\,
    \frac{\bm{\epsilon}_{\lambda} \dot
    (\bm{x}\!-\!\bm{z})}{(\bm{x}\!-\!\bm{z})^2}.
    \eeq
Similarly the probability amplitude for emission from the antiquark
line of the dipole will be
$\cal{M}_{\lambda}^a(k^+,\bm{z}\-\bm{y})$. Then the differential
emission probability in the interval $\dif^2\bm{z}$ and from $k^+$
to $k^+ \!- \dif k^+$ (we anticipate that the change in the
probability will be positive when $k^+$ decreases and therefore the
longitudinal phase space ``opens up'') will be given by
    \beq\label{PequalM2}
    \dif P = -
    (2\pi)^2\sum_{a,\lambda}
    \left| \cal{M}_{\lambda}^a(k^+,\bm{x}\!-\!\bm{z})
    + \cal{M}_{\lambda}^a(k^+,\bm{z}\!-\!\bm{y})\right|^2
    \dif^2 \bm{z}\, \dif k^+.
    \eeq
The factor $(2\pi)^2$ arises when we transform the (square of the)
wavefunction of the initial state in momentum space to the one in
coordinate space. By using $t^a t^a = (N_c^2 - 1)/(2 N_c) \to N_c/2$
to sum over color indices in the large-$N_c$ limit, and
$\sum_{\lambda}\epsilon_{\lambda}^{\i} \epsilon_{\lambda}^{*\j} =
\delta^{\i\j}$ to sum over polarization indices, we finally arrive
at
    \beq\label{AppSplitProb}
    \dif P =
        \atpi\,
        \frac{(\bm{x}-\bm{y})^2}
        {(\bm{x}-\bm{z})^2(\bm{z}-\bm{y})^2}\,
        \dif^2 \bm{z}\, \dif Y.
    \eeq
In the above equation, we have introduced the rapidity increment
$\dif Y$. Notice that this factor arises from the longitudinal phase
space $-\dif k^+ /k^+ = - \dif x/x = \dif \ln (1/x) = \dif Y$.
\section{The Elementary Dipole-Dipole Scattering Amplitude}\label{Appdipdip}

Here we will calculate the imaginary part of the dipole-dipole
scattering amplitude in the two-gluon exchange approximation. In
this limit only tree diagrams need to be considered and one can
simply use classical field equations. Let us assume that the dipole
$(\bm{u},\bm{v})$ creates a field $\alpha^a$ which is ``seen'' from
the dipole $(\bm{x},\bm{y})$. In general, the amplitude for the
dipole $(\bm{x},\bm{y})$ to scatter off a classical field $\alpha^a$
is determined by
    \beq\label{T0alpha}
    T(\bm{x}\bm{y}|\alpha) =
    \frac{g^2}{4 N_c}\,
    \big[\alpha^a(\bm{x}) - \alpha^a(\bm{y}) \big]^2.
    \eeq
When this field is created by a source $\rho^a$, it can be
determined by the solution of the Poisson equation in the Coulomb
gauge, more precisely
    \beq\label{Poisson}
    \lap{z} \alpha^a(\bm{z}) = - \rho^a(\bm{z})\,\,
    \Longrightarrow\,\,
    \alpha^a(\bm{z}) =
    -\frac{1}{4 \pi}
    \int\limits_{\bm{w}}
    \ln\big[(\bm{z}-\bm{w})^2 \mu^2\big] \rho^a(\bm{w}).
    \eeq
Then Eq.~(\ref{T0alpha}) becomes
    \beq\label{T0rho}
    T(\bm{x}\bm{y}|\rho) =
    \frac{g^2}{64 \pi^2 N_c}
    \left[
    \int\limits_{\bm{w}}
    \ln \bigg[\frac{(\bm{x}-\bm{w})^2}{(\bm{y}-\bm{w})^2}\bigg]\,
    \rho^a(\bm{w})
    \right]^2.
    \eeq
In the problem under consideration the source consists of the quark
and the antiquark of the dipole $(\bm{u},\bm{v})$, and therefore its
charge is given by
    \beq\label{rhodipole}
    \rho^a(\bm{w}) = g \,t^a
    \left[ \delta(\bm{w}-\bm{u}) -
    \delta(\bm{w} - \bm{v}) \right].
    \eeq
Substituting the expression (\ref{rhodipole}) into Eq.~(\ref{T0rho})
and by making use of
    \beq\label{tata}
    t^a t^a = \frac{N_c^2-1}{2 N_c},
    \eeq
we finally arrive at
    \beq\label{A0Nc}
    T_0(\bm{x}\bm{y}|\bm{u}\bm{v}) =
    \frac{\alpha_s^2}{8}\,
    \frac{N_c^2-1}{N_c^2}\,
    \ln^2 \left[\frac{(\bm{x}-\bm{v})^2 (\bm{y}-\bm{u})^2}
    {(\bm{x}-\bm{u})^2 (\bm{y}-\bm{v})^2}
    \right].
    \eeq
Notice that this is invariant under the exchange of the two dipoles,
i.e.~$\bm{x}\bm{y} \leftrightarrow \bm{u}\bm{v}$ and under the
exchange of the quark and the antiquark within the same dipole,
i.e.~$\bm{x} \leftrightarrow \bm{y}$ and/or $\bm{u} \leftrightarrow
\bm{v}$. The latter will not be true, in general, when we relax the
two-gluon exchange approximation.

One can also integrate over the impact parameter of the scattering,
i.e.~over the relative distance $\bm{b}=(\bm{x}\!+\bm{y} \- \bm{u}
\-\bm{v})/2$ between the two dipoles, average over the orientation
of the two dipoles, and multiply by a factor of 2 to obtain a total
cross section
    \beq\label{sigmadd}
    \sigma_{\rm DD}(r_1,r_2) =
    2 \pi \alpha_s^2\, r_<^2
    \bigg( 1 + \ln\frac{r_>}{r_<}\bigg).
    \eeq
In the above equation $r_1$ and $r_2$ are the sizes of the two
dipoles $(\bm{x},\bm{y})$ and $(\bm{u},\bm{v})$, $r_< = {\rm
min}(r_1,r_2)$ and $r_> = {\rm max}(r_1,r_2)$, while we have also
set the color factor equal to its large-$N_c$ value. When either of
the dipole sizes vanishes, this total cross section vanishes due to
color transparency.

\section{The BFKL Equation: From Dipole Densities
to Scattering Amplitudes}\label{AppBFKL}

The BFKL equation for the dipole density in the wavefunction of the
target reads
    \beq\label{AppnBFKL}
    \frac{\del \lan n_{\bm{u}\bm{v}} \ran}{\del Y}=
    \frac{\abar}{2\pi}
    \int\limits_{\bm{z}}
    \big[{\cal M}_{\bm{u}\bm{z}\bm{v}} \lan n_{\bm{u}\bm{z}} \ran
    +{\cal M}_{\bm{z}\bm{v}\bm{u}} \lan n_{\bm{z}\bm{v}} \ran
    -{\cal M}_{\bm{u}\bm{v}\bm{z}} \lan n_{\bm{u}\bm{v}} \ran \big].
    \eeq
In the two-gluon exchange approximation, the amplitude for a dipole
$(\bm{x},\bm{y})$ to scatter of the target is given by a linear
transformation of the dipole density, namely
    \beq\label{Tfuncn}
    \lan T_{\bm{x}\bm{y}} \ran=
    \int\limits_{\bm{u}\bm{v}}
    T_0(\bm{x}\bm{y}|\bm{u}\bm{v})\, \lan n_{\bm{u}\bm{v}} \ran,
    \eeq
with $T_0(\bm{x}\bm{y}|\bm{u}\bm{v})$ the elementary dipole-dipole
scattering amplitude at lowest order in perturbation theory, and
when the color-factor is set to its large-$N_c$ value, one has
(cf.~Eq.~(\ref{A0Nc}))
    \beq\label{T0}
    T_0(\bm{x}\bm{y}|\bm{u}\bm{v}) =
    \frac{\alpha_s^2}{8}
    \ln^2 \left[\frac{(\bm{x}-\bm{v})^2 (\bm{y}-\bm{u})^2}
    {(\bm{x}-\bm{u})^2 (\bm{y}-\bm{v})^2}
    \right].
    \eeq
Differentiating Eq.~(\ref{Tfuncn}) with respect to $Y$, using
Eq.~(\ref{AppnBFKL}) and making some trivial interchanges of
variables we obtain
    \beq\label{TBFKLtemp}
    \!\!\!
    \frac{\del \lan T_{\bm{x}\bm{y}} \ran }{\del Y}=
    \frac{\abar}{2\pi}\!
    \int\limits_{\bm{u}\bm{v}\bm{z}}\!
    \cal{M}_{\bm{u}\bm{v}\bm{z}}
    \big[
    T_0(\bm{x}\bm{y}|\bm{u}\bm{z})
    +T_0(\bm{x}\bm{y}|\bm{z}\bm{v})
    \-T_0(\bm{x}\bm{y}|\bm{u}\bm{v})
    \big]
    \lan n_{\bm{u}\bm{v}} \ran.
    \eeq
Let us concentrate on the integration over $\bm{z}$, and make a
change of variables through the conformal transformation
    \beq\label{conformal}
    z \to \frac{z + c_1}{c_2\, z -1},
    \eeq
with the parameters $c_1$ and $c_2$ given by
    \beq\label{cs}
    c_1 = - \frac{x^{-1} - y^{-1} + u^{-1} - v^{-1}}
    {(x u)^{-1} - (y v)^{-1}}, \quad
    c_2 = \frac{x - y + u - v}
    {x u - y v},
    \eeq
and where we have defined the complex variable $z=z_1 + \rmi\, z_2$,
with $\bm{z} = (z_1,z_2)$ (and similarly for
$\bm{x},\bm{y},\bm{u},\bm{v}$). This particular transformation for
$\bm{z}$ does not change the functional form of the integrand,
composed of the dipole kernel, the square bracket and the
integration measure $\dif^2\bm{z}$, but simply performs the mapping
$\bm{x}\leftrightarrow \bm{u}$ and $\bm{y} \leftrightarrow \bm{v}$.
Therefore Eq.~(\ref{TBFKLtemp}) becomes
    \beq\label{TBFKLtemp2}
    \!\!\!
    \frac{\del \lan T_{\bm{x}\bm{y}} \ran}{\del Y}=
    \frac{\abar}{2\pi}\!
    \int\limits_{\bm{u}\bm{v}\bm{z}}\!
    \cal{M}_{\bm{x}\bm{y}\bm{z}}
    \big[
      T_0(\bm{x}\bm{z}|\bm{u}\bm{v})
    + T_0(\bm{z}\bm{y}|\bm{u}\bm{v})
    \- T_0(\bm{x}\bm{y}|\bm{u}\bm{v})
    \big]
    \lan n_{\bm{u}\bm{v}} \ran,
    \eeq
and now by making use of Eq.~(\ref{Tfuncn}), we finally arrive at
the equation describing the scattering amplitude of a single
projectile dipole off the target in the BFKL approximation which
reads
    \beq\label{TBFKL}
    \frac{\del \lan T_{\bm{x}\bm{y}} \ran}{\del Y}=
    \frac{\abar}{2\pi}
    \int\limits_{\bm{z}}
    \cal{M}_{\bm{x}\bm{y}\bm{z}}
    \big[\lan T_{\bm{x}\bm{z}} \ran +
    \lan T_{\bm{z}\bm{y}} \ran
    - \lan T_{\bm{x}\bm{y}} \ran
    \big].
    \eeq

\section{Eigenvalues of the BFKL Equation}\label{Appeigen}

Here we shall find the eigenvalues of the BFKL equation in the
simplified case where the scattering amplitude satisfies
$T_{\bm{x}\bm{y}} = T(r)$, where $r=|\bm{r}|\equiv|\bm{x}-\bm{y}|$
is the size of the dipole. The integration measure on the right hand
side of Eq.~(\ref{TBFKL}) may be written as
    \beq\label{d2z}
    \dif ^2 \bm{z} =
    2\pi r_1 r_2 \dif r_1 \dif r_2
    \int\limits_{0}^{\infty} \dif \ell\,\ell\,
    J_0(\ell r)
    J_0(\ell r_1)
    J_0(\ell r_2),
    \eeq
with $r_1$ and $r_2$ the sizes of the child dipoles
$(\bm{x},\bm{z})$ and $(\bm{z},\bm{y})$ respectively, and where
$J_0$ is a Bessel function of the first kind. Then the right hand
side of Eq.~(\ref{TBFKL}) (divided by $\abar$) becomes
    \beq\label{KonTtemp}
    \cal{K}_{\rm BFKL}^{\epsilon} \otimes T(r) \equiv
    \!\! \int\limits_{r_1 r_2 \ell}
    \!\!
    \frac{1}{r_1^{1-2\epsilon}}\,
    \frac{1}{r_2^{1-2\epsilon}}\,
    r^2 \ell\,
    \aln J_0(\ell r)
    J_0(\ell r_1)
    J_0(\ell r_2)
    \nn
    \aln \times [T(r_1) + T(r_2) - T(r)],
    \eeq
with $\epsilon$ a positive regularization constant which at the end
will be set equal to zero. Since there is no mass scale appearing in
the BFKL equation and the kernel combined with the integration
measure has no dimension, one expects the eigenfunctions to be pure
powers of the dipole size $r$. Indeed, with $T(r) = r^{2(1-\gamma)}$
one can successively integrate over $r_1$, $r_2$ and $\ell$ in
Eq.~(\ref{KonTtemp}) to obtain
    \beq\label{Konrtemp}
    \hspace{-0.40cm}
    \cal{K}_{\rm BFKL}^{\epsilon} \!\!\otimes\! r^{2(1\!-\gamma)} \!\!=\!
    \frac{\Gamma(\epsilon)}
    {\Gamma(1\!-\!\epsilon)}\!
    \bigg[
    \frac{\Gamma(\gamma\!-\!2\epsilon)
    \Gamma(1\!+\!\epsilon\!-\!\gamma)}
    {\Gamma(\gamma\!-\!\epsilon)
    \Gamma(1\!+\!2\epsilon\!-\! \gamma)}
    \!-\!\frac{\Gamma\big(\frac{1}{2}\!-\!\epsilon\big)}
    {4^{\epsilon} \Gamma\big(\frac{1}{2}\!+\!\epsilon\big) }\bigg]
    r^{2(1\!-\gamma) + 4 \epsilon}.
    \eeq
Taking the limit $\epsilon \to 0$ one finds
    \beq\label{Konr}
    \cal{K}_{\rm BFKL}\otimes r^{2(1-\gamma)}
    = \chi(\gamma)\,r^{2(1-\gamma)},
    \eeq
where
    \beq\label{chi}
    \chi(\gamma)=
    2\, \psi(1) - \psi(\gamma) - \psi(1-\gamma),
    \eeq
with $\psi(\gamma)$ the logarithmic derivative of the
$\Gamma$-function, i.e.
    \beq\label{Gamma}
    \psi(\gamma) \equiv \frac{\dif \ln \Gamma(\gamma)}{\dif \gamma}.
    \eeq

An alternative, and perhaps easier, derivation may be given by
noticing that the eigenvalue may be written as
    \beq\label{chialt1}
    \chi(\gamma) =
    \frac{1}{\pi}
    \int\limits_{\bm{z}}
    \frac{\bm{x}^2}{\bm{z}^2(\bm{x}-\bm{z})^2}
    \left(\frac{\bm{z}^{2-2\gamma}}{\bm{x}^{2-2\gamma}}
    -\frac{\bm{x} \dot \bm{z}}{\bm{x}^2} \right),
    \eeq
where we have set $\bm{y}=\bm{0}$, we have used the fact that the
first two terms in the BFKL equation give the same result and we
have split the virtual term into two parts using $\bm{x}^2 =
\bm{x}\dot \bm{z} - \bm{x} \dot (\bm{z} \- \bm{x})$, where both
parts contribute equally. Now we can make the change of variable $z
= x u$ to obtain
    \beq\label{chialt2}
    \chi(\gamma) \aln =
    \frac{1}{\pi}
    \int\limits_{0}^{1} \dif u\,
    \int \limits_{0}^{2\pi} \dif \phi\,
    \frac{u^{1-2\gamma} + u^{2\gamma-1}- 2\cos\phi}
    {1 - 2 u \cos \phi +u^2 }
    \nn \aln =
    2 \int\limits_{0}^{1} \dif u\,
    \frac{u^{1-2\gamma} + u^{2\gamma-1}- 2 u}
    {1 - u^2 },
    \eeq
where we have transformed the integration over $u$ from $1$ to
$\infty$ to an integration from $0$ to $1$, by letting $u \to 1/u$.
Now the integration over $u$ leads to Eq.~(\ref{chi}).

\section{The JIMWLK Hamiltonian}\label{AppJIMWLK}

In this Appendix we ``sketch'' the derivation of the JIMWLK
Hamiltonian. Let us decompose the field $A^{\mu}$ which is
associated with the wavefunction of the energetic hadron as
    \beq\label{AppAmu}
    A^{\mu} = B^{\mu} + a^{\mu} + \delta A^{\mu},
    \eeq
where $B^{\mu}$ is the background field generated by the source
$\rho$, $a^{\mu}$ represents the semi-fast modes with longitudinal
momenta such that $b \Lambda^{+} \ll |p^+| \ll \Lambda^{+}$ and with
$b \ll 1$, while $\delta A^{\mu}$ contains modes with momenta $k^+
\lesssim b \Lambda^{+}$. Since we want to measure expectation values
of observables at scales below or equal to $b \Lambda^{+}$, we need
to integrate the semi-fast modes $a^{\mu}$. Therefore we define an
effective action as
    \beq\label{AppSeffdef}
    S_{\rm eff} =
    - \rmi \ln \int\limits_{b \Lambda^+}^{\Lambda^+}
    \cal{D}\alpha \exp \left( \rmi S\,[A,\rho] \right),
    \eeq
where the action $S\,[A,\rho]$ generates the classical field
equations in the presence of a source $\rho$ (cf.~Eq.~(\ref{YMEq}))
when $\delta A^{\mu}=0$. The background field $B^{\mu}$ is
determined by the solution to the Poisson equation. In the
light-cone (LC) gauge, where only the transverse components are
non-zero, i.e.~$B^{\mu} = \delta^{\mu\i}B^{\i}$, we have
    \beq\label{AppBfrho}
    D_{\nu} F^{\nu +} =
    - D_{\i}\, \del^{+} B^{\i} = \rho(x^-,\bm{x}).
    \eeq
In writing the above equation we have used that $F^{-+} = - \del^{+}
\delta A^{-} \simeq 0$, which follows from the fact that $\delta
A^{-}$ contains modes with very small longitudinal momenta and
therefore it varies very slowly with $x^{-}$. We can write the
solution to Eq.~(\ref{AppBfrho}) as
    \beq\label{AppBfU}
    B^{\i} = \frac{\rmi}{g}\,
    \tilde{U}(x^-,\bm{x})\, \del^{\i}_{\bm{x}}\, \tilde{U}^{\dag}(x^-,\bm{x}),
    \eeq
where the Wilson line $\tilde{U}^{\dag}$ is given by
    \beq\label{AppWilson}
    \tilde{U}^{\dag}(x^-,\bm{x}) =
    {\rm P} \exp \Bigg[
    \rmi\, g \int\limits_{-\infty}^{x^-}
    \dif z^- \alpha^a(z^-,\bm{x})\, T^a
    \Bigg],
    \eeq
and similarly for $U$. In the above equation $\alpha \equiv
\hat{B}^+$ is the background field in the Coulomb gauge and, with
$\hat{\rho}$ the corresponding Coulomb gauge source, it satisfies
    \beq\label{AppPoisson}
    \lap{x}\, \alpha^a(x^-,\bm{x}) = -\hat{\rho}^a(x^-,{\bm{x}}).
    \eeq
Notice that $B^{\i}$ is a two-dimensional pure gauge, i.e.~$F^{\i\j}
= 0$. Now we expand the action $S$ around $A^{\mu}_0 \equiv
\delta^{\mu\i} B^{\i} + \delta^{\mu -} \delta A^{-}$ and to second
order in the semi-fast modes $a^{\mu}$, that is
    \beq\label{AppSexpansion}
    S = S_0 + \frac{\delta S}{\delta A^{\i}}\, a^{\i} +
    \frac{1}{2}\,a^{\mu}\, G_{\mu\nu}^{-1}\, a^{\nu}.
    \eeq
The coefficient of the linear term may be written as
    \beq\label{AppdSdA}
    \frac{\delta S}{\delta A^{\i}}  =
    D_{\nu}F^{\nu \i} =
    D_{\j}F^{\j \i} + D^{+}F^{- \i} + D^{-} F^{+ \i} \simeq
    2 D^+ F^{-\i},
    \eeq
where we have used the fact that $F^{\i\j}=0$ and also the
approximate equality $D^{-} F^{+ \i} \simeq  D^+ F^{-\i}$. The
latter arises from $\del^+ \delta A^{-} \simeq 0$ which was
justified earlier just after Eq.~(\ref{AppBfrho}). In the quadratic
term in Eq.~(\ref{AppSexpansion}) $G_{\mu\nu}$ is the propagator in
the presence of the background field $B^{\i}$ (when calculating this
propagator, the field $\delta A^{-}$ can be set equal to zero to the
order of accuracy). Now we can perform the integration over the
semi-fast modes $a^{\mu}$ to obtain the (change of the) effective
action. It is given by the four-dimensional double integral
    \beq\label{AppSeff1}
    \Delta S_{\rm eff} =
    -\,\frac{1}{2} \int\limits_{xy}
    (2 D^+ F^{-\i})_x \,G^{\i\j}_{xy}\, (2 D^+ F^{-\i})_y.
    \eeq
Up to now we have been working in the LC gauge. However, the above
expression is gauge invariant and it is more convenient to calculate
it in the Coulomb gauge (we shall immediately drop the hat denoting
the Coulomb gauge quantities so far, like in
Eq.~(\ref{AppPoisson})). In this gauge the propagator reads
    \beq\label{AppProp}
    \hspace{-0.65cm} G^{\i\j}_{xy} = -
    \frac{\rmi\, \Delta Y}{4\pi}\,
    \delta^{\i\j}\,\delta_{\bm{x}\bm{y}}\!
    \left[ \Theta(x^- \- y^-) \tilde{U}^{\dag}_{x^-y^-}(\bm{x})+
    \Theta(y^- \- x^-) \tilde{U}_{y^-x^-}(\bm{x}) \right],
    \eeq
with $\Delta Y = \ln(1/b)$ representing the differential enhancement
in the longitudinal phase space and where the Wilson line $U^{\dag}$
is given by Eq.~(\ref{AppWilson}) but with the lower limit replaced
by $y^-$ (and similarly for the Wilson line $U$). Using the fact
that this propagator satisfies $D_{x}^{+} G_{xy}^{\i\j}=0$, it is
straightforward to show that the integrand in Eq.~(\ref{AppSeff1})
is a total derivative with respect to both $x^-$ and $y^-$, so that
the result of the integration comes only from the ``surface'' terms.
Furthermore, since the propagator is independent of the light-cone
time, we can immediately integrate over $x^+$ and $y^+$. Using
$F^{-\i} = -\del^{\i} \delta A^{-}$ and defining $A^{-}(x^-,\bm{x})
= \int \dif x^+ \delta A^{-}(x)$ we arrive at
    \beq\label{AppSeff2}
    \hspace{-1cm}\frac{\Delta S_{\rm eff}}{\Delta Y} =
    \frac{\rmi}{2\pi}
    \int\limits_{\bm{x}}[\aln
    \del^{\i} A_{\bm{x}}^{-}(\infty)\,
    \del^{\i} A_{\bm{x}}^{-}(\infty) +
    \del^{\i} A_{\bm{x}}^{-}(-\infty)\,
    \del^{\i} A_{\bm{x}}^{-}(-\infty)
    \nn \aln -
    \del^{\i} A_{\bm{x}}^{-}(\infty)\,
    \tilde{V}^{\dag}_{\bm{x}}\,
    \del^{\i} A_{\bm{x}}^{-}(-\infty)\-
    \del^{\i} A_{\bm{x}}^{-}(-\infty)\,
    \tilde{V}_{\bm{x}}\,
    \del^{\i} A_{\bm{x}}^{-}(\infty)],
    \eeq
where the Wilson lines are determined by Eq.~(\ref{WilsonVa}). Now
the evolution Hamiltonian can be obtained via the replacement $A^{-}
\to -\rmi \delta/\delta \rho$. Using the Poisson equation $\nabla^2
\alpha = -\rho$ we can express $\delta/\delta \rho$ in terms of
$\delta/\delta \alpha$, and recalling that the functional
derivatives will act at the end-points of Wilson lines we can infer
that
    \beq\label{Appdda}
    \frac{\delta}{\delta \alpha^a_{\bm{x}}(-\infty)}
    =\frac{\delta}{\delta \alpha^b_{\bm{x}}(\infty)}
    (\tilde{V}^{\dag}_{\bm{x}})^{ba} =
    \tilde{V}^{ab}_{\bm{x}}
    \frac{\delta}{\delta \alpha^b_{\bm{x}}(\infty)},
    \eeq
so that we can express the functional derivatives at $x^- = -\infty$
in terms of those at $x^- = \infty$. Then we obtain the JIMWLK
Hamiltonian in its ``standard'' form
    \beq\label{AppHK}
    H = -\, \frac{1}{(2\pi)^3}\!
    \int\limits_{\bm{u}\bm{v}\bm{z}}\!
    K_{\bm{u}\bm{v}\bm{z}}\,
    \frac{\delta}{\delta \alpha^a_{\bm{u}}}
    \left[ 1
    +\tilde{V}_{\bm{u}}^{\dag} \tilde{V}_{\bm{v}}
    -\tilde{V}_{\bm{u}}^{\dag} \tilde{V}_{\bm{z}}
    -\tilde{V}_{\bm{z}}^{\dag} \tilde{V}_{\bm{v}}
    \right]^{ab}\!
    \frac{\delta}{\delta \alpha^b_{\bm{v}}},
    \eeq
where the kernel $K_{\bm{u}\bm{v}\bm{z}}$ is given by
    \beq\label{AppK}
    K_{\bm{u}\bm{v}\bm{z}}=
    \frac{(\bm{u}-\bm{z})\dot
    (\bm{z}-\bm{v})}{(\bm{u}-\bm{z})^2(\bm{z}-\bm{v})^2}.
    \eeq
When the Hamiltonian acts on the gauge-invariant observables defined
in Eq.~(\ref{Ogen}) one can let $2 K_{\bm{u}\bm{v}\bm{z}} \to
\cal{M}_{\bm{u}\bm{v}\bm{z}}$ to arrive at the simplified form given
in Eq.~(\ref{HJIMWLK}) \cite{HIIM05}.

\section{The Balitsky Equations}\label{AppBalitsky}

Here we derive the first two Balitsky equations starting from the
JIMWLK Hamiltonian. The latter is given by
    \beq\label{AppHJIMWLK}
    H = -\, \frac{1}{16 \pi^3}\!
    \int\limits_{\bm{u}\bm{v}\bm{z}}\!
    \cal{M}_{\bm{u}\bm{v}\bm{z}}
    \left[ 1
    +\tilde{V}_{\bm{u}}^{\dag} \tilde{V}_{\bm{v}}
    -\tilde{V}_{\bm{u}}^{\dag} \tilde{V}_{\bm{z}}
    -\tilde{V}_{\bm{z}}^{\dag} \tilde{V}_{\bm{v}}
    \right]^{ab}
    \frac{\delta}{\delta \alpha^a_{\bm{u}}}\,
    \frac{\delta}{\delta \alpha^b_{\bm{v}}},
    \eeq
while the evolution equation for a generic operator $\cal{O}$ is
determined by
    \beq\label{Oevol}
    \frac{\del \lan \cal{O} \ran}{\del Y}=
    \lan H \cal{O}\ran.
    \eeq
First we would like to find the action of the JIMWLK Hamiltonian on
the $S$-matrix
    \beq\label{Sxy}
    S_{\bm{x}\bm{y}}=
    \frac{1}{N_c}\, \tr \big(V^{\dag}_{\bm{x}} V_{\bm{y}}\big),
    \eeq
which describes the scattering of a single dipole off the hadronic
target. The action of the functional derivative on the Wilson lines
is
    \beq\label{donVdag}
    &&\frac{\delta}{\delta \alpha^a_{\bm{u}}}\,
    V_{\bm{x}}^{\dag} =
    i g \delta_{\bm{x}\bm{u}}\, t^a V_{\bm{x}}^{\dag},\\
    \label{donV}
    &&\frac{\delta}{\delta \alpha^a_{\bm{u}}}\,
    V_{\bm{x}} =
    - i g  \delta_{\bm{x}\bm{u}} V_{\bm{x}}\, t^a.
    \eeq
Now we easily find that
    \beq\label{ddonVV}
    \frac{\delta}{\delta \alpha^a_{\bm{u}}}\,
    \frac{\delta}{\delta \alpha^b_{\bm{v}}}\,
    S_{\bm{x}\bm{y}} &&\!\!\!\!\!\rightarrow
    \frac{g^2}{N_c}\,
    \left[
    \delta_{\bm{x}\bm{u}} \delta_{\bm{y}\bm{v}}\,
    \tr\big(t^b t^a V_{\bm{x}}^{\dag} V_{\bm{y}} \big)
    +
    \delta_{\bm{x}\bm{v}} \delta_{\bm{y}\bm{u}}\,
    \tr\big(t^a t^b V_{\bm{x}}^{\dag} V_{\bm{y}} \big)
    \right]
    \nn
    &&\!\!\!\!\!\rightarrow
    \frac{2 g^2}{N_c}\,
    \delta_{\bm{x}\bm{u}} \delta_{\bm{y}\bm{v}}\,
    \tr\big(t^b t^a V_{\bm{x}}^{\dag} V_{\bm{y}} \big),
    \eeq
where we have dropped terms proportional to $\delta_{\bm{u}\bm{v}}$
since the kernel $\cal{M}_{\bm{u}\bm{v}\bm{z}}$ vanishes when
$\bm{u}=\bm{v}$, and we have anticipated that both terms in the
square bracket in the above equation will contribute the same. Let
us consider the first term in the square bracket of
Eq.~(\ref{AppHJIMWLK}). Using $t^a t^a = (N_c^2-1)/(2 N_c)$, we find
the first contribution to $H S_{\bm{x}\bm{y}}$
    \beq\label{first}
    {\rm first} =
    -\frac{N_c^2-1}{2 N_c^2}\, \atpi\,
    \int\limits_{\bm{z}}
    \cal{M}_{\bm{x}\bm{y}\bm{z}}\,
    S_{\bm{x}\bm{y}}.
    \eeq
Now consider the contribution coming from the second term in
Eq.~(\ref{AppHJIMWLK}). In order to transform the adjoint Wilson
lines to fundamental ones, we shall make use of
    \beq\label{adjtofun}
    \big(\tilde{V}^{\dag}\big)^{ba} t^b =
    \tilde{V}^{ab} t^b = V^{\dag} t^a\, V.
    \eeq
Then it is straightforward to show that the contribution of the
second term is the same as the one of the first, that is
    \beq\label{second}
    {\rm second} =
    -\frac{N_c^2-1}{2 N_c^2}\, \atpi\,
    \int\limits_{\bm{z}}
    \cal{M}_{\bm{x}\bm{y}\bm{z}}\,
    S_{\bm{x}\bm{y}}.
    \eeq
The third term involves
    \beq
    \big(\tilde{V}_{\bm{x}}^{\dag}\big)^{ac}\,
    \tilde{V}_{\bm{z}}^{cb}\,
    \tr\big(t^b \,t^a\, V_{\bm{x}}^{\dag} V_{\bm{y}} \big) =
    \tr\big(V_{\bm{z}}^{\dag}\, t^c\, V_{\bm{z}}
    V_{\bm{x}}^{\dag} \,t^c\, V_{\bm{y}} \big),
    \eeq
and by using the identity
    \beq\label{Fierz1}
    \tr \big(t^a A \,t^a B \big)
    =\frac{1}{2}\,\tr(A) \tr(B)
    -\frac{1}{2 N_c}\, \tr(AB),
    \eeq
which arises from the Fierz identity
    \beq\label{Fierz}
    \big(t^a\big)^{\i \j}
    \big(t^a\big)^{k\ell}
    =\frac{1}{2}\,
    \delta^{\i \ell} \delta^{\j k}
    -\frac{1}{2 N_c}\,
    \delta^{\i \j} \delta^{k \ell},
    \eeq
we find that the corresponding contribution (which is also equal to
the one coming from the fourth term) is
    \beq\label{third}
    {\rm third} =
    {\rm fourth} =
    \frac{1}{2}\,
    \atpi\,
    \int\limits_{\bm{z}}
    \cal{M}_{\bm{x}\bm{y}\bm{z}}
    \left[S_{\bm{x}\bm{z}} S_{\bm{z}\bm{y}}
    -\frac{1}{N_c^2}\, S_{\bm{x}\bm{y}}\right].
    \eeq
Combining Eqs.~(\ref{first}), (\ref{second}) and
(\ref{third})\footnote{Notice the exact cancelation of the
contributions which are subdominant at large-$N_c$.} we finally
arrive at the first Balitsky equation
    \beq\label{AppBal1}
    \frac{\del \lan S_{\bm{x}\bm{y}} \ran}{\del Y}=
    \atpi\,
    \int\limits_{\bm{z}}
    \cal{M}_{\bm{x}\bm{y}\bm{z}}
    \left[\lan S_{\bm{x}\bm{z}} S_{\bm{z}\bm{y}}\ran
    -\lan S_{\bm{x}\bm{y}} \ran \right].
    \eeq

Now we would like to scatter two dipoles $(\bm{x}_1,\bm{y}_1)$ and
$(\bm{x}_2,\bm{y}_2)$ off the hadron. The corresponding $S$-matrix
is given by
    \beq\label{S12}
    S^{(2)}_{\bm{x}_1\bm{y}_1;\bm{x}_2\bm{y}_2}=
    S_{\bm{x}_1\bm{y}_1} S_{\bm{x}_2\bm{y}_2},
    \eeq
with $S_{\bm{x}\bm{y}}$ the $S$-matrix for a single dipole as
determined by Eq.~(\ref{Sxy}). In order to find the evolution
equation for this operator we need again to act with the JIMWLK
Hamiltonian. When both functional derivatives act on the same
dipole, then the other dipole is just a spectator and the evolution
equation can be trivially obtained as a result of the Leibnitz rule
and the first Balitsky equation. We have
    \beq\label{Bal2lead}
    \frac{\del \lan  S_{\bm{x}_1\bm{y}_1}S_{\bm{x}_2\bm{y}_2} \ran}
    {\del Y}=
    \aln\, \atpi\,
    \int\limits_{\bm{z}}
    \cal{M}_{\bm{x}_1\bm{y}_1\bm{z}}\,
    \lan (S_{\bm{x}_1\bm{z}} S_{\bm{z}\bm{y}_1} \-
    S_{\bm{x}_1\bm{y}_1})
    S_{\bm{x}_2\bm{y}_2}\ran
    \nn
    \aln + \,\{1 \leftrightarrow 2\}
    +\cal{O}\left(N_c^{-2}\right),
    \eeq
where we have anticipated that the remaining terms will be
suppressed at large-$N_c$, something which will be verified in what
follows. Let us consider the terms which arise when each of the
functional derivatives acts on a different dipole. For example,
there will be a term coming from the action of $\delta/\delta
\alpha^a_{\bm{u}}$ on $V^{\dag}_{\bm{x}_1}$ and the action of
$\delta/\delta \alpha^b_{\bm{v}}$ on $V_{\bm{y}_2}$. For this
particular contribution we have
    \beq\label{ddonVVVV}
    \frac{\delta}{\delta \alpha^a_{\bm{u}}}\,
    \frac{\delta}{\delta \alpha^b_{\bm{v}}}\,
    S_{\bm{x}_1\bm{y}_1} S_{\bm{x}_2\bm{y}_2}
    \rightarrow
    \frac{g^2}{N_c}\,
    \delta_{\bm{x}_1\bm{u}} \delta_{\bm{y}_2\bm{v}}\,
    \tr \big(t^a V_{\bm{x}_1}^{\dag} V_{\bm{y}_1} \big)
    \tr \big(t^b V_{\bm{x}_2}^{\dag} V_{\bm{y}_2} \big).
    \eeq
Using the identity
    \beq\label{Fierz2}
    \tr \big(t^a A \big)
    \tr \big(t^a B \big)
    =\frac{1}{2}\,\tr(A B)
    -\frac{1}{2 N_c}\, \tr(A) \tr(B),
    \eeq
which arises from the Fierz identity (\ref{Fierz}) we find that the
contribution of the first term in the square bracket in
Eq.~(\ref{AppHJIMWLK}) is
    \beq\label{firstS2}
    {\rm first} =
    -\frac{1}{4 N_c^3} \atpi
    \int\limits_{\bm{z}}
    \cal{M}_{\bm{x}_1\bm{y}_2\bm{z}}
    \Big[\aln    \tr \big(V^{\dag}_{\bm{x}_1} V_{\bm{y}_1}
    V^{\dag}_{\bm{x}_2} V_{\bm{y}_2}\big)
    \nn
    \aln -\frac{1}{N_c}\,
    \tr \big(V^{\dag}_{\bm{x}_1} V_{\bm{y}_1}\big)
    \tr \big(V^{\dag}_{\bm{x}_2} V_{\bm{y}_2}\big)
    \Big].
    \eeq
Similarly, and by using Eq.~(\ref{adjtofun}), we find that the
contribution of the second term is
    \beq\label{secondS2}
    {\rm second} =
    -\frac{1}{4 N_c^3} \atpi
    \int\limits_{\bm{z}}
    \cal{M}_{\bm{x}_1\bm{y}_2\bm{z}}
    \Big[\aln
    \tr \big(V_{\bm{y}_1} V^{\dag}_{\bm{x}_1}
    V_{\bm{y}_2} V^{\dag}_{\bm{x}_2}\big)
    \nn
    \aln -\frac{1}{N_c}\,
    \tr \big(V^{\dag}_{\bm{x}_1} V_{\bm{y}_1}\big)
    \tr \big(V^{\dag}_{\bm{x}_2} V_{\bm{y}_2}\big)
    \Big],
    \eeq
while the third and the fourth term give
    \beq\label{thirdS2}
    {\rm third} =
    {\rm fourth} =
    \frac{1}{4 N_c^3} \atpi
    \int\limits_{\bm{z}}
    \cal{M}_{\bm{x}_1\bm{y}_2\bm{z}}
    \Big[
    \aln \tr \big(
    V^{\dag}_{\bm{x}_1}
    V_{\bm{z}}
    V^{\dag}_{\bm{x}_2}
    V_{\bm{y}_2}
    V^{\dag}_{\bm{z}}
    V_{\bm{y}_1} \big)
    \nn
    -\frac{1}{N_c}\,
    \aln \tr \big(V^{\dag}_{\bm{x}_1} V_{\bm{y}_1}\big)
    \tr \big(V^{\dag}_{\bm{x}_2} V_{\bm{y}_2}\big)
    \Big].
    \eeq
Putting Eqs.~(\ref{firstS2}), (\ref{secondS2}) and (\ref{thirdS2})
together we obtain
    \beq\label{oneS2}
    \frac{1}{4 N_c^3} \atpi
    \int\limits_{\bm{z}}
    \cal{M}_{\bm{x}_1\bm{y}_2\bm{z}}
    \Big[
    \aln 2\,\tr \big(
    V^{\dag}_{\bm{x}_1}
    V_{\bm{z}}
    V^{\dag}_{\bm{x}_2}
    V_{\bm{y}_2}
    V^{\dag}_{\bm{z}}
    V_{\bm{y}_1} \big)
    - \tr \big(
    V^{\dag}_{\bm{x}_1}
    V_{\bm{y}_1}
    V^{\dag}_{\bm{x}_2}
    V_{\bm{y}_2}\big)
    \nn
    \aln - \tr \big(V_{\bm{y}_1}
    V^{\dag}_{\bm{x}_1}
    V_{\bm{y}_2}
    V^{\dag}_{\bm{x}_2}\big)
    \Big].
    \eeq
There are seven more terms like the one in Eq.~(\ref{oneS2}). Three
terms are obtained when $\delta/\delta \alpha^a_{\bm{u}}$ acts on
the first dipole (and therefore $\delta/\delta \alpha^b_{\bm{v}}$
acts on the second dipole). They can be read from Eq.~(\ref{oneS2})
with the replacement $\cal{M}_{\bm{x}_1\bm{y}_2\bm{z}} \to
\cal{M}_{\bm{y}_1\bm{x}_2\bm{z}}, -\cal{M}_{\bm{x}_1\bm{x}_2\bm{z}},
-\cal{M}_{\bm{y}_1\bm{y}_2\bm{z}}$. The last four terms are obtained
when $\delta/\delta \alpha^a_{\bm{u}}$ acts on the second dipole
(and therefore $\delta/\delta \alpha^b_{\bm{v}}$ acts on the first
dipole) and they can be simply read from the first four terms with
the replacement $1 \leftrightarrow 2$. Putting everything together
we finally find that the non-leading (in the number of colors)
contribution to the evolution of $\lan
S_{\bm{x}_1\bm{y}_1}S_{\bm{x}_2\bm{y}_2} \ran$ is given by
    \beq\label{Bal2nonlead}
    \frac{\del \lan  S_{\bm{x}_1\bm{y}_1}S_{\bm{x}_2\bm{y}_2} \ran}
    {\del Y}\Big|_{\rm NL}=
    \aln\,
    \frac{1}{2 N_c^3}
    \atpi\,
    \int\limits_{\bm{z}}
    \big[
    \cal{M}_{\bm{x}_1\bm{y}_2\bm{z}}
    +\cal{M}_{\bm{y}_1\bm{x}_2\bm{z}}
    \-\cal{M}_{\bm{x}_1\bm{x}_2\bm{z}}
    \-\cal{M}_{\bm{y}_1\bm{y}_2\bm{z}}
    \big]
    \nn
    \aln \times \big[
    \tr \big(
    V^{\dag}_{\bm{x}_1} V_{\bm{z}} V^{\dag}_{\bm{x}_2}
    V_{\bm{y}_2} V^{\dag}_{\bm{z}} V_{\bm{y}_1} \big)
    +
    \tr \big(
    V^{\dag}_{\bm{x}_2} V_{\bm{z}} V^{\dag}_{\bm{x}_1}
    V_{\bm{y}_1} V^{\dag}_{\bm{z}} V_{\bm{y}_2} \big)
    \nn
    \aln -
    \tr \big(
    V^{\dag}_{\bm{x}_1} V_{\bm{y}_1} V^{\dag}_{\bm{x}_2} V_{\bm{y}_2}\big)
    - \tr \big(
    V_{\bm{y}_1} V^{\dag}_{\bm{x}_1} V_{\bm{y}_2} V^{\dag}_{\bm{x}_2}\big)
    \big].
    \eeq
One sees that the terms in Eq.~(\ref{Bal2nonlead}) are of
non-dipolar structure and of order $\cal{O}(1/N_c^2)$ when compared
to the terms in Eq.~(\ref{Bal2lead}), as one had anticipated. The
second Balitsky equation is obtained by identifying the antiquark of
the first dipole with the quark of the second dipole. With the
slight change of notation $\bm{x}_1 \to \bm{x}$, $\bm{y}_1 =
\bm{x}_2 \to \bm{z}$, $\bm{y}_2 \to \bm{y}$ and $\bm{z} \to \bm{w}$,
we can write this second equation as
    \beq\label{AppBal2}
    \!\!\!\!\!\!\frac{\del \lan  S_{\bm{x}\bm{z}}S_{\bm{z}\bm{y}} \ran}
    {\del Y}=
    \atpi\,
    \int\limits_{\bm{w}}
    \aln \cal{M}_{\bm{x}\bm{z}\bm{w}}\,
    \lan (S_{\bm{x}\bm{w}} S_{\bm{w}\bm{z}}- S_{\bm{x}\bm{z}})
    S_{\bm{z}\bm{y}} \ran
    \nn
    \!\!\!\!\!\!+\atpi\,
    \int\limits_{\bm{w}}
    \aln
    \cal{M}_{\bm{z}\bm{y}\bm{w}}\,
    \lan S_{\bm{x}\bm{z}}
    (S_{\bm{z}\bm{w}} S_{\bm{w}\bm{y}} - S_{\bm{z}\bm{y}} )\ran
    \nn
    \!\!\!\!\!\!\aln\hspace{-2.2cm}
    +\frac{1}{2 N_c^2}\,
    \atpi\,
    \int\limits_{\bm{w}}
    (\cal{M}_{\bm{x}\bm{y}\bm{w}}
    \!-\!\cal{M}_{\bm{x}\bm{z}\bm{w}}
    \!-\!\cal{M}_{\bm{z}\bm{y}\bm{w}})
    \lan
    Q_{\bm{x}\bm{z}\bm{w}\bm{y}}+
    Q_{\bm{x}\bm{w}\bm{z}\bm{y}}
    \ran,
    \eeq
where the quadrupole operator $Q$ is
    \beq\label{AppQop}
    Q_{\bm{x}\bm{z}\bm{w}\bm{y}}
    \equiv
    \frac{1}{N_c}
    \left[
    \tr \big( V^{\dag}_{\bm{x}} V_{\bm{w}} V^{\dag}_{\bm{z}}
    V_{\bm{y}} V^{\dag}_{\bm{w}} V_{\bm{z}} \big)
    -\tr \big( V^{\dag}_{\bm{x}} V_{\bm{y}} \big)
    \right].
    \eeq

\section{Solving the BK equation in the High Density Region}\label{AppBKSol}

Let us derive the limiting form of the $S$-matrix for dipole-hadron
scattering in the region $\Lam^2 \ll Q^2 \ll Q_s^2$, with $1/Q$ the
dipole size, as determined from the solution of the BK equation,
that is from the factorized form of Eq.~(\ref{Bal1b}). Since we are
looking for the solution in a regime where the $S$-matrix approaches
its black-disk limit, i.e.~$S \to 0$, we can neglect the quadratic
in $S$ term in the BK equation. In order to do this properly, we
need to restrict the region of integration in the transverse
coordinates according to
    \beq\label{Appbound}
    1/Q_s^2 \ll (\bm{x}-\bm{z})^2,
    (\bm{z}-\bm{y})^2 \ll 1/Q^2.
    \eeq
The lower limit arises from the fact that $Q_s^2$ is the boundary
determining the transition from a region where the scattering is
weak to a region where it becomes strong, while the upper limit
determines the dominant logarithmic contribution to the r.h.s.~of
the BK equation. Thus, we easily find that
    \beq\label{Appdlogs}
    \frac{\del \ln \lan S \ran}{\del Y} =
    -\abar \int\limits_{1/Q_s^2}^{1/Q^2}
    \frac{\dif z^2}{z^2} = -\abar \ln(Q_s^2/Q^2).
    \eeq
Using the leading behavior for the energy dependence of $Q_s$ given
in Eq.~(\ref{lambda}), we can write the derivative with respect to
$Y$ as
    \beq\label{AppddY}
    \frac{\del}{\del Y} =
    \frac{\del \ln(Q_s^2/Q^2)}{\del Y}\,
    \frac{\del}{\del \ln(Q_s^2/Q^2)} =
    \abar\, \frac{\chi(\gamma_s)}{1\-\gamma_s}\,
    \frac{\del}{\del \ln(Q_s^2/Q^2)}.
    \eeq
Then it is trivial to solve Eq.~(\ref{Appdlogs}) to obtain
\cite{MS96,LT00,IM01}
    \beq\label{AppTsat}
    \lan S \ran \approx
    \exp \left[
    -\frac{1-\gamma_s}{2 \chi(\gamma_s)}\,
    \ln^2 \frac{Q_s^2}{Q^2}
    \right].
    \eeq
It is very likely that this form remains the same when one takes
into account the effects of fluctuations, but with a coefficient in
the exponent which is reduced by a factor of 2 \cite{Im04b}.

\section{The Diffusion to the Ultraviolet}\label{Apptwosteps}

In this Appendix we shall show the significance of the high momenta
contributions to the evolution. Since at these high momenta the
system is dilute, we shall deal only with the BFKL equation and for
simplicity we will even suppress the effects of the absorptive IR
boundary. So let us consider the amplitude for a dipole of size
$1/Q$ to scatter off a dipole of size $1/\mu$. The general
(rotationally symmetric and integrated over impact parameter)
solution is given by Eq.~(\ref{BFKLsol}) with the replacement $r \to
1/Q$. Now we would like to find the solution for $\abar Y \gg 1$ and
close to the momentum $Q_0^2$ defined by
    \beq\label{AppQ0}
    Q_0^2(Y) =
    \mu^2 \exp \left[
    \frac{\abar \chi(\gamma_s)}{1-\gamma_s} Y
    \right].
    \eeq
Notice that this is a line parallel to the saturation one if we
neglect the prefactors. In the diffusion approximation, we can write
the solution to the BFKL equation as
    \beq\label{AppBFKLsol}
    T(Q,Y) = \frac{2\pi\alpha_s^2}{\mu^2}\,
    T_0(\gamma_s)
    \left(
    \frac{Q_0^2}{Q^2}
    \right)^{1-\gamma_s}\!\!
    \frac{1}{\sqrt{\pi D_s Y}}\,
    \exp \left[ -\frac{\ln^2(Q^2/Q_0^2)}{D_s Y} \right].
    \eeq
Now let us try to obtain the same solution, by performing two global
evolution steps, from $0$ to $Y_1$ and then from $Y_1$ to $Y$. Then
one can write the solution, which we denote as $\tilde{T}$, in the
form
    \beq\label{AppTbar}
    \tilde{T}(Q,Y)= \aln\int\limits_{0}^{\infty}
    \frac{\dif Q_1^2}{Q_1^2}\,
    T(Q_1,Y_1)
    \nn
    \aln\times \int\limits_{\cal{C}} \frac{\dif \gamma}{2\pi\rmi}
    \exp \left[
    \abar \chi(\gamma) (Y\-Y_1) \-
    (1-\gamma) \ln \left(Q^2/Q_1^2\right)
    \right],
    \eeq
which has a simple interpretation; the amplitude at $Y_1$ and for a
given momentum $Q_1$ is evolved to $Y$, and then an integration is
performed over the initial condition (at $Y_1$). Performing the
integration over $\gamma$ in the diffusion approximation and using
the solution given above in Eq.~(\ref{AppBFKLsol}) with $Q \to Q_1$
and $Y\to Y_1$, we obtain
    \beq\label{AppTbar2}
    \tilde{T}(Q,Y) = \aln
    \frac{2 \pi \alpha_s^2}{\mu^2}\,
    T_0(\gamma_s)
    \left(
    \frac{Q_0^2}{Q^2}
    \right)^{1-\gamma_s}\!\!
    \frac{1}{\sqrt{\pi D_s Y_1}}\,
    \frac{1}{\sqrt{\pi D_s (Y\-Y_1)}}
    \nn
    \aln\times\int\limits_{0}^{\infty}
    \frac{\dif Q_1^2}{Q_1^2}\,
    \exp \left[ -\frac{\ln^2(Q^2/Q_1^2)}{D_s (Y \- Y_1)}
    -\frac{\ln^2(Q_1^2/Q_0^2)}{D_s Y_1}\right].
    \eeq
It is straightforward to show that the integration over $Q_1$ will
lead to a result identical to the one in Eq.~(\ref{AppBFKLsol}). But
say we want to find the region in $Q_1$, that contributes the most
in order to get Eq.~(\ref{AppBFKLsol}). Imposing an ultraviolet
cutoff at $Q_{\rm UV}$ and setting, for example, $Y_1=Y/2$, we find
    \beq\label{AppTratio}
    \frac{\tilde{T}(Q,Y)}{T(Q,Y)} =
    1- \frac{1}{2}\,
    {\rm erfc} \left[
    \frac{2\ln(Q_{\rm UV}^2/Q_0^2) - \ln(Q^2/Q_0^2)}{\sqrt{D_s
    Y}}\right],
    \eeq
where ${\rm erfc}(x)$ is the complimentary error function, for which
we recall that
    \beq\label{erfc}
    {\rm erfc}(x)=
    \begin{cases}
        \displaystyle{2-\frac{\exp(-x^2)}{\sqrt{\pi}x}} &
        \text{ for\,  $x \ll -1$}
        \\*[0.1cm]
        \displaystyle{1} &
        \text{ for\,  $x=0$}
        \\*[0.1cm]
        \displaystyle{\frac{\exp(-x^2)}{\sqrt{\pi}x}} &
        \text{ for\,  $x \gg 1$}.
    \end{cases}
    \eeq
If we want to calculate the amplitude, say close to $Q_0$, the
second term in the argument of the error function can be neglected.
Therefore, the two results $T(Q,Y)$ and $\bar{T}(Q,Y)$ will agree so
long as $\ln(Q_{\rm UV}^2/Q_0^2) \gg \sqrt{D_s Y} $. Thus, even if
at the final rapidity $Y$ we are interested in the amplitude very
close to the ``central line'' $Q_0$, at the intermediate steps of
evolution all values of $Q$ that extend up to the diffusion radius
contribute to the final result.

\section{The Langevin Equation and the Hierarchy}\label{Applangevin}

Let us show in a simplified zero-dimensional case, how a particular
Langevin equation is equivalent to an infinite hierarchy describing
splitting and merging processes. Consider the equation
    \beq\label{langevineq}
    \frac{\dif n}{\dif Y}
    = \alpha\, n -\beta \,n^2 +
    \sqrt{\gamma\, n}\, \nu \equiv A+B\nu,
    \eeq
where $\nu(Y)$ is a Gaussian white noise; $\langle \nu(Y) \rangle
=0$ and $\langle \nu(Y)\,\nu(Y')\rangle = \delta(Y-Y')$, and with
all other higher noise correlators vanishing. This Langevin equation
should be understood with the Ito prescription for the
discretization of time, namely, if one writes $Y = \j \,\epsilon$,
where $\j$ is a non-negative integer and $\epsilon$ the time step,
Eq.~(\ref{langevineq}) should read
    \beq\label{langevindisc}
    \frac{n_{\j+1}-n_{\j}}{\epsilon} =
    A_{\j} + B_{\j}\, \nu_{\j+1},
    \eeq
with $\langle \nu_\j \rangle = 0$ and $\langle \nu_\i \nu_\j \rangle
= (1/\epsilon)\, \delta_{\i\j}$. Then it is a simple exercise to
show that the evolution of the expectation value of an arbitrary
function $F(n)$ is determined by the equation
    \beq\label{dFdY}
    \frac{\dif \lan F(n) \ran}{\dif Y} =
    \lan A F'(n) \ran +
    \frac{1}{2}\, \lan B^2 F''(n) \ran.
    \eeq
With $F(n) = n^{\kappa}$ and with $A$ and $B$ as defined in
Eq.~(\ref{langevineq}), the above equation leads to
    \beq\label{nkevol}
    \frac{\dif \lan n^{\kappa}\ran}{\dif Y} =
    \alpha\, \kappa\, \lan n^{\kappa} \ran
    -\beta\, \kappa\, \lan n^{\kappa+1} \ran
    + \gamma\,\frac{\kappa (\kappa\-1)}{2}\,\lan n^{\kappa-1} \ran.
    \eeq
The generalization to the, two-dimensional and with non-local
vertices, large-$N_c$ QCD case is straightforward.

\nocite{BT04}

%\bibliographystyle{unsrt}
%\bibliography{refs}

\end{document}